\documentclass[11pt]{article}
\usepackage{epstopdf}   
\usepackage[normalem]{ulem}                                                    
\usepackage[a4paper,hmarginratio=1:1,vmarginratio=2:3,totalwidth=15.2cm,totalheight=22.6cm]{geometry}
\usepackage{bm,epstopdf,epsfig,amsmath,amssymb,
amsfonts,colordvi,wrapfig,comment,cancel,verbatim,slashed}
\usepackage{graphicx,graphics}
\usepackage[font=md,captionskip=8pt]{subfig}
\usepackage[usenames,dvipsnames]{color}
\usepackage[noadjust]{cite}
\usepackage{xcolor} 
\usepackage[utf8]{inputenc}
\usepackage{setspace}

\usepackage[utf8]{inputenc}

\newcommand{\w}{b}  
\newcommand{\Om}{\omega}  

\newcommand{\genG}{\mathsf\Gamma}
\newcommand{\genO}{\mathsf w}

\numberwithin{equation}{section}

 \usepackage{hyperref}

\long\def\symbolfootnote[#1]#2{\begingroup
\def\thefootnote{\fnsymbol{footnote}}\footnote[#1]{#2}\endgroup}

\begin{document} 
\begin{flushright}
\end{flushright}

\thispagestyle{empty}
\vspace{3.cm}
\begin{center}

%

%
%
 
%
%

  {\Large \bf  The Gauge Theory of Weyl Group and its Interpretation \\ \vspace{0.3cm} as Weyl Quadratic Gravity
}  

 \vspace{1.5cm}
 
 {\bf  Cezar Condeescu}\,\,   and  {\bf  Andrei Micu}\,\,
 \symbolfootnote[1]{E-mail: ccezar@theory.nipne.ro, amicu@theory.nipne.ro.}
 
\bigskip\bigskip

{\small Department of Theoretical Physics, National Institute of Physics
 \smallskip 

 and  Nuclear Engineering (IFIN - HH), Bucharest, 077125 Romania}
\end{center}

\medskip

\begin{abstract}
  \begin{spacing}{0.99}
    \noindent
    \newline In this paper we give an extensive description of Weyl quadratic gravity as the gauge theory of the Weyl group. The previously discovered (vectorial) torsion/non-metricity equivalence is shown to be built-in as it corresponds to a redefinition of the generators of the Weyl group. We present a generalisation of the torsion/non-metricity duality which includes, aside from the vector, also a traceless 3-tensor with two antisymmetric indices and vanishing skew symmetric part. 
    A discussion of this relation in the case of minimally coupled matter fields is given. We further point out that a Rarita-Schwinger field can couple minimally to all the components of torsion and some components of non-metricity. Alongside we present the same gauge construction for the Poincar\'e and conformal groups. We show that even though the Weyl group is a subgroup of the conformal group, the gauge theory of the latter is actually only a special case of Weyl quadratic gravity. 
\end{spacing}
\end{abstract}

\newpage

\tableofcontents

\pagebreak 

\section{Introduction}
\label{sec1}

Einstein's theory of General Relativity (GR) has been very successful in predicting/explaining a  plethora of gravitational phenomena from the deviation of light in gravitational fields and black-holes to the recently detected gravitational waves. It is the cornerstone of the standard model of big-bang cosmology. Nonetheless there are reasons to go beyond it, either in finding a UV completion of the theory or in trying to account in a different way for the observed dark matter/dark energy. There are many generalisations possible for GR. From a theoretical point of view it is based on the unique symmetric connection that is metric compatible ({\it i.e.} the Levi-Civita connection) and the corresponding (pseudo-)Riemannian curvature tensor. In general, an affine connection need not be symmetric nor metric compatible, the departures from the two being measured by the torsion and non-metricity tensors, respectively. The framework for working with a (possibly dynamical) connection with torsion and non-metricity has been set in the formulation of Metric Affine Gravity (MAG); see \cite{Hehl:1994ue, Iosifidis:2019dua} for reviews. There has been a good deal of activity recently on the particular case of teleparallel gravities where one sets the curvature to zero. It comes essentially in two variants, Teleparallel Gravity (TG) with only non-zero torsion  and Symmetric Teleparallel Gravity (STG) with only non-zero non-metricity  (see \cite{Bahamonde:2021gfp} for a review including applications to cosmology). This have lead to the interesting finding of the (classically) equivalent reformulations of GR in terms of pure torsion or pure non-metricity \cite{Trinity}. 

On the other hand, in high energy physics, gauge theories have had tremendous success in describing the electromagnetic, weak and strong interactions and are fundamental in the Standard Model of particle physics. Interestingly, gravity theories can also be formulated as gauge theories of space-time symmetries, though often (as is the case with GR) with non-standard Lagrangians ({\it i.e.} not necessarily quadratic in the gauge curvatures). In this sense, a gauge principle is behind the descriptions of all known interactions. The Coleman-Mandula theorem \cite{Coleman:1967ad}, under certain assumptions, restricts the list of possible space-time (bosonic) gauge symmetries to three possibilities: the Poincar\'e group, the Weyl group\footnote{The Weyl group comprises the transformations of the Poincar\'e group extended with dilatations (but not including special conformal transformations).} and the (full) conformal group. Fermionic space-time symmetries \cite{Haag:1974qh} constitute an exception to the theorem and thus, supersymmetric extensions of the three groups are also allowed as space-time gauge symmetries of the S-matrix, yielding various supergravity theories. The gauging of the Poincar\'e group would yield naturally a theory of quadratic gravity (see \cite{Utiyama,Kibble} for early work and \cite{Alvarez-Gaume:2015rwa,Obukhov:2018bmf,Salvio:2018crh} for more recent accounts), if one assumes a standard gauge theory Lagrangian. They have the interesting property of being renormalisable \cite{Stelle:1976gc} but however suffer in general from unitarity problems, usually ghosts arising from the presence of higher derivatives. It remains to be seen if one can live with them \cite{Hawking:2001yt} or if solutions can be found to eliminate the ghosts \cite{Bender:2007wu,Maldacena:2011mk,Mannheim:2021oat,Hell:2023rbf}. 

Historically speaking, the first gauge theory ever written is attributed to Weyl \cite{Weyl1,Weyl2,Weyl3} and corresponds to the gauging of the Weyl group \cite{Charap,Bregman} (see also \cite{Lasenby:2015dba,Scholz,DGnew}, and the thesis \cite{Jia:2024ujz} for reviews). This theory (Weyl quadratic gravity) will be the main focus of our paper though for comparison purposes we will deal with both the Poincar\'e and the conformal cases. Traditionally, Weyl gravity has been formulated in terms of a (dilatation or local scale) gauge invariant symmetric connection which depends, in addition to the metric, on the gauge field of dilatations. It turns out that the corresponding connection has vectorial non-metricity (semi-metricity) which is rather cumbersome to work with. The idea that the gauge field of dilatations could actually be instead identified  with vectorial torsion has been mentioned, for instance, in \cite{Nieh:1981xk,Obukhov:1982zn} or more recently in \cite{Karananas:2015eha,Klemm,Sauro:2022hoh, Barker:2024goa}. 
The equivalence between vectorial non-metricity and vectorial torsion in Weyl gravity has been shown in \cite{Condeescu} by explicitly constructing the action in the two pictures. A projective transformation relates the two connections and the action is actually invariant (up to a redefinition of the dimensionless couplings). On one hand, our paper can be seen as an extended version of \cite{Condeescu} including now all the details and subtleties of gauging space-time symmetries. On the other hand, we extend the previous results in various ways. First, we show here  that the vectorial torsion/non-metricity equivalence/duality is built-in the theory. This happens because the projective transformation that relates the two connections is equivalent at the level of the gauge algebra to a redefinition of the generators, and hence one is guaranteed {\it apriori} to construct the same gauge theory. The presence of dilatations is essential here, as in general torsion and non-metricity have different properties. We want to stress that this equivalence is different than \cite{Trinity} as in our case the curvature tensor is always non-zero\footnote{In our case, as in general relativity, curvature encodes the dynamics of the gravitational field, while torsion or non-metricity describe different degrees of freedom which have their own kinetic terms. }
and the action quadratic.
Moreover, we also consider minimal couplings to matter fields. We show here that a spin $3/2$ fermion (in contrast to spin $1/2$) can actually couple to the dilatation gauge field via the spin connection. Additional dilatation covariant torsional degrees of freedom can be added to the theory via the constraint on the curvature of translations which fixes the connection, as happens for example in supergravity. In the context of this generalised Weyl gravity we show that a more general torsion/non-metricity duality is possible. In addition to a vector, the decompositions of torsion and non-metricity have another irreducible component in common, that of a traceless three index two form with zero completely antisymmetric part. The two components, we mentioned can be mapped one into the other by a transformation that generalises the projective transformation. Starting from a non-metric spin connection one simply defines the torsion spin connection as the antisymmetric part of the non-metric connection. The first Cartan structure equation can equivalently be written in terms of the two connections (non-metric and torsion). 
 
In order to make the case for Weyl gravity, we mentioned here the phenomenological potential of the theory. It has been shown that it is spontaneously broken to Einstein gravity \cite{Ghilen0,Ghilen1} via a Stueckelberg mechanism with a massive dilatation gauge field and generating dynamically the Planck mass and the cosmological constant. The Standard Model can be embedded in Weyl geometry without additional degrees of freedom \cite{SMW}. Applications to inflation \cite{WI3,WI1,Prokopec-inflation, Lucat}, galaxy rotation curves \cite{Craciun:2023bmu,Burikham:2023bil } and black-hole solutions \cite{Yang:2022icz} have been studied/found. Finally, it is anomaly free \cite{DG1} and thus, one can take it seriously at the quantum level (see also \cite{Englert, Reiher, Prokopec, DGnew}). As mentioned before, being a higher derivative theory, Weyl gravity is prone to having ghosts in its spectrum. On the other hand, it is already known that Weyl gauge symmetry improves the behavior in this respect as the ghost mode propagated by $R^2$ is eaten in the Stueckelberg mechanism which gives a mass to the Weyl boson \cite{Ghilen0,Ghilen1}. However, a proper analysis of the full theory including the Weyl tensor squared has not been done. It is likely that this term gives rise to ghosts, though in the case of quadratic conformal gravity, based solely on this term, there are claims otherwise \cite{Mannheim:2021oat}.

The dilatation symmetry is very powerful in the sense that the resulting quadratic Lagrangian contains all the terms compatible with the symmetry. This is to be contrasted with the Poincar\'e case where an infinity of terms of arbitrary high powers of curvatures and derivatives are allowed by the symmetry. Even if not present classically, one expects them to be generated at the quantum level. We make a clear requirement that local symmetries should have associated propagating gauge bosons in order to be true gauge theories. This is to be contrasted with the cases were one merely imposes a (local) symmetry \cite{Edery:2014nha, Karananas:2021gco}. We expect that in quantum gravity all symmetries are truly gauged.\footnote{It was recently realised in \cite{Noris} that restricted (Weyl) symmetries as discussed in \cite{Edery:2014nha} are not really consistent (non-invariant equations of motion) and one needs to introduce gauge fields for all local symmetries.}

We also discuss in detail the gauging of the full conformal group \cite{KTN1,KTN2} which contains special conformal transformations in addition to Poincar\'e and dilatations. Seen as a gauge theory, conformal gravity is peculiar since, even though it has more generators, it actually has less propagating gauge bosons than Weyl gravity. Both the dilatation and special conformal gauge fields turn out to be non-dynamical \cite{KTN1, KTN2,Wheeler2} and thus, the gauged version is equivalent to merely imposing the conformal symmetry on a purely metric Lagrangian. Thus, it is not a true gauge theory. We also show in our paper that conformal gravity can be obtained from Weyl gravity by taking the pure gauge limit (for dilatations). Hence Weyl theory is more general and contains conformal gravity as a limit. It is the only true gauge theory that implements Weyl invariance (defined as local rescalings of the metric).        

Our paper is organised as follows. In Section \ref{sec2} we briefly review general aspects of gauge theory which we then apply to the Poincar\'e group as a warm-up exercise. The main part of the section is dedicated to Weyl gravity which is analysed in detail in the two frames with torsion and non-metricity both on the tangent space and in space-time. The manifestly covariant formulation that connects naturally the two is also considered here. In Section \ref{sec3} we apply gauge theory methods to the conformal group and compare with the Weyl case. In Section \ref{sec4} we consider a generalized version of Weyl gravity with a connection containing covariant torsion in addition. We show that a more general torsion non-metricity duality is possible in this framework. In the end we present our conclusions.

In the paper, most of the notations are introduced along the way. There is a general pattern for symbols we use, which we briefly explain here, but which is presented in more details in the Appendix. Purely Riemannian quantities are denoted by a {\it ring}, $\mathring{}$, accent, $\mathring X$. Objects appearing in the torsion picture, detailed below, are \emph{bare} objects, i.e. $X$, while a {\it tilde}, $\tilde{}$, is used to denote objects in the non-metric formulation, $\tilde X$. Finally a {\it hat}, $\hat{}$, is used to denote gauge covariant (Weyl or conformal) objects, $\hat X$. Covariant derivatives are denoted by $\nabla$ in space-time and by $D$ on the tangent space, in each case with the accents explained above depending on the explicit picture we work in. A glossary for a quick reminder of the symbols and notations is given at the end in Appendix \ref{glossary}.

\section{Gauge theory of the Weyl group}
\label{sec2}

\subsection{General aspects of gauge theories}

In this section we present briefly the general formalism of gauge theories which we shall apply later for the case of (bosonic) space-time gauge symmetries allowed by the Coleman-Mandula theorem \cite{Coleman:1967ad}: the Poincar\'e group, the Weyl group and the conformal group with the main focus on the Weyl case.

Consider a (matrix) Lie group $G$ with associated Lie algebra $\mathfrak{g}$. The generators $T_A$ of the algebra $\mathfrak{g}$ satisfy the general commutation relations 
\begin{equation}
  [T_A,T_B] = f_{AB}{}^C T_C \, ,
\end{equation}
with structure constants $f_{AB}{}^C$. The gauge field $B$ is a Lie algebra valued one-form defining a connection on the associated $G$-bundle
\begin{equation}
B = B^AT_A = B_\mu^A T_A dx^\mu \, .
\end{equation}   
An element $h \in G$ of the Lie group induces a finite gauge transformation having the following expression
\begin{equation}
B\mapsto B' = hBh^{-1}+h dh^{-1} \, .
\label{gaugeB}
\end{equation}
The gauge covariant field strength $R$ is the $\mathfrak{g}$-valued two-form curvature of the bundle defined from the connection $B$ as
\begin{equation}
R= dB+B \wedge B \, .
\label{curvatureR}
\end{equation}
It is easy to see from eqs.\,\eqref{gaugeB} and \eqref{curvatureR} that under a gauge transformation defined by $h$ we have
\begin{equation}
R\mapsto R' = h Rh^{-1} \, .
\label{gaugeR}
\end{equation}
In practice, one makes use of the infinitesimal versions of eqs.\,\eqref{gaugeB} and \eqref{gaugeR}\footnote{Our conventions about gauge theories are similar to \cite{Freedman} but with opposite signs for the gauge transformations. More details about the formalism of gauging space-time symmetries and subtleties related to local translations can be found in the same reference.}.  For this purpose, consider $h$ of the form $h=e^{\epsilon^A T_A}$  (thus an element of the connected component of the identity) and expand to linear order in the gauge functions $\epsilon^A$ to obtain
\begin{align}
\delta_\epsilon B &= - d\epsilon - [B,\epsilon]\, ,  & \delta_\epsilon R = [\epsilon,R] \, ,
\end{align}
where $\epsilon = \epsilon^A T_A$. It is convenient in applications to pass to components. To summarise, for every generator $T_A$ we have introduced a gauge field $B_\mu^A$ which under an infinitesimal variation $\delta_{\epsilon}$ transforms as 
\begin{equation}
  \delta_\epsilon B_\mu^A = - \partial_\mu \epsilon^A
  - \epsilon^C B_\mu^B f_{BC}{}^A \, .
\label{gauge-var}
\end{equation}
The gauge covariant field strengths (curvatures) $R= \frac12 R_{\mu \nu}^A T_A dx^\mu \wedge dx^\nu$ (which we denote sometimes with $R_{\mu \nu}(T^A)$) have the following expressions in components
\begin{equation}
  R_{\mu \nu}^A = 2 \partial_{[\mu} B_{\nu]}^A + B_\nu^C B_\mu^B f_{BC}{}^A \, .
\label{field-strength}
\end{equation}
It is not hard to check from \eqref{gaugeR} that
these field strengths transform infinitesimally as
\begin{equation}
  \delta_\epsilon R_{\mu \nu}^A  =-\epsilon^C R_{\mu \nu}^B f_{BC}{}^A \, .
\label{fs-var}
\end{equation}
At this level, an action for a pure gauge theory quadratic in the curvatures can be constructed as \cite{KTN2}
\begin{equation}
  \label{gauge-action}
  S_{gauge} = \int d^dx  \sqrt{g}\,  Q^{\mu \nu \rho \sigma}_{AB} R_{\mu \nu}^A R^B_{\rho \sigma}\, ,
\end{equation}
where $g:= |\det g_{\mu \nu}|$ and $Q^{\mu\nu\rho \sigma}_{AB}$ depends on the theory considered. Since the groups that we consider always contain Poincar\'e as a subgroup, $Q$ can be, for example, of the form $Q_{AB}^{\mu \nu \rho \sigma} \sim g^{\mu \rho} g^{\nu \sigma} \mathcal{C}_{AB}$ or $Q_{AB}^{\mu \nu \rho \sigma} \sim \epsilon^{\mu \nu \rho \sigma} \mathcal C_{AB}$ with $C_{AB} = f_{AC}{}^D f_{BD}{}^C$ being the Cartan bilinear form. Notice that other terms beyond the Yang-Mills ones are possible for gravity, for example $R^2$ (with $R$ the Ricci scalar) is obtained from $Q_{ab,cd}^{\mu \nu \rho \sigma} \sim e_{[a}^\mu e_{b]}^\nu e_{[c}^\rho e_{d]}^\sigma$ where we employed the adjoint indices for the Lorentz group $A = [ab], B=[cd]$.

One can couple additionally matter fields to the pure gauge action above. When this is done via the gauge covariant derivative, denoted by $\mathcal D_\mu$, it is called minimal coupling. Let $\phi^i$ be fields (quantities) that transform covariantly, that is  
\begin{equation}
  \label{matter-trsf}
  \delta_\epsilon \phi^i = \epsilon^A T_A\phi^i \, , 
\end{equation}
then the corresponding covariant derivatives for these fields are defined as
\begin{equation}
  \label{gauge-derivative}
  \mathcal D_\mu \phi^i  = \partial_\mu \phi^i+ B_\mu^A T_A \phi^i \, ,
\end{equation}
where the generators $T_A$ will be taken in the appropriate representation under which the fields $\phi^i$ transform. The commutator of two covariant derivatives will be proportional to the curvature
\begin{equation}
[\mathcal D_\mu, \mathcal D_\nu] \phi^i = R_{\mu \nu}^A T_A \phi^i \, .
\end{equation}
One can actually take the equation above as the definition of the field strengths. Notice also that $\mathcal D_\mu$ is well defined on any covariant quantity, in particular the field strengths. By acting on them one can obtain the Bianchi identity
\begin{equation}
\label{gauge-bianchi}
\mathcal D_{[\mu} R_{\nu \rho]}^A = 0\, .
\end{equation}
As advertised, we are interested in space-time gauge symmetries which include local translations. These need to be treated differently, in that we shall use a covariant derivative $\hat{\mathcal D}_\mu$ which does not covariantise with respect to local translations. To understand this, note that for a scalar field which is a singlet under all other symmetries except translations, the covariant derivative introduced in \eqref{gauge-derivative} is identically zero \cite{Freedman}.
One can see indeed that we have
\begin{align}
\mathcal D_\mu \phi &= 0\, , & \hat{\mathcal D}_\mu \phi = \partial_\mu \phi \neq 0 \, .
\end{align}
Moreover, the constraint we shall impose later on the field strength of translations will implement general coordinate transformations instead of local translations.
The two covariant derivatives above are, in general, related by
the generator of translations $P_a$ (in our case $P_a= - \partial_a)$
\begin{equation}
  \hat{\mathcal D}_\mu := \mathcal D_\mu - e_\mu^a P_a \equiv \hat D_\mu \, ,
\end{equation}
where we anticipated that the derivative $\hat{\mathcal D}_\mu$ is going to be the gauge covariant derivative on the tangent space $\hat D_\mu$.

\subsection{Gauge theory of the Poincar\'e group}

Gauging space-time symmetries has its motivation in trying to interpret/construct
general relativity and other theories of gravity as gauge theories and thus, bringing them closer to the Standard Model of particle physics. Even though this works in principle, for theories of gravity a number of subtleties arise when considered as gauge theories. As a warm-up exercise we first illustrate them on the well-known example of gauging the Poincar\'e symmetry.\footnote{A new approach in this direction was recently put forward in \cite{Addazi}.}
The algebra of the Poincar\'e group is given by the commutation relations
\begin{equation}
\begin{split}
  \label{Poincare}
  &  [P_{a}, M_{bc}] =  \eta_{a b} P_{c} - \eta_{ac}P_b\, , 
  \qquad [P_a, P_b] = 0 \, ,  \\
& [M_{ab}, M_{cd}] = \eta_{a c}M_{d b} - \eta_{bc}M_{da} - \eta_{ad} M_{cb} + \eta_{bd} M_{ca} \, ,
\end{split}
\end{equation}
where $P_a$ denotes the generator of translations and
$M_{ab} = M_{[ab]}$ is the generator of Lorentz transformations and $\eta_{ab} = \text{diag}(+,-,\ldots,-)$  is the Minkowski metric. We
define these generators on the tangent space (otherwise one would have to work with structure functions) and therefore all the
latin indices $(a,b, \ldots)$ denote tangent space indices. The non-zero structure constants can be read from the above commutators
\begin{equation}
\label{P-structure}
f_{[ab][cd]}{}^{[ef]} = 2 \left(\eta_{cb} \delta^{ef}_{ad} - \eta_{db} \delta^{ef}_{ac}
  - \eta_{ca} \delta^{ef}_{bd} + \eta_{da} \delta^{ef}_{bc} \right)\, , \qquad
f_{a[bc]}{}^d  =  \eta_{ab} \delta_{c}^d - \eta_{ac} \delta_b^d \, .
\end{equation}
Above, for brevity, we denoted
$\delta^{ab}_{cd} \equiv \tfrac12(\delta^a_c \delta^b_d - \delta^b_c
\delta^a_d)$. As we have seen in the previous section, for every generator of the group we associate a gauge field. Thus, for
translations $P_a$ we introduce the field $e^a_\mu$ which will be identified
{\it later on} with the vielbein, while for the Lorentz transformations $M_{ab}$ we
introduce $\Om_\mu{}^{ab}$ which we identify with the spin connection. 
Under infinitesimal variations $\delta_\epsilon$ these fields transform according to \eqref{gauge-var}
\begin{equation}
  \label{eovar}
  \begin{aligned}
    \delta_\epsilon e_\mu^a & = - \partial_\mu \xi^a - \Om_\mu{}^a{}_b \xi^b
    - \lambda^a{}_b e_\mu^b = -D_\mu \xi^a - \lambda^a{}_b e_\mu^b\, , \\
    \delta_\epsilon \Om_\mu{}^{ab} & = - \partial_\mu \lambda^{ab} - \lambda^a{}_c \Om_\mu{}^{cb} \, ,
  \end{aligned}
\end{equation}
where $\epsilon = \xi^a P_a + \frac12 \lambda^{ab} M_{ab}$ and thus $\xi^a$ is the parameter for local translations while $\lambda^{ab}$ is the parameter for local Lorentz transformations. Above, we made use of the usual Lorentz covariant derivative $D_\mu$ on the tangent space defined by $\Om_\mu{}^{ab}$, which in the case of the Poincar\'e algebra coincides with the gauge covariant derivative $\hat D_\mu$ (where as mentioned in the previous section, local translations are left out). The curvatures of these fields can be constructed as in \eqref{field-strength}
\begin{align}
   R_{\mu \nu}(P^a) &= \partial_\mu e_\nu^a - \partial_\nu e_\mu^a
  + \Om_\mu{}^{ab} e_{\nu b} -  \Om_\nu{}^{ab} e_{\mu b} \, , \label{p-1} \\
  R_{\mu \nu}(M^{ab}) &= \partial_\mu \Om_\nu{}^{ab} -
  \partial_\nu \Om_\mu{}^{ab} + \Om_\mu{}^a{}_c \Om_\nu{}^{cb}
  - \Om_\nu{}^a{}_c \Om_\mu{}^{cb} \label{p-2} \, ,
\end{align}
and, under \eqref{eovar} are guaranteed to transform covariantly
\eqref{fs-var}
\begin{equation}
\label{variation-poincare}
  \begin{aligned}
    \delta_\epsilon R_{\mu \nu}(P^a) & = -\xi_b R_{\mu \nu}(M^{ab}) 
   + \lambda^a{}_b R_{\mu \nu}(P^b)\, , \\
    \delta_\epsilon R_{\mu \nu}(M^{ab}) & =  - 2 \lambda^{[a}{}_c R_{\mu \nu}(M^{b] c})\, .
  \end{aligned}
\end{equation}

We can now make contact with gravity theories. The two curvatures that we have introduced have a clear interpretation in the Cartan formalism. First, $R_{\mu \nu}(P^a)$ is nothing but the definition of the
torsion tensor,  $D e^a = T^a$. Second, $R_{\mu \nu}(M^{ab})$ is precisely the ``Riemann'' curvature tensor $R^{ab}{}_{\mu \nu}$ expressed in terms of the spin connection. Hence we have the following identifications
\begin{align}
R_{\mu \nu}(P^a) &= T_{\mu \nu}{}^a\, , & R_{\mu\nu}(M^{ab}) = R^{ab}{}_{\mu \nu}\, .
\label{constraint-torsion}
\end{align}
In general relativity, the torsion is taken to be identically zero. Therefore, in order to reproduce it by the gauging of the Poincar\'e group one needs to impose the following constraint on the gauge curvature of local translations
\begin{equation}
  \label{constraint}
  R_{\mu \nu}(P^a) = 0 \, .
\end{equation}
This can be solved in the usual way for the spin connection and we
find the well-known solution
\begin{equation}
  \label{spinLC}
  \mathring \Om_\mu{}^{ab} = 2e^{\nu[a} \partial_{[\mu} e_{\nu]}^{b]}
  - e^{\nu[a}e^{b]\sigma}e_{\mu c} \partial_\nu e_\sigma^c \, ,
\end{equation}
which via the standard vielbein postulate \eqref{postulate} written as $\mathring \Gamma_{\mu \nu}^\rho = e_a^\rho \mathring D_\mu e_\nu^a$ corresponds to the Levi-Civita connection
\begin{equation}
\label{Levi-Civita}
\mathring \Gamma_{\mu \nu}^\rho = \frac12 g^{\rho \lambda} (\partial_\mu g_{\nu \lambda} + \partial_\nu g_{\mu \lambda} - \partial_\lambda g_{\mu \nu}) \, .
\end{equation}
Alternatively, one could keep the spin connection as a dynamical field and solve its equation of motion (as in Palatini gravity). This, however would have to be done case by case (adding matter fields would in general modify the spin connection in this approach). Practically, it is much more convenient to impose a constraint independently of the theory. Aside from fixing the spin connection, imposing the constraint \eqref{constraint} or the more general one with torsion in \eqref{constraint-torsion} allows us to express local translations in terms of general coordinate transformations with parameters
$\xi^\mu = e_a^\mu \xi^a$. Indeed, starting from the variation of $e_\mu^a$ in \eqref{eovar} under local translations and using, for the time being, the general constraint with torsion \eqref{constraint-torsion} written equivalently as $D_\mu e_\nu^a - D_\nu e_\mu^a = T_{\mu \nu}{}^a$, we find
\begin{equation}
  \label{gcte}
   \delta_\xi e_\mu^a =- D_\mu (e_\nu^a \xi^\nu ) = -\xi^\nu D_\nu e_\mu^a
  - e_\nu^a  D_\mu \xi^\nu - \xi^\nu T_{\mu \nu}{}^a \equiv \delta_{cgct} e_\mu^a \, ,
\end{equation}
where we defined the notion of covariant general coordinate transformation $\delta_{cgct}$ (also called (gauge) covariant diffeomorphism). These are actually the symmetries implemented into the theory rather than the local translations we started with, in \eqref{eovar}. Notice that the variation above is different from the usual Lie derivative $\mathcal L_\xi e_\mu^a = - \xi^\nu \partial_\nu e_\mu^a - e_\nu^a \partial_\mu \xi^\nu$ and it corresponds indeed to a covariantised version ($D_\mu$ instead of $\partial_\mu$) in the presence of torsion. It is instructive to understand the transformation above when the torsion is set to zero. For $T_{\mu \nu}{}^a=0$, equation \eqref{gcte} is nothing but a combination between a standard diffeomorphism, implemented by the Lie derivative $\mathcal L_\xi e_\mu^a$ and a Lorentz transformation $\Lambda^a{}_b :=-\xi^\nu \Om_\nu{}^a{}_b$.
Thus {\it after} imposing the constraint \eqref{constraint-torsion} one can identify the gauge field of translations with the vielbein.

In the tangent space approach the metric is a composite field  $g_{\mu \nu} = \eta_{ab} e_\mu^a e_\nu^b$  and  its variation under covariant general coordinate transformations is 
\begin{equation}
  \label{gctg}
  \delta_\xi g_{\mu \nu}= - \xi^\rho  D_\rho g_{\mu \nu}
  - g_{\rho \nu}  D_\mu \xi^\rho - g_{\mu \rho} D_\nu \xi^\rho- \xi^\rho T_{\mu \rho \nu} - \xi^\rho T_{\nu \rho \mu}
  = - \nabla_\mu \xi_\nu
  -  \nabla_\nu \xi_\mu
  \, ,
\end{equation}
where in the last term we introduced a space-time covariant derivative with non-zero torsion $\nabla_\mu$ arising as the solution of the constraint \eqref{constraint-torsion} and $D_\mu g_{\nu \rho} = \partial_\mu g_{\nu \rho}$. If one sets the torsion $T_{\mu \nu}{}^a$ to zero, and thus in the context of Einstein's general relativity, the covariant diffeomorphism of the metric coincides precisely with the transformation under standard diffeomorphisms given by the Lie derivative
\begin{equation}
\delta_\xi g_{\mu \nu} = \mathcal L_\xi g_{\mu \nu} = - \mathring \nabla_\mu \xi_\nu - \mathring \nabla_\nu \xi_\mu \, ,
\end{equation} 
where now the space-time covariant derivative $\mathring \nabla$ is defined by the Levi-Civita connection.

Finally, let us write down the Bianchi identities for the Poincar\'e gauging resulting from eq.\,\eqref{gauge-bianchi}. An important point to stress here is that one has to use the gauge covariant derivative $\mathcal D_\mu$ which includes also a covariantisation with respect to translations. From the variations of the field strengths \eqref{variation-poincare} it is easy to see that $\mathcal D_\mu$ acts as
\begin{align}
\mathcal D_\mu  R_{\nu \rho}(P^a) &= \partial_\mu R_{\nu \rho}(P^a) - e_\mu^b R_{\nu \rho}(M^a{}_b)+\Om_\mu{}^a{}_b R_{\nu \rho}(P^b)\, ,\\
\mathcal D_\mu R_{\nu \rho}(M^{ab}) & = \partial_\mu R_{\nu \rho}(M^{ab}) - 2 \Om_\mu{}^{[a}{}_c R_{\nu \rho}(M^{b]c})\, .
\end{align}
After employing the notation in eq.\,\eqref{constraint-torsion}, the Bianchi identities \eqref{gauge-bianchi} in this case acquire the familiar form
\begin{align}
\label{bianchi-1}
R^a{}_{[\mu \nu \rho]} &= D_{[\mu} T_{\nu \rho]}{}^a\, ,  & D_{[\mu} R^{ab}{}_{\nu \rho]} & = 0 \, ,
\end{align}
in the Cartan formalism where $D_\mu$ is the tangent space covariant derivative. The equations above can also be written in terms of the space-time covariant derivative $\nabla_\mu$ with torsion $T_{\mu \nu}{}^\rho$. One finds
\begin{align}
R^\sigma{}_{ [\mu \nu \rho]} & =\nabla_{[\mu} T_{\nu \rho]}{}^\sigma - T_{[\mu \nu}{}^\tau T_{\rho] \tau }{}^\sigma\, , &  \nabla_{[\mu} R^{\sigma \tau}{}_{\nu \rho]} &= T_{[\mu \nu}{}^\lambda R^{\sigma \tau}{}_{\rho] \lambda} \, .
\end{align}
Hence, we have shown how one can recover general relativity or more general gravity theories with torsion from gauging the Poincar\'e group. We have not mentioned a Lagrangian up to this point. General relativity is obtained from the Ricci scalar $\mathring R$ and thus its Lagrangian is linear in the curvature. In a standard gauge theory one would naturally build a theory quadratic in the curvatures as in eq.\,\eqref{gauge-action}. Proceeding in this way one can construct quadratic gravity as a gauge theory which in turn is actually renormalisable \cite{Stelle:1976gc} but non-unitary (due to the presence of higher derivatives). Taking into account the symmetries of the Riemann tensor \eqref{Rcomb} and the possible independent terms which can be written out of it \eqref{Riemann2}, the most general quadratic Lagrangian which can be written has the form \cite{Lasenby:2015dba}
\begin{equation}
  \label{actionP}
  \begin{split}
  S_p &= \int d^d x \sqrt{g}\,  \big( \alpha_1 R_{\mu \nu \rho \sigma} R^{\mu \nu \rho \sigma} + \alpha_2 R_{\mu \nu \rho \sigma} R^{\mu \sigma \nu \rho} + \alpha_3 R_{\mu \nu \rho \sigma} R^{\rho \sigma \mu \nu}  \\
  &  +  \alpha_4 R^2 + \alpha_5 R_{\mu \nu} R^{\mu \nu} + \alpha_6 R_{\mu \nu}R^{ \nu \mu} +  \beta_1 T_{\mu \nu \rho} T^{\mu \nu \rho} + \beta_2 T_{\mu \nu \rho} T^{\mu \rho \nu } +  \beta_3 T_\mu T^\mu \big) \, ,
\end{split}
\end{equation}
where $T_\mu$ denotes the trace of the torsion tensor $T_\mu = T_{\mu \nu}{}^\nu$.
Notice however, that the Poincar\'e symmetry allows an infinity of terms in the Lagrangian with any number of derivatives. Even if these terms are not present classically, it is expected that they are generated by quantum corrections. The proliferation of possible terms makes the Poincar\'e theory very difficult to deal with at quantum level. As we will see later on, extending the space-time gauge group can reduce drastically the number of terms. In the following theories that we examine, we will impose the constraint in eq.\,\eqref{constraint} though later on for the Weyl case we generalise the arguments to include the more general \eqref{constraint-torsion}.

\subsection{Gauge theory for the Weyl group}

From the short list of space-time gauge symmetries allowed by the Coleman--Mandula theorem we consider now the gauging of the Weyl group. This is actually a subgroup of the conformal group
which is generated by the Poincar\'e group extended with dilatations,
but does not include special conformal transformations.

There are two ways to look at this theory. On one hand, dilatations can be thought
of as an extra symmetry which is imposed on top of the Poincar\'e
symmetry reflecting that the Weyl group is a semi-direct product $ISO(1,d-1) \rtimes \mathbb{R}$ and thus allowing us to keep the structure group of the frame bundle to be, as in the Poincar\'e theory, the Lorentz group. One the other hand, one can include dilatations in the structure group of the frame bundle which will now be a subgroup of the general linear group $GL(d)$, \cite{Sauro:2022hoh}.  In the first case the metric is preserved by transformations of the structure group,
while in the second case the metric is not preserved. We shall therefore
name the first case as the torsion case (this will become obvious why shortly) while the second case will be named the non-metric case.

\subsubsection{The torsion case}

The gauge algebra that we start from is
\begin{equation}
\begin{split}
  \label{gauge-algebra}
  &  [P_{a}, M_{bc}] =  \eta_{a b} P_{c} - \eta_{ac}P_b\, , \qquad   [D,P_a] = P_a\, ,
  \qquad [P_a, P_b] = 0 \, , \qquad    [D, M_{ab}] = 0 \, , \\
& [M_{ab}, M_{cd}] = \eta_{a c}M_{d b} - \eta_{bc}M_{da} - \eta_{ad} M_{cb} + \eta_{bd} M_{ca}\, ,
\end{split}
\end{equation}
 with $P_a$ and $M_{ab}$ defined before generating the Poincar\'e subalgebra (actually ideal) and $D$ the generator of dilatations. The non-zero structure constants of the Lie algebra
above are then inferred to be
\begin{equation}
\label{structure}
\begin{split}
  &f_{[ab][cd]}{}^{[ef]} = 2\left(\eta_{cb} \delta^{ef}_{ad} - \eta_{db} \delta^{ef}_{ac}
  - \eta_{ca} \delta^{ef}_{bd} + \eta_{da} \delta^{ef}_{bc} \right)\, ,\\
&f_{a[bc]}{}^d  =  \eta_{ab} \delta_{c}^d - \eta_{ac} \delta_b^d\, , \qquad f_{\# a}{}^b = \delta_a^b \, ,
\end{split}
\end{equation}
with the indices $a, [bc], \#$ referring to $P_a, M_{bc}$ and $D$
respectively. On top of the fields $e_\mu^a$ and $\Om_\mu{}^{ab}$ from the pure Poincar\'e case, we also introduce a gauge field for dilatations,
$\w_\mu$.

The infinitesimal transformations of the fields under the
Weyl group can be deduced from the general formula \eqref{gauge-var}
and from the specific structure constants
\eqref{structure}
\begin{align}
  \delta_\epsilon e_\mu^a & = - \partial_\mu \xi^a - \xi_b \Om_\mu{}^{ab}
  - \xi^a \w_\mu + \lambda^a{}_b e_\mu^b + \lambda_D e_\mu^a \, ,\label{var-1}\\
  \delta_\epsilon \Om_{\mu}{}^{ab} & =- \partial_\mu \lambda^{ab}
  - 2 \lambda^{[a}{}_c \Om_\mu{}^{b]c} \, , \label{var-2}\\
   \delta_\epsilon \w_\mu & = - \partial_\mu \lambda_D \, , \label{var-3}
\end{align}
with gauge parameter $\epsilon$ given by
\begin{equation}
  \label{Wvar}
  \epsilon \equiv \epsilon^A T_A = \xi^a P_a
  + \frac12 \lambda^{ab} M_{ab} + \lambda_D D \, .
\end{equation}
The corresponding field strengths, also
determined by the structure constants in eq.\,\eqref{structure}, have the form
\begin{align}
    R_{\mu \nu} (P^a) &= 2 D_{[\mu} e_{\nu]}^a +2 \w_{[\mu} e_{\nu]}^a \, ,\label{translations}\\
R_{\mu\nu}(M^{ab}) & =  \partial_\mu \Om_\nu{}^{ab} -
  \partial_\nu \Om_\mu{}^{ab} + \Om_\mu{}^a{}_c \Om_\nu{}^{cb}
  - \Om_\nu{}^a{}_c \Om_\mu{}^{cb} = R^{ab}{}_{\mu \nu} \, ,\label{rotations}\\
R_{\mu \nu}(D) & =\partial_\mu \w_\nu - \partial_\nu \w_\mu =  F_{\mu \nu}\, . \label{dilatations}
\end{align}
For later use
we introduced the notation $F_{\mu \nu}$ for the field strength
of the dilatation gauge field $\w_\mu$, and $R^{a}{}_{b\mu \nu}$ for
the usual two-form curvature tensor defined from the commutator of
the tangent space (Lorentz) covariant derivatives
\begin{equation}
  R^a{}_{b\mu \nu} : = e_b^\sigma [D_\mu, D_\nu] e_\sigma^a \, .
\end{equation}
The general formalism of gauge theories guarantees that the field strengths transform covariantly under \eqref{Wvar}
\begin{align}
  \delta_\epsilon R_{\mu \nu}(P^a) &=
   -\xi_b R_{\mu \nu}(M^{ab}) - \xi^a R_{\mu \nu}(D)
   + \lambda^a{}_b R_{\mu \nu}(P^b)
   + \lambda_D R_{\mu \nu}(P^a) \label{field-1} \, ,\\
  \delta_\epsilon R_{\mu \nu}(M^{ab}) & =
  - 2 \lambda^{[a}{}_c R_{\mu \nu}(M^{b] c})\label{field-2} \, ,\\
  \delta_\epsilon R_{\mu \nu}(D) & = 0 \, . \label{field-3}
\end{align}
Until this point in the construction, the vielbein and the spin connection were completely independent fields. As we discussed in detail in the
previous section we now impose the constraint \eqref{constraint} in
order to recover (covariant) general coordinate transformations from gauged
translations. This will also allow us to express the spin connection $\Om_\mu{}^{ab}$ in terms of the vielbein $e_\mu^a$ and the Weyl field $\w_\mu$.
Notice that it is consistent to impose the constraint $R_{\mu \nu}(P^a)=0$ since $R_{\mu \nu}(P^a)$ transforms into itself under Lorentz rotations and dilatations. It is instructive to see explicitly how imposing \eqref{constraint} on gauged translations yields covariant (also with respect to dilatations) general coordinate transformations.
Compared to the Poincar\'e case, the field strength of local
translations in eq.\,\eqref{translations} contains an extra term
depending on $\w_\mu$. The constraint \eqref{constraint} can again
be interpreted as the first Cartan structure equation where the
additional term plays the role of torsion
\begin{equation}
  2 D_{[\mu} e_{\nu]}^a = - 2 \w_{[\mu} e_{\nu]}^a \equiv T_{\mu\nu}{}^a \, .
  \label{cartan-torsion}
\end{equation} 
By solving the equation above for the spin connection we find
\begin{equation}
  \Om_\mu{}^{ab} = \mathring{\Om}_\mu{}^{ab} +2 e_\mu^{[a} e^{b]\nu} \w_\nu\, ,
  \label{spin-invariant}
\end{equation}
where the $\mathring{\Om}_\mu{}^{ab}$ was given in eq.\,\eqref{spinLC}.

It is important to notice that the spin connection above is invariant
under dilatations. This is guaranteed from the fact that in
\eqref{var-2} the spin connection does not change under dilatations
and the constraint from which we solved for it is also invariant under
dilatations.\footnote{This can also be checked explicitly
  using \eqref{spin-invariant} and the corresponding transformations
  \eqref{var-1} and \eqref{var-3}.}
Therefore, this connection is appropriate to covariantise derivatives for
objects which have vanishing Weyl charge on the tangent space.
For objects which have non-vanishing Weyl weight on the tangent space
we need to use the true covariant derivative of the gauge theory as
given by \eqref{gauge-derivative}
\begin{equation}
  \label{Dhat}
  \hat D_\mu X = D_\mu  X + q_X \w_\mu X \, ,
\end{equation}
where $X$ denotes an arbitrary object of Weyl weight $q_X$, which can carry any number of
(tangent-space) indices and $D_\mu$ is the tangent-space covariant
derivative defined with the torsion connection $\Om_\mu{}^{ab}$.
As an example that will be useful later, we illustrate the action of $\hat D_\mu$ on the vielbein 
\begin{equation}
  \hat D_\mu e_\nu^a = \partial_\mu e_\nu^a + \Om_\mu{}^a{}_b e_\nu^b
  + \w_\mu e_\nu^a = D_\mu e_\nu^a + \w_\mu e_\nu^a \, .
\end{equation}

With the new constraint we can redo the calculations in the previous
section, \eqref{gcte} and \eqref{gctg}. We find that under translations
the vielbein transforms as
\begin{equation}
  \label{cgcte}
  \delta_\xi e_\mu^a = -\xi^\nu \hat D_\nu e_\mu^a
  - e_\nu^a \hat D_\mu \xi^\nu \equiv \delta_{cgct} e_\mu^a \, ,
\end{equation}
and the metric as 
\begin{equation}
  \label{cgctg}
  \delta_\xi g_{\mu \nu}= - \xi^\rho \hat D_\rho g_{\mu \nu}
  - g_{\rho \nu} \hat D_\mu \xi^\rho - g_{\mu \rho} \hat D_\nu \xi^\rho
  = \mathcal L_\xi g_{\mu \nu} - 2 \xi^\rho \w_\rho g_{\mu \nu}=
  - \hat \nabla_\mu \xi_\nu - \hat \nabla_\nu \xi_\mu \, ,
\end{equation}
where the Lie derivative of the metric can be written as $\mathcal L_\xi g_{\mu \nu} = - \xi^\rho \partial_\rho g_{\mu \nu}
  - g_{\rho \nu} \partial_\mu \xi^\rho - g_{\mu \rho} \partial_\nu \xi^\rho$. 
Thus, the transformation of the metric is not exactly the Lie derivative but a covariant version of it. This should not be surprising as the parameters for translations are not singlets under
the gauge group and therefore the ordinary Lie derivative does not
transform covariantly under the gauge group. Furthermore, in eq.\,\eqref{cgctg} we also have defined a space-time gauge covariant derivative $\hat \nabla_\mu$ which on a vector $V^\rho$ with arbitrary charge satisfies
  \begin{equation}
  \label{hat-nabla}
 \hat \nabla_\mu V^\rho = e_a^\rho \hat D_\mu V^a \, .
 \end{equation}
This operator will be the basis of the space-time gauge covariant formulation which is discussed later on. As we mentioned at the beginning, in this section we considered a
$SO(1,d-1)$ structure group of the frame bundle on top of which we
imposed dilatation invariance. This results in a spin connection which
is valued in the $so(1,d-1)$ Lie algebra and therefore $\Om_\mu{}^{ab}$ is antisymmetric in the indices 
$a, b$. This is also equivalent to saying
that the connection is metric. It also preserves the completely antisymmetric tensor $\epsilon_{a_1 \ldots a_d}$. Indeed, it can be checked explicitly that
\begin{align}
\label{g-and-e}
  \hat D_\mu \eta_{ab} = D_\mu \eta_{ab} &= 0 \, , & \hat D_\mu\epsilon_{a_1 \ldots a_d} = D_\mu \epsilon_{a_1 \ldots a_d} & = 0 \, ,
\end{align}
where we have used eq.\,\eqref{Dhat} and that $\eta_{ab}$ and $\epsilon_{a_1 \ldots a_d}$ have zero Weyl charge. Thus, also the gauge covariant derivative is metric. In order to complete our discussion on the tangent space let us derive also the Bianchi identities. Recall that for these, one needs to use the derivative $\mathcal D_\mu$ which covariantises also with respect to translations. From the field strengths variations eqs.\,\eqref{field-1}-\eqref{field-3} one obtains
\begin{align}
\mathcal D_\mu R_{\nu \rho}(P^a) &= \partial_\mu R_{\nu \rho}(P^a) - e_\mu^b R_{\nu \rho}(M^a{}_b) - e_\mu^a R_{\nu \rho}(D) + \Om_\mu{}^a{}_b R_{\nu \rho}(P^b) + \w_\mu R_{\nu \rho}(P^a)\, ,\nonumber\\
\mathcal D_\mu R_{\nu \rho}(M^{ab}) & = \partial_\mu R_{\nu \rho}(M^{ab}) - 2 \Om_\mu{}^{[a}{}_c R_{\nu \rho}(M^{b]c})\, . \label{der-1}
\end{align}
After antisymmetrising the indices $\mu, \nu, \rho$, and imposing the constraint $R_{\mu \nu}(P^a) = 0$, we obtain the identities
\begin{align}
\label{bianchi-torsion}
R^a{}_{[\mu \nu \rho]} &= -e_{[\mu}^aF_{\nu \rho]} \, , & D_{[\mu}R^{ab}{}_{\nu \rho]} & = 0 \, ,
\end{align}
where the notation in eqs.\,\eqref{rotations}-\eqref{dilatations} has been used. The above form is indeed standard since for the torsion under consideration \eqref{cartan-torsion} one has $D_{[\mu} T_{\nu \rho]}{}^a = - e_{[\mu} F_{\nu \rho]}$.
Finally, notice that, as we have already mentioned, the connection $\Om_\mu{}^{ab}$ in eq.\,\eqref{spin-invariant} has
torsion \eqref{cartan-torsion}. Therefore, the case discussed in this
section is the metric formulation with torsion of the gauge theory of
the Weyl group.

\subsubsection{The non-metricity case}

Let us consider in this section the case when the structure group of the frame
bundle includes dilatations and hence it is isomorphic to $SO(1,d-1)\rtimes \mathbb{R} \subset GL(d)$. In order to understand the embedding in the general linear group it is useful to introduce the new
generators $\tilde M_{ab}$ as
\begin{equation}
  \label{tildeM}
  \tilde M_{ab} = M_{ab} + \frac1d \eta_{ab} D\, ,
\end{equation}
which are no longer antisymmetric in the indices $a,b$. The
commutation relations in eq.\,\eqref{gauge-algebra} become
\begin{equation}
  \begin{aligned}
    & \big[ P_a, \tilde M_{bc} \big] = \eta_{ab}P_c - \eta_{ac} P_b
    - \frac1d\eta_{bc} P_a \, , \\
    & \big[ \tilde M_{ab}, \tilde M_{cd} \big] = \eta_{a c} \tilde M_{[d b]}
    - \eta_{bc}\tilde M_{[da]} - \eta_{ad} \tilde M_{[cb]} + \eta_{bd} \tilde M_{[ca]}
     \, .
  \end{aligned}
\end{equation}
The structure constants in the new basis of generators can be inferred from above to be given by
\begin{align}
\label{f-tilde}
\tilde f_{a\{bc\}}{}^d &= f_{a [bc]}{}^d - \frac1{d}\eta_{bc} \delta_a^d\, , & \tilde f_{\{ab\} \{cd\}}{}^{\{ef\}} = f_{[ab][cd]}{}^{[ef]} \, ,
\end{align}
reflecting the fact that $[\tilde M_{(ab)}, \tilde M_{(cd)}] = 0$ generates an Abelian subalgebra. Now for the generators $P_a$ we introduce the gauge fields $e_\mu^a$
as before, while for $\tilde M_{ab}$ we introduce the new spin
connection $\tilde \Om_\mu{}^{ab}$, but there is no explicit Weyl field since it is now contained in $\tilde \Om_\mu{}^{ab}$ as its trace  
in the pair $a,b$ (see \eqref{traceO} below). Notice that some care is necessary when expanding sums over the generators $\tilde M_{ab}$ in order to count them only once. For instance we have the identity (to be taken as definition of the sum)
\begin{equation}
\tilde \Om_\mu{}^{ab} \tilde M_{ab} := \frac12 \tilde \Om_\mu{}^{[ab]} \tilde M_{[ab]} + \tilde \Om_\mu{}^{(ab)} \tilde M_{(ab)}\, ,
\end{equation}
where the symmetric part reduces to the trace since $\tilde M_{(ab)} \sim \eta_{ab}$ and hence no factor of $1/2$. 

With the new structure constants in eq.\,\eqref{f-tilde} one can show that the curvatures in this case become
\begin{equation}
  \begin{aligned}
    & R_{\mu \nu}(P^a) = \partial_\mu e_\nu^a - \partial_\nu e_\mu^a
    +  \tilde \Om_\mu{}^a{}_b e_\nu^b - \tilde \Om_\nu{}^a{}_b e_\mu^b \, , \\
    & R_{\mu \nu} (\tilde M^{ab}) = \partial_\mu \tilde \Om_\nu{}^{ab} -
    \partial_\nu \tilde \Om_\mu{}^{ab} + \tilde \Om_\mu{}^a{}_c
    \tilde \Om_\nu{}^{cb} - \tilde \Om_\nu{}^a{}_c \tilde \Om_\mu{}^{cb}
    = \tilde R^{ab}{}_{\mu \nu}\, ,
  \end{aligned}
\end{equation}
where one uses that $\tilde \Om_\mu{}^{ab}$ admits a decomposition into antisymmetric part and trace part according to $\tilde \Om_\mu{}^{ab} = \tilde \Om_\mu{}^{[ab]}+\tfrac1{d} \eta^{ab} \tilde \Om_\mu{}^c{}_c$ (and the trace part does not contribute to the term $\tilde \Om^a{}_c \wedge \tilde \Om^{cb}$). Again, $\tilde R^{ab}{}_{\mu \nu}$ is also given by the commutator of two
tangent space covariant derivatives defined by the new connection $\tilde \Om_\mu{}^{ab}$
\begin{equation}
  \tilde R^a{}_{b\mu \nu} : = e_b^\sigma [\tilde D_\mu, \tilde D_\nu] e_\sigma^a \, ,
\end{equation}
and thus it has a clear geometric interpretation as curvature tensor for a connection with vectorial non-metricity (see eq.\,\eqref{non-metricity1} below). The variations of the field strengths under $\epsilon = \xi^a P_a + \tilde \lambda^{ab} \tilde M_{ab}$ are given by
\begin{align}
  \delta_\epsilon R_{\mu \nu}(P^a) &=
   -\xi_b R_{\mu \nu}(\tilde M^{ab}) 
   + \tilde \lambda^a{}_b R_{\mu \nu}(P^b) \, , \label{nm-1}\\
  \delta_\epsilon R_{\mu \nu}(\tilde M^{ab}) & =
  -  \tilde \lambda^{[ac]}\eta_{cd} R_{\mu \nu}(\tilde M^{[b d]})+ \tilde \lambda^{[bc]}\eta_{cd} R_{\mu \nu}(\tilde M^{[a d]}) \, . \label{nm-2}
\end{align}
Defining the Weyl field as the trace
\begin{equation}
  \label{traceO}
  \w_\mu = \frac1d \eta_{ab} \tilde \Om_\mu{}^{ab}\, ,
\end{equation}
we can solve for $\tilde \Om_\mu{}^{ab}$ from the constraint $R_{\mu \nu}(P^a)=0$.
We find
\begin{equation}
  \label{tildeom}
  \tilde \Om_\mu{}^{ab} = \Om_\mu{}^{ab} + \w_\mu \eta^{ab} = \mathring{\Om}_\mu{}^{ab} +2 e_\mu^{[a} e^{b]\nu} \w_\nu + \w_\mu \eta^{ab} \, ,
\end{equation}
and the fact that it can be written in terms of $\Om_\mu{}^{ab}$
should not be surprising as with \eqref{traceO}, $R_{\mu \nu}(P^a)$ is
the same as the one in the previous section and therefore we solve the
same constraint, but for a different field.

Since \eqref{tildeom} is no longer antisymmetric in $a,b$, the connection is not metric anymore
\begin{equation}
\label{non-metricity1}
  \tilde D_\mu \eta_{ab} = \partial_\mu \eta_{ab}
  -\eta_{cb} \tilde \Om_\mu{}^c{}_a    - \eta_{ac}
\tilde \Om_\mu{}^c{}_b   = -2 \w_\mu \eta_{ab} \, ,
\end{equation}
and therefore, the non-metricity is given by the Weyl field.
Some care is required when raising and lowering indices when working with the non-metric connection. For instance 
\begin{equation}
\tilde D_\mu \eta^{ab} = \partial_\mu \eta^{ab} + \tilde \Om_\mu{}^a{}_c \eta^{cb} + \tilde \Om_\mu{}^b{}_c \eta^{ac} = + 2 \w_\mu \eta^{ab}\, ,
\end{equation}
yields a flipped sign after raising the indices on the Minkowski metric. 
Similarly, for the $\epsilon$-tensor, one obtains  
\begin{equation}
\tilde D_\mu \epsilon_{abcd} = -4 \w_\mu \epsilon_{abcd}\, .
\end{equation}

Let us observe that contrary to \eqref{cartan-torsion} the constraint equation for the field strength of translations can now be written as
\begin{equation}
\label{Cartan-non-metricity}
  R_{\mu \nu}(P^a) = 2 \tilde D_{[\mu} e_{\nu]}^a = 0 \, ,
\end{equation}
and therefore the spin connection $\tilde \Om_\mu{}^{ab}$ is torsionless. Finally, let us write down the Bianchi identities for the non-metricity case. From the variations eqs.\,\eqref{nm-1}-\eqref{nm-2} one obtains
\begin{align}
\mathcal D_\mu R_{\nu \rho}(P^a) &= \partial_\mu R_{\nu \rho}(P^a) - e_\mu^b R_{\nu \rho}(\tilde M^a{}_b) + \tilde \Om_\mu{}^a{}_b R_{\nu \rho}(P^b)\, ,\\
\mathcal D_\mu R_{\nu \rho}(\tilde M^{ab}) & = \partial_\mu R_{\nu \rho}(\tilde M^{ab}) -  \tilde \Om_\mu{}^{[ac]} \eta_{cd} R_{\nu \rho}(\tilde M^{[bd]}) +  \tilde \Om_\mu{}^{[bc]} \eta_{cd} R_{\nu \rho}(\tilde M^{[ad]})\, .
\end{align}
After using the curvature constraint $R_{\mu \nu}(P^a) = 0$ and expressing in terms of $\tilde R^{a}{}_{b \mu \nu}$ and $\tilde D_\mu$, the antisymmetrisation of the equations above can be brought to the following form
\begin{align}
\label{bianchi-nm}
\tilde R^a{}_{[\mu \nu \rho]} & = 0\, , & \tilde D_{[\mu} R^a{}_{|b| \nu \rho]} & = 0 \, ,
\end{align} 
where the index $b$ is excluded from the antisymmetrisation and its position is important since $\tilde D_\mu$ does not commute with the raising and lowering of (tangent space) indices. 

To conclude, we have shown that the same theory of gauged Weyl
group admits two geometric interpretations: one in
terms of a metric connection with torsion and the other one in terms
of a torsion-free non-metric connection. The associated curvature tensors $R^a{}_{b\mu \nu}$ and $\tilde R^a{}_{b\mu \nu}$, though having different geometric properties, are both covariant under dilatations. The distinction between the
two interpretations is whether we interpret the term $\w_{[\mu} e_{\nu]}^a$
as torsion ({\it i.e.} on the RHS of the Cartan structure equation \eqref{cartan-torsion}) or we
interpret it as part of the spin connection \eqref{Cartan-non-metricity}. We shall discuss more
this relation in the coming sections. For now we note that in
principle one can split the term $\w_{[\mu} e_{\nu]}^a$ in two pieces, say two halves, and
interpret one half as torsion and the other half absorbed in the
spin connection. In this way one can obtain a connection with both
non-metricity and torsion. However such an interpretation does not seem
to have any practical advantage and therefore we shall not discuss it
further.

\subsubsection{Space-time interpretation}\label{s3.2}

So far we discussed the gauge theory of the Weyl group only at the level of the tangent space. We now go
further and describe the space-time interpretation and construct
actions for the theory built so far. We shall discuss the two
cases above in parallel.

Let us start by writing the affine connections of the two cases using
the standard formula (equivalent to the vielbein postulate) 
\begin{align}
  \Gamma_{\mu \nu}^\rho & \equiv e_a^\rho D_\mu e_\nu^a \, , & \tilde\Gamma_{\mu \nu}^\rho & \equiv e_a^\rho \tilde D_\mu e_\nu^a \, ,
\end{align}
where in order to find either $\Gamma$ or $\tilde \Gamma$ we shall use the tangent space derivatives from each of the cases.
We find
\begin{equation}
  \label{affine-1}
  \begin{split}
    \Gamma_{\mu \nu}^\rho &= \mathring \Gamma_{\mu \nu}^\rho
    +  \delta_\mu^\rho \w_\nu  - g_{\mu\nu} \w^\rho \, ,   \\
      \tilde \Gamma_{\mu \nu}^\rho &= \mathring \Gamma_{\mu \nu}^\rho
  +  \delta_\mu^\rho \w_\nu  +  \delta^\rho_\nu \w_\mu - g_{\mu\nu} \w^\rho  \, .
  \end{split}
\end{equation}  
Clearly, $\Gamma_{\mu \nu}^\rho$ is metric compatible $\nabla_\mu g_{\nu \rho} =0$
with torsion $T_{\mu\nu}{}^\rho \equiv 2\Gamma^\rho_{[\mu \nu]} =2
\delta_{[\mu}^\rho \w_{\nu]}$, while $\tilde \Gamma_{\mu \nu}^\rho$ is
non-metric, $\tilde \nabla_\mu g_{\nu \rho} = -2 \w_\mu g_{\nu\rho}$,
but is symmetric in $\mu, \nu$ and therefore torsion-free.
Contrary to the tangent space case,
$\tilde \Gamma_{\mu \nu}^\rho$ is invariant under local dilatations, while
$\Gamma_{\mu \nu}^\rho$ transforms as a gauge field.

For later reference, notice the relation between $\Gamma$ and $\tilde\Gamma$
- they differ precisely by a projective transformation
\begin{equation}
\label{projective}
  \tilde\Gamma_{\mu\nu}^\rho=\Gamma_{\mu\nu}^\rho +\delta_\nu^\rho \w_\mu\, .
\end{equation}
This is the same as the relation between spin connections found in the first part of eq.~\eqref{tildeom}. At the level of the gauge algebra it merely corresponds to a redefinition of generators \eqref{tildeM} and thus it is a built-in symmetry of the theory. 

The Riemann tensor associated to the torsion connection $\Gamma$
\begin{equation}
\label{riemann-2}
  R^\rho{}_\sigma{}_{\mu \nu} = \partial_\mu \Gamma^\rho_{\nu \sigma}
  - \partial_\nu \Gamma^\rho_{\mu \sigma} + \Gamma^\rho_{\mu \tau}
  \Gamma^\tau_{\nu \sigma} - \Gamma^\rho_{\nu \tau} \Gamma^\tau_{\mu \sigma} \, ,
\end{equation}
can also be computed directly from the (gauge covariant) tangent space
curvature \eqref{rotations} by making use of the vielbein
\begin{equation}
   R^\rho{}_\sigma{}_{\mu \nu}  = e_a^\rho e_\sigma^b R^a{}_{b \mu \nu} \, .
\end{equation}
Either way, after separating the dependence on the Weyl gauge field $\w_\mu$, one obtains the following equations for the curvature tensor and its contractions
\begin{align}
    R_{ \rho \sigma \mu \nu } & = \mathring{R}_{ \rho \sigma \mu \nu }
    + \left[g_{\mu \sigma} \mathring{\nabla}_\nu \w_\rho
      - g_{\mu \rho} \mathring{\nabla}_\nu  \w_\sigma
      + g_{\nu \rho}  \mathring{\nabla}_\mu
      \w_\sigma - g_{\nu \sigma}\mathring{\nabla}_\mu \w_\rho  \right] \nonumber\\[2pt]
    &+ \w^2(g_{\mu \sigma} g_{\nu \rho} - g_{\mu \rho} g_{\nu \sigma})
    + \w_\mu(\w_\rho g_{\nu \sigma} - \w_\sigma g_{\nu \rho})
    + \w_\nu(\w_\sigma g_{\mu \rho} - \w_\rho g_{\mu \sigma})\, , \label{riem-1} \\[2pt]
    R_{\mu \nu} & = \mathring{R}_{\mu \nu} - (d-2) \mathring{\nabla}_\nu
    \w_\mu- g_{\mu \nu} \mathring{\nabla}_\rho \w^\rho + (d-2)\w_\mu \w_\nu
    - (d-2)g_{\mu \nu} \w_\rho \w^\rho\, , \label{riem-2}  \\
    R &= \mathring{R} - 2(d-1)\, \mathring{\nabla}_\mu \w^\mu
    - (d-1)(d-2)\, \w_\mu \w^\mu \, ,  \label{riem-3}
\end{align}
where $\mathring{R}_{ \rho \sigma \mu \nu }, \mathring R_{\mu \nu} $ and $\mathring R$ are expressed in terms of the Levi-Civita connection \eqref{Levi-Civita}. Similarly, inserting $\tilde \Gamma$ in \eqref{riemann-2} we find the effect of a projective transformation on the curvature tensors \cite{DG1,Tann,Jia }
\begin{equation}
  \begin{aligned}
    \label{correspondence}
    \tilde R_{\rho \sigma \mu\nu}  &=  R_{\rho \sigma \mu \nu}
    + g_{\rho \sigma} F_{\mu \nu} \, ,
    \qquad
     & \tilde R_{\mu \nu} & =  R_{\mu \nu} + F_{\mu \nu}\, , \qquad  &
     \tilde R & =  R \, ,
\end{aligned}
\end{equation}
where $F_{\mu \nu}$ is the Abelian field strength of dilatations defined in eq.\,\eqref{dilatations}.

It is important to note that in the torsion case we have an additional
curvature, corresponding to dilatations $F_{\mu \nu}$, whereas in the non-metricity case, this curvature ``is hidden'' inside
$\tilde R_{\rho \sigma \mu \nu}$ and manifests itself in the fact that
$\tilde R_{\rho \sigma \mu \nu}$ is no longer antisymmetric in the first indices (as it
happens in the torsion case), but has a component proportional to the
metric
\begin{equation}
  \label{tracetR}
  \tilde R_{(\rho \sigma) \mu \nu} =g_{\rho \sigma} F_{\mu \nu}\, ,
\end{equation}
as can be seen from the first equation in \eqref{correspondence}. Notice that both $R^\rho{}_{\sigma \mu \nu}$ and $\tilde R^\rho{}_{\sigma \mu \nu}$ are covariant under general coordinate transformations and dilatations and hence are equally good to work with. In order to bridge our work with the existing literature on the subject let us infer here the transformations under finite gauge dilatations. Let $h=e^{\lambda_D D}$, then from eq.\,\eqref{gaugeB} with $B=(e_\mu^a P_a + \tfrac12 \Om_\mu{}^{ab} M_{ab} + \w_\mu D) dx^\mu$ one obtains
\begin{equation}
B' = \left[e^{\lambda_D} e_\mu^a P_a + \frac12 \Om_\mu{}^{ab} M_{ab}+(\w_\mu- \partial_\mu \lambda_D) D \right] dx^\mu \, ,
\end{equation}
and hence finite dilatations act as
\begin{align}
  \label{Weyl1}
  e_\mu^a &\mapsto e^{\lambda_D} e_\mu^a\, , & \Om_\mu{}^{ab} &\mapsto \Om_\mu{}^{ab}\, , & \tilde \Om_\mu{}^{ab} &\mapsto \tilde \Om_\mu{}^{ab} - \eta^{ab} \partial_\mu \lambda_D\, , & \w_\mu \mapsto \w_\mu - \partial_\mu \lambda_D\, ,
\end{align}
reflecting the fact that one spin connection is invariant and the other transforms as a gauge field with roles reversed with respect to $\Gamma$ and $\tilde \Gamma$. The metric, which from the tangent space perspective is a composed object, $g_{\mu \nu} = e_\mu^a e_\nu^b \eta_{ab}$, transforms as
\begin{equation}
  \label{Weyl2}
  g_{\mu \nu} \mapsto e^{2 \lambda_D} g_{\mu \nu} \, ,
\end{equation}
while the transformation under finite dilatations of the curvatures can be inferred from eq.\,\eqref{gaugeR} 
\begin{align}
R^\rho{}_{\sigma \mu \nu} &\mapsto R^\rho{}_{\sigma \mu \nu}\, , & R_{\mu \nu} & \mapsto R_{\mu \nu}\, , & R & \mapsto e^{-2 \lambda_D} R\, ,\\
\tilde R^\rho{}_{\sigma \mu \nu} &\mapsto \tilde R^\rho{}_{\sigma \mu \nu}\, , & \tilde R_{\mu \nu} & \mapsto \tilde R_{\mu \nu}\, , & \tilde R & \mapsto e^{-2 \lambda_D} \tilde R \, ,
\end{align}
and for the curvature of translations 
\begin{equation}
R_{\mu \nu}(P^a) \mapsto e^{\lambda_D} R_{\mu \nu}(P^a) \, ,
\end{equation}
confirming that the constraint is invariant also under finite dilatations. In deriving the above one uses the identity $e^{\lambda_D D} P_a e^{-\lambda_D D} = e^{\lambda_D} P_a$ which can be proven from a version of the Baker-Campbell-Hausdorff formula, namely $e^X Ye^{-X} = \sum_{n=0}^\infty \tfrac{1}{n!} \text{ad}_X^n Y$ with $\text{ad}_X Y := [X,Y]$. 

For completeness we list here the Bianchi identities in space-time. From eqs.\,\eqref{bianchi-torsion} and \eqref{bianchi-nm} one obtains
\begin{align}
R_{\sigma [\mu \nu\rho]} & = - g_{\sigma [\mu} F_{\nu \rho]}\, , & \nabla_{[\mu} R^{\sigma \tau}{}_{\nu \rho]} &= 2 \w_{[\mu} R^{\sigma \tau}{}_{\nu \rho]}\, . \label{bianchi-st1}\\
\tilde R_{\sigma [\mu \nu\rho]} & = 0\, , & \tilde \nabla_{[\mu} \tilde R^{\sigma}{}_{|\tau|}{}_{\nu \rho]} &= 0 \, . \label{bianchi-st2}
\end{align}
Finally, let us notice that, even though the derivative operators $\nabla_\mu$ and $\tilde \nabla_\mu$ associated to the space-time connections $\Gamma$ and $\tilde \Gamma$ can be used to define dilatation invariant curvatures
\begin{align}
[\nabla_\mu, \nabla_\nu] V^\rho
  &= R^\rho{}_{\sigma \mu \nu} V^\sigma - T_{\mu \nu}{}^\sigma \nabla_\sigma V^\rho \, , \label{commutator-torsion}\\
  [\tilde \nabla_\mu, \tilde \nabla_\nu] V^\rho
  &= \tilde R^\rho{}_{\sigma \mu \nu} V^\sigma \, , \label{commutator-non-metricity}
\end{align}
they do not always yield a dilatation covariant quantity (e.g. $\nabla_\mu R$ or $\tilde \nabla_\mu \tilde R$). Thus, either picture, torsion or non-metric, needs to be completed with a (space-time) gauge covariant derivative $\hat \nabla_\mu$ which is the subject of the next subsection.

\subsubsection{The gauge covariant case}

In the previous sections we focused on the possible geometric
interpretations of the gauge theory of the Weyl group. We can also
choose a more gauge theoretical description where the accent is put on
gauge invariance/covariance.

The object which is the most interesting in this case is the space-time gauge
covariant derivative $\hat \nabla_\mu$ introduced earlier \eqref{hat-nabla}. On an arbitrary $(r,p)$-tensor $X$ its action can be expressed either in terms of $\nabla_\mu$ associated to torsion or either in terms of $\tilde \nabla_\mu$ associated to non-metricity
\begin{equation}
\label{general-formula}
  \hat \nabla_\mu X = \nabla_\mu X + q_X \w_\mu X = \tilde \nabla_\mu X + \tilde q_X \w_\mu X\, ,
\end{equation}
where we have suppressed the indices on X, which should read $X^{\mu_1 \ldots \mu_r}{}_{\nu_1 \ldots \nu_p}$. Above, $q_X$ denotes the tangent space Weyl charge/weight of $X$ and $\tilde q_X$ denotes the space-time Weyl charge of X. Taking into account the charge of the vielbein, yields the following connection between $q_X$ and $\tilde q_X$
\begin{equation}
 q_X = \tilde q_X+r-p  \, .
\end{equation}
The tangent space Weyl weight $q_X$ is an intrinsic charge as it is independent of the position of the indices. This is due to the fact that $\eta_{ab}$ has Weyl charge zero. On the other hand, the space-time charge $\tilde q_X$ will depend on the position of the indices since the metric $g_{\mu \nu}$ has Weyl weight $+2$. 

Note that due to the explicit presence of the Weyl charge in the derivative $\hat \nabla$, there is no affine connection associated to this gauge covariant derivative \footnote{The covariant derivative has been introduced before by other authors, \cite{Dirac}}. This can be most easily seen for scalar fields. Given a covariant derivative corresponding to an affine connection, the covariant derivative of all scalar fields has precisely the same form, $\partial_\mu$. On the other hand, $\hat \nabla$ may have different expressions on scalars, $\hat \nabla_\mu = \partial_\mu + q \w_\mu$, depending of their Weyl charge, $q$.

Another crucial remark is that this gauge covariant derivative
preserves the metric. This is easy to verify from the above formula as
the tangent space charge of the metric $q_g$ vanishes, and the
covariant derivative $\nabla_\mu$ is constructed with a metric
connection as explained in the previous sections. Furthermore, it is also compatible with the space-time completely antisymmetric tensor $\sqrt{g}\, \epsilon_{\mu \nu \rho \sigma}$. Hence, analogous to \eqref{g-and-e} we have
\begin{align}
\hat \nabla_\mu g_{\nu \rho} & = 0\, , & \hat \nabla_\mu \left( \sqrt{g} \, \epsilon_{\nu \rho \sigma \delta} \right) & = 0 \, .
\end{align}
Notice that the property above, also holds for the covariant derivative with torsion $\nabla_\mu g_{\nu \rho} = 0$ and $\nabla_\mu \left( \sqrt{g} \, \epsilon_{\nu \rho \sigma \delta} \right) = 0$. Instead, for the non-metric connection one has
\begin{align}
\tilde \nabla_\mu g_{\nu \rho} &= - 2 \w_\mu g_{\nu \rho} \, , & \tilde \nabla_\mu \left( \sqrt{g} \, \epsilon_{\nu \rho \sigma \delta} \right) = -4 \w_\mu  \sqrt{g} \, \epsilon_{\nu \rho \sigma \delta} \, .
\end{align}
The space-time gauge covariant derivative $\hat \nabla_\mu$ allows one to write the second Bianchi identities eqs.\,\eqref{bianchi-st1}-\eqref{bianchi-st2} in the simple way
\begin{align}
\hat \nabla_{[\mu} R^{\sigma \tau}{}_{\nu \rho]} & = 0\, , & \hat \nabla_{[\mu} \tilde R^{\sigma \tau}{}_{\nu \rho]} & = 0 \, .
\end{align}
The commutator of two gauge covariant derivatives can be computed from
the general formula \eqref{general-formula}  with eqs.\,\eqref{commutator-torsion}-\eqref{commutator-non-metricity}. Hence on a vector $V^\rho$ with charges $q_V$ and $\tilde q_V$ on the tangent space and space-time respectively we obtain 
\begin{equation}
  \label{hatcomm}
  \big [ \hat \nabla_\mu, \hat \nabla_\nu \big ] V^\rho =
   R^\rho{}_{\sigma \mu \nu}
  V^\sigma+ q_V F_{\mu \nu} V^\rho = \tilde R^\rho{}_{\sigma \mu \nu}
  V^\sigma+ \tilde q_V F_{\mu \nu} V^\rho  \, ,
\end{equation}
illustrating on the commutator the equivalence of (vectorial) torsion and non-metricity for the Weyl gauge theory. Notice that the formula above is standard for a gauge theory since on the RHS one has, as usual, a sum over the field strengths and no torsion. This comes again to confirm that torsion
or non-metricity which we encountered before, are simple artifacts of the
specific geometric picture we choose to work with. Working just with
$\hat \nabla$ we do not see any sign of torsion or non-metricity.

In view of the equivalence (vectorial) torsion/non-metricity in Weyl gravity, a natural question arises about whether the norm of vectors is changed under parallel transport. It is well-known that parallel transport with a metric connection, like $\Gamma$ in our case, even though it has torsion, always preserves the norm (independently of the Weyl charge). Indeed, assuming parallel transport, {\it i.e.} $dx^\mu \nabla_\mu V^\rho= 0$ then from metric compatibility it immediately follows that
\begin{equation}
d|V|^2 = dx^\mu \nabla_\mu |V^2| = 0 \, . \label{transport-1}
\end{equation}
On the other hand, parallel transport with the connection $\tilde \Gamma$ will, in general, change the norm of a vector according to
 \begin{equation}
d|V|^2 = dx^\mu \tilde \nabla_\mu |V^2| = -2 \w_\mu  |V|^2 dx^\mu \, , \label{transport-2}
\end{equation}
where we assumed parallel transport with $\tilde \Gamma$, that is $dx^\mu \tilde \nabla_\mu V^\nu= 0$. How do the two reconcile? The answer lies in the fact that we are dealing with a gauge theory. As such, physical  observables \emph{must be} gauge invariant. Therefore, parallel transport within Weyl theory makes physical sense only with the gauge covariant derivative $\hat \nabla_\mu$ which we have seen it is compatible with the metric. Now, assuming covariant parallel transport
\begin{equation}
  \label{covparallel}
  dx^\mu \hat \nabla_\mu V^\nu = 0\, ,
\end{equation}
it immediately follows from the metricity of $\hat \nabla $ that the gauge covariant variation of the norm of a vector is equal to zero, that is
\begin{equation}
dx^\mu \hat \nabla_\mu |V|^2 = 0\, . \label{cov-norm}
\end{equation}
One can then expand $\hat \nabla$ using the general formula \eqref{general-formula} together with the fact that the Weyl charge of the norm squared $|V|^2 = g_{\mu \nu} V^\mu V^\nu = \eta_{ab} V^a V^b$ is equal to $2 q_V$
\begin{equation}
\hat \nabla_\mu |V|^2 = \partial_\mu |V|^2 + 2 q_V \w_\mu |V|^2 \, .
\end{equation}  
Therefore, the equation \eqref{cov-norm} can equivalently be written as
\begin{equation}
  d |V|^2 \equiv dx^\mu \partial_\mu |V|^2 = - 2 q_V \w_\mu |V|^2 dx^\mu\, ,
\end{equation}
which differs, in general, from either \eqref{transport-1} or \eqref{transport-2}. In particular, it follows that the norm of vectors with zero tangent space Weyl charge (geometric vectors) is preserved under parallel transport in Weyl quadratic gravity. Note that only for such geometric vectors their norm is gauge invariant and therefore can be a physical observable. Thus, the relation above solves the apparent inconsistency between
the metric and torsion formulations and settles the (historical) issue of vectors
changing their norm under parallel transport in gauge theories of the
Weyl group (similar arguments can be found in \cite{Hobson:2020doi,Hobson:2021iwy,Ghilencea:2022lcl}, \cite{Condeescu}).

\subsubsection{The vielbein postulate}
The vielbein postulate associates a (tangent space) spin connection to a given (space-time) affine connection and vice versa. Up to now we assumed the standard postulate of the form
\begin{equation}
\label{postulate1}
\partial_\mu e_\nu^a + \Om_\mu{}^a{}_b e_\nu^b - e_\rho^a \Gamma_{\mu \nu}^\rho = 0 \, ,
\end{equation}   
and a similar one with $\tilde \Om_\mu{}^a{}_b$ and $\tilde \Gamma_{\mu \nu}^\rho$. Its application to Weyl geometry yields a pairing of a dilatation invariant metric spin connection $\Om_\mu{}^a{}_b$ with a metric connection with torsion  $\Gamma_{\mu \nu}^\rho$ that transforms as a gauge field. In the other formulation, the same postulate pairs a non-metric spin connection $\tilde \Om_\mu{}^a{}_b$ that transforms like a gauge field to a dilatation invariant non-metric connection $\tilde \Gamma_{\mu \nu}^\rho$. Thus, the standard vielbein postulate preserves the metricity/non-metricity property of the connection when passing from the tangent space to space-time but not the transformation under gauged dilatations. Using eq.~\eqref{projective}, one can equivalently write the vielbein postulate above in the following form
\begin{equation}
\label{postulate2}
(\partial_\mu + \w_\mu) e_\nu^a + \Om_\mu{}^a{}_b e_\nu^b - e_\rho^a \tilde \Gamma_{\mu \nu}^\rho = 0 \, , 
\end{equation}  
where one covariantises the partial derivative on the vielbein with respect to dilatations. Putting the respective  term $\w_\mu e_\nu^a$ on the RHS implies that one of the connections, $\tilde \Gamma_{\mu \nu}^\rho$, is non-metric. The above writing of the vielbein postulate ensures that one is working with dilatation invariant connections both on the tangent space and in space-time. However, a metric connection on the tangent space will be paired with a non-metric one in space-time. This is actually the usual choice in the ``traditional'' non-metric formulation of Weyl gravity. A third equivalent writing of the vielbein postulate is possible in which both connections transform like gauge fields. This is ensured by using a ``de-covariantised'' derivative $\partial_\mu - \w_\mu$ on the vielbein which also transforms like a gauge field under dilatations. Hence, using eq.~\eqref{tildeom}, one can also write
\begin{equation}
\label{psotulate3}
(\partial_\mu - \w_\mu)  e_\nu^a + \tilde \Om_\mu{}^a{}_b e_\nu^b  -e_\rho^a \Gamma_{\mu \nu}^\rho = 0 \, .
\end{equation}  
It is a rather unnatural choice, but valid nonetheless. Again the above, matches a non-metric spin connection to a metric connection with torsion. Finally, all the different postulates above are equivalent  re-writings of the same equation illustrating the torsion/non-metricity equivalence in Weyl gravity.

\subsection{Weyl gauge theory -- the quadratic action}

We are now in position to write the most general quadratic action
for the Weyl gauge theory. We shall see explicitly that both cases
discussed above (torsion and non-metricity) give rise to the same
action which is actually the same as the action for Weyl quadratic
gravity.

The only objects we have at our disposal are the curvatures constructed
so far as they transform covariantly. They
are going to appear quadratically in the action with indices contracted
either with the metric
$g_{\mu \nu}$ or the completely antisymmetric $\epsilon$-density
$\epsilon_{\mu_1 \ldots \mu_d}$ (or their tangent space counterparts). Parity violating terms will be considered later on and hence for the moment we assume parity invariance.

To write the Weyl-invariant action, we need the Weyl charges of the
various fields and field strengths under gauged dilatations\footnote{These
  are easily determined by the variations given in
  eqs. \eqref{var-1}-\eqref{var-3} and eqs. \eqref{field-1}-\eqref{field-3}.}.
They are
\begin{table}[h]
  \label{charges}
  \centering
  \begin{tabular}{ccccccccccccc} 
 $e_\mu^a$ & $e_a^\mu$ & $\eta_{ab}$ & $g_{\mu\nu}$ & $g^{\mu\nu}$ & $\Om_\mu{}^{ab}$ & $\sqrt{g}$
    & $\epsilon^{\mu_1...\mu_d}$ & $\epsilon_{\mu_1...\mu_d}$ & $R^{a}{}_b{}_{\mu\nu}$ & $R_{\mu\nu}$ & $R$ & $F_{\mu\nu}$
    \\[2mm]
    \hline
    1 & -1 & 0 & 2 & -2 & 0 & $d$ & 0 & 0& 0& 0 & -2 & 0
    \\[2mm]
  \end{tabular} \, .
\end{table}

In Riemannian geometry, the structure of the indices for the Riemann
tensor only leave us with a few possibilities for a quadratic action.
When using a connection with torsion or non-metricity more care
is needed in order to correctly identify all the independent possibilities. 

In the case of a connection with torsion, the Riemann tensor itself
contains more degrees of freedom as now it lacks the symmetry of
exchanging the first with the second pair of indices. However in the
case of the gauge theory of the Weyl group we have the following relation\footnote{
It can be shown directly from eq.\,\eqref{riem-1} or in general by using the Bianchi identity \eqref{bianchi-1} and the second combinatorial identity in eq.\,\eqref{Rcomb}.}  
\begin{equation}
  \label{symm-1}
  R_{ \rho \sigma \mu \nu} - R_{ \mu \nu \rho \sigma} = -F_{\mu \rho} g_{\sigma \nu}
  + F_{\mu \sigma} g_{\rho \nu} + F_{\nu \rho} g_{\sigma \mu}
  - F_{\nu \sigma} g_{\rho \mu} \, ,
\end{equation} 
It follows immediately that the Ricci tensor is no longer symmetric
\begin{equation}
\label{symm-2}
  R_{\mu \nu} - R_{\nu \mu} = (d-2) F_{\mu \nu} \, .
\end{equation}
We see that this lack of symmetry does not actually introduce new
degrees of freedom as we already had $F_{\mu \nu}$ as an independent
curvature for dilatations. Moreover, the Bianchi identity which
vanishes in the Riemannian case now reads
\begin{equation}
  R^a{}_{[ \mu \nu \rho]} = D_{[\mu} T_{\nu \rho]}{}^a
  = -e_{[\mu}^a F_{\nu \rho]} \, ,
\end{equation}
and again we see that no new degrees of freedom are introduced.
We can therefore write the following identities which are
shown to hold true in $d$ dimensions:
\begin{align}
  R_{\rho \sigma \mu \nu } R^{ \rho \sigma \mu \nu} &=R_{\rho \sigma \mu \nu }
    R^{  \mu \nu \rho \sigma}
    + 2(d-2) F_{\mu \nu} F^{\mu \nu} \, , \label{id-1}\\
  R_{\rho \sigma \mu \nu } R^{\mu \sigma \nu \rho } &=
    \frac12\left( 3R_{[\rho \sigma \mu ]\nu  }-
     R_{\mu \sigma \nu \rho } \right) R^{\mu \sigma \nu \rho } =
     - \frac12 R_{\rho \sigma \mu \nu } R^{\rho \sigma \mu \nu }
     +\frac{d-2}{2} F_{\mu \nu} F^{\mu \nu} \, ,\label{id-2}
\end{align}
and
\begin{align}
  \label{id-3}
  R_{\mu \nu}R^{\mu \nu} &= R_{\mu \nu} R^{\nu \mu} + \frac{(d-2)^2}{2}F_{\mu \nu} F^{\mu \nu} \, ,
    \\
   R_{\mu \nu}F^{\mu \nu} &= \frac12 (d-2) F_{\mu \nu} F^{\mu \nu} \label{id-4} \, .
\end{align}
From above, we see that there are four independent terms from the
contractions with the metric.  They can be chosen to be
$R^2, R_{(\mu \nu)} R^{(\mu \nu)}, R_{\mu \nu \rho \sigma} R^{\mu \nu
  \rho \sigma}$ and $F_{\mu \nu} F^{\mu \nu}$. The same conclusion can
be reached in the non-metric case as well\footnote{Similar formulas as eqs.\,\eqref{id-1}-\eqref{id-4} can be written in the non-metric case as well by making use of \eqref{correspondence}.}. We have already shown that
the difference in the Riemann tensor compared to the torsion case is
the trace on the first indices of $\tilde R_{\rho \sigma \mu \nu}$,
\eqref{tracetR}, which is just the dilatational curvature
$F_{\mu \nu}$ in the torsion case. Therefore, after separating this trace
from $\tilde R_{\rho \sigma \mu \nu}$ we have the same setup as in the
torsion case.
In addition to the above discussion, in four
dimensions, a linear combination of the four possible terms gives rise to the Chern-Euler-Gauss-Bonnet term $G$,
which for a connection with torsion is given by
\begin{equation}
\label{Euler}
  G = R^2 - 4R_{\mu \nu} R^{\nu \mu} + R_{\mu \nu \rho \sigma}
  R^{\rho \sigma \mu \nu} \, ,
\end{equation} 
which is actually defined with the help of the $\epsilon$-density (see eq.\,\eqref{Euler-class}). Notice the position of the contracted indices which is essential in
making $G$ a topological invariant (total derivative in the action)
for a metric connection with torsion in four dimensions.  In $d$ dimensions,
$G$ is no longer a topological invariant. However, it is Weyl
covariant as it is quadratic in the covariant curvatures. The inclusion of $G$ is important when considering the theory at the quantum level, in particular in showing the absence of Weyl anomalies \cite{DG1}. 
A convenient choice of independent quadratic terms each invariant
under gauged dilatations is given in the following
action\footnote{Other invariant terms (either topological or
    linear in the curvature) that are total derivatives in four
    dimensions are discussed in Section \ref{aD}.}
\begin{equation}
\begin{split}
  S &= \int d^4x \sqrt{g}\,  \left[a_0\, R^2+b_0\, F_{\mu \nu} F^{\mu \nu} + c_0\,
    C_{ \rho \sigma \mu \nu} C^{\rho \sigma \mu \nu }  + d_0\, G \right] \\
    & = \int d^4x \sqrt{g}\,  \left[a_0\, \tilde R^2+b'_0\, F_{\mu \nu} F^{\mu \nu} + c_0\,
    \tilde C_{ \rho \sigma \mu \nu} \tilde C^{\rho \sigma \mu \nu }  + d_0\, \tilde G \right] \, ,
\end{split}
\label{action}
\end{equation}
where we expressed the action in terms of torsion curvatures in the first line and non-metric curvatures in the second line
\footnote{Strictly speaking, in the non-metric picture one should make use of $\tilde R^\rho{}_{ \rho \mu \nu}$ instead of $F_{\mu \nu}$.}.
In the above, $\tilde G$ is defined as
\begin{equation}
  \label{nm-GB}
  \tilde G = \tilde R^2 - 4\tilde R_{\mu \nu} \tilde R^{\nu \mu} + \tilde R_{\mu \nu \rho \sigma}
  \tilde R^{\rho \sigma \mu \nu} - 4(d-1) F_{\mu \nu}F^{\mu \nu} =G\, ,
\end{equation} 
and $C_{\rho \sigma \mu \nu }$ is the Weyl tensor associated to the curvature $R_{\rho \sigma \mu \nu}$
\begin{equation}
\label{weyl-tensor}
  C_{ \rho \sigma \mu \nu}= R_{ \rho \sigma \mu \nu}
  + \frac1{d-2} \left(R_{ \sigma \mu} g_{\nu \rho} - R_{ \rho \mu} g_{\nu \sigma}
    + R_{ \rho \nu} g_{\mu \sigma} - R_{ \sigma \nu} g_{\mu \rho} \right) 
  +  R \frac{(g_{\mu \rho} g_{\nu \sigma} - g_{\mu \sigma} g_{\nu \rho})}{(d-1) (d-2)} \, ,
\end{equation}
which is traceless
\begin{align}
C^\rho{}_{\mu \rho \nu} & = 0\, , & C_\mu{}^\rho{}_{\nu \rho} & = 0\, , & C^\rho{}_{\rho \mu \nu} & = 0\, , & C_{\mu \nu}{}^\rho{}_\rho & = 0 \, .
\end{align} 
For $\tilde C_{\rho \sigma \mu \nu}$, as found usually in the literature, we adopt the same definition as for $C_{\rho \sigma \mu \nu}$ but now using $\tilde R_{\rho \sigma \mu \nu}$ and its contractions and we find
\begin{equation}
\tilde C_{\rho \sigma \mu \nu} = C_{\rho \sigma \mu \nu} - \frac1{d-2} \left(F_{\mu \sigma} g_{\nu \rho} -F_{\mu \rho} g_{\nu \sigma} + F_{\nu \rho} g_{\mu \sigma} - F_{\nu \sigma} g_{\mu \rho} \right) + g_{\rho \sigma} F_{\mu \nu} \, .
\end{equation}
Note that this is not the correct Weyl tensor corresponding to $\tilde R_{\rho \sigma \mu \nu}$ as it is not completely traceless\footnote{The completely traceless tensor in the case of non-metric connections is given in appendix, eq. \eqref{real-weyl}.}
\begin{align}
\tilde C^\rho{}_{\mu \rho \nu} & = 0\, , & \tilde C_\mu{}^\rho{}_{\nu \rho} & = -2 F_{\mu \nu}\, , & \tilde C^\rho{}_{\rho \mu \nu} & = d\, F_{\mu\nu}\, , & \tilde C_{\mu \nu}{}^\rho{}_\rho & = 0 \, .
\end{align}
Moreover, in the torsion formulation the Weyl tensor is symmetric under the exchange of the first and last pair of indices $C_{\rho \sigma \mu \nu} = C_{\mu \nu \rho \sigma}$ while in the non-metric one it obeys
\begin{equation}
\tilde C_{\rho \sigma \mu \nu} - \tilde C_{\mu \nu \rho \sigma} = g_{\rho \sigma} F_{\mu \nu} - g_{\mu \nu} F_{\rho \sigma} \, .
\end{equation}
In fact, $\tilde C_{\rho \sigma \mu \nu}$ defined in this way is the result of the projective transformation \eqref{projective} acting on $C_{\rho \sigma \mu \nu}$. This can be easily checked using eq.\,\eqref{correspondence} which encodes the effect of the aforementioned transformation on the curvature tensors. Actually, the action in the second line of eq.\,\eqref{action} can be obtained from the first line by the effect of the projective transformation \eqref{projective} term by term \footnote{Notice that a projective invariance was also found in the case of $f(R) $ gravity \cite{Iosifidis, Sotiriou}.}. Indeed, under $\Gamma_{\mu \nu}^\rho \mapsto \Gamma_{\mu \nu}^\rho + \delta_\nu^\rho \w_\mu$ we have
\begin{align}
R &\mapsto \tilde R\ (= R)\, , & F_{\mu \nu} &\mapsto F_{\mu \nu}\, , & C_{\rho \sigma \mu \nu} &\mapsto \tilde C_{\rho \sigma \mu \nu}\, , & G &\mapsto \tilde G\ (=G)\, , 
\end{align}
illustrating the projective invariance of Weyl gravity. As noted before, this invariance can be seen {\it apriori} from the fact that the gauge algebra redefinition of generators \eqref{tildeM} is equivalent to the projective transformation \eqref{projective} and thus it is not an accidental symmetry of the action but rather built-in Weyl gravity.

For completeness, let us give here the explicit expansions of the Weyl tensors squared in terms of the curvatures
\begin{align}
C_{\rho \sigma \mu \nu } C^{\rho \sigma \mu \nu}  = R_{\rho \sigma \mu \nu} R^{\rho \sigma \mu \nu}-\frac{4}{d-2} R_{\mu \nu} R^{\mu \nu} + \frac{2}{(d-1)(d-2)} R^2 = C_{\rho \sigma \mu \nu } C^{\mu \nu \rho \sigma }
\end{align} 
and for the non-metric case
\begin{equation}
 \tilde C_{ \rho \sigma \mu \nu} \tilde C^{\rho \sigma \mu \nu }  = C_{ \rho \sigma \mu \nu} C^{\rho \sigma \mu \nu }+ \frac{d^2-2d+4}{d-2} F_{\mu \nu} F^{\mu \nu}  \neq \tilde C_{\rho \sigma \mu \nu } \tilde C^{ \mu \nu \rho \sigma} \, .
\end{equation}
This shows that passing from torsion to non-metricity redefines the coupling $b_0$ to $b'_0$ by the coefficient of the $F_{\mu \nu}F^{\mu \nu}$ term above. An advantage of working with the torsion basis for curvatures is that the corresponding Weyl tensor reduces actually to the Riemannian Weyl tensor $\mathring C_{\rho \sigma \mu \nu}$ and hence it is independent of the dilatation gauge field $\w_\mu$. Indeed, the following identity holds
\begin{equation}
\label{weyl-torsion}
C_{\rho \sigma \mu \nu} = \mathring C_{\rho \sigma \mu \nu}\, .
\end{equation}
Finally, Eq.\,\eqref{action} is exactly the action of Weyl quadratic gravity which we have shown that it possesses a projective invariance \eqref{projective}, which in turn, gives rise to an equivalence between vectorial torsion and vectorial non-metricity. Therefore, we can say that gauging the Weyl group uniquely leads to the
Weyl quadratic theory of gravity.
It is  worth writing the version of this action in  $d$ dimensions
relevant for computing quantum corrections.
Since $R$ transforms covariantly, we do not really need a compensator field to
maintain Weyl gauge symmetry in $d$ dimensions, and write instead the following
analytical continuation \cite{DG1}
\begin{equation}
  S = \int d^dx \sqrt{g}\, \left[a_0\, R^2  +b_0\,
    F_{\mu \nu} F^{\mu \nu} + c_0\,
    \mathring C_{ \rho \sigma \mu \nu} \mathring C^{\rho \sigma \mu \nu } + d_0\, G \right]
  (R^{2})^{(d-4)/4} \, ,
\end{equation}
where we have already used eq.\,\eqref{weyl-torsion}.

\subsection{Total derivative terms} \label{aD}

We discuss here the terms that are total derivatives in four
dimensions and that may be added to the Weyl gravity action when
considering the theory at the quantum level (via dimensional
regularization) or in higher dimensions or on topologically
non-trivial space-times. For completeness, we also mention the parity
violating terms.

\subsubsection{Topological terms}
In four dimensions, one can construct two topological invariants at
quadratic order in the curvature two form $R^a{}_b$. The corresponding
terms are also invariant under gauged dilatations. We first consider
the case of a metric connection. The first invariant is the Euler
class $e(R)$ which, when integrated over the manifold, gives the Euler
characteristic. It is given by
\begin{equation}
\label{Euler-class}
  e(R) = \frac{1}{32 \pi^2} \epsilon_{abcd} R^{ab} \wedge R^{cd}
  = \frac{1}{32 \pi^2} d^4x \sqrt{g}\, \left(R^2
    - 4 R_{\mu \nu}R^{\nu \mu} + R_{\mu \nu \rho \sigma}
    R^{\rho \sigma \mu \nu} \right) \, ,
\end{equation}
where the position of the summed indices is essential when using a
connection with non-zero torsion. The expression above is a total
derivative for a metric connection in four space-time dimensions. In
arbitrary dimension, one simply defines the Euler-Gauss-Bonnet
combination to be given by\footnote{In arbitrary even dimension the
  Euler class is of order $d/2$ in the curvature and hence it cannot
  be considered in a quadratic action unless $d=4$.}
\begin{equation}
  G = R^2 - 4 R_{\mu \nu}R^{\nu \mu} + R_{\mu \nu \rho \sigma}
  R^{\rho \sigma \mu \nu} \, ,
\end{equation}
and it is no longer a total derivative in $d\neq 4$. For a general non-metric
connection the expression $\epsilon_{abcd} \tilde R^{ab} \wedge \tilde R^{cd}$ no longer gives the Euler class/characteristic \cite{Milnor}. However, as we show below, for the special case of vectorial non-metricity this is still true. Indeed, under a projective transformation
$\Om_\mu{}^a{}_b \mapsto \tilde \Om_\mu{}^a{}_b = \Om_\mu{}^a{}_b +
\w_\mu \delta^a_b$ the curvature two form transforms as
\begin{equation}
  R^a{}_b \mapsto \tilde R^a{}_b = R^a{}_b + \delta^a_b F \, .
\end{equation}
With the equation above it is now easy to see that the defining expression
of the Euler class is actually invariant under the projective
transformation
\begin{equation}
  \epsilon_{abcd} R^{ab} \wedge R^{cd}
  = \epsilon_{abcd} \tilde R^{ab} \wedge \tilde R^{cd} \, .
\end{equation}
The contraction of the two $\epsilon$-densities will yield, in general, an
expression of the form \cite{Babourova:1996id}
\begin{equation}
  \tilde G = \tilde R^2 - \left(\tilde R^{(1)}_{\mu \nu}
  + \tilde R^{(2)}_{\mu \nu} \right) \left(\tilde R^{(1) \nu \mu}
  + \tilde R^{(2)\nu \mu} \right)
  +\tilde R_{\mu \nu \rho \sigma} \tilde R^{\rho \sigma \mu \nu} = G \, ,
\label{GB-nonmetricity}
\end{equation}
where the traces of the curvature tensors are defined in \eqref{nm-traces}.
In the case of Weyl non-metricity they read
\begin{align}
  \label{traces-weyl}
  \tilde R^{(1)}_{\mu \nu} &= R_{\mu \nu}+F_{\mu \nu} \, , & \tilde R^{(2)}_{\mu \nu} &= R_{\mu \nu}-F_{\mu \nu} \, , & \frac1d \tilde R^{(3)}_{\mu \nu} &= F_{\mu \nu} \, .
\end{align}
The expression in \eqref{GB-nonmetricity} is hence the definition of the Euler-Gauss-Bonnet
term for the case of Weyl non-metricity and it is not hard to see that it corresponds to the definition already given in eq.~\eqref{nm-GB}. It is guaranteed to be a
topological invariant/total derivative in four dimensions since it is
equal to the $G$ given before for the metric case. \\
The second topological invariant, is the first Pontryagin class
$p_1(R)$ with the definition
\begin{equation}
  p_1(R) = -\frac{1}{8 \pi^2} \text{tr} (R \wedge R)
  = \frac1{32\pi^2} d^4x \, \epsilon^{\mu \nu \rho \sigma}
  R^a{}_{b \mu \nu} R^b{}_{a \rho \sigma} \, .
\end{equation}
Since it contains only one $\epsilon$-density the term above is odd
under parity and it is usually dropped from the action. It
can only be considered in four dimensions since only in this case it
is proportional to the volume form.\\
Finally, for the Pontryagin class one can see that it is not invariant
under the projective transformation. However, the extra term that is
generated is also a topological invariant. Indeed, we have
\begin{equation}
  \text{tr} (\tilde R \wedge \tilde R) = \text{tr} (R \wedge R)
  + d\,  F\wedge F \, ,
\end{equation}
where $d$ is the space-time dimension (and not the exterior derivative!). Actually, there are two invariants \cite{Babourova:1996id}
$\text{tr} (\tilde R \wedge \tilde R)$ above and
$\text{tr}\tilde R \wedge \text{tr} \tilde R = d^2 F \wedge F$ since in this case $\tilde R^{(ab)} \neq 0$. Notice that the trace is taken over the tangent space indices, $\text{tr} \tilde R = \tilde R^a{}_a$.

\subsubsection{Linear terms}

So far we concentrated on gauge theory type actions which are quadratic in curvatures. However, we can also write terms which are linear in curvatures and are still Weyl invariant.
 They are build making use of the gauge covariant
derivative $\hat \nabla$. There are three such terms even under parity
one may consider \cite{Wheeler2}, namely
\begin{equation}
  \hat \Box R = g^{\mu \nu}\hat\nabla_\mu \hat \nabla_\nu R\ ,
  \qquad \hat \nabla_\mu \hat \nabla_\nu R^{\mu \nu} \ ,
  \qquad \hat \nabla_\mu \hat \nabla_\nu F^{\mu \nu} \, ,
\end{equation}
and two terms odd under parity
\begin{equation}
  \frac{1}{\sqrt{g}} \epsilon^{\mu \nu \rho \sigma} \hat \nabla_\mu
  \hat\nabla_\nu R_{\rho \sigma} \sim \frac{1}{\sqrt{g}}
  \epsilon^{\mu \nu \rho \sigma} \hat \nabla_\mu \hat \nabla_\nu F_{\rho \sigma} \, ,
\end{equation}
which when multiplied by $\sqrt{g}$ yield a dilatation invariant in
four dimensions. The last two terms are not independent since one has
$R_{[\mu \nu]} =  \tfrac12(d-2) F_{\mu \nu}$. Note that all the terms above are of the form
\begin{equation}
  \label{current}
  \hat \nabla_\mu J^\mu \, ,
\end{equation}
with $J^\mu$ a current with Weyl charge $w_J=-4$. In the context of Weyl geometry we can show \cite{Wheeler2}
\begin{equation}
  \int d^d x \sqrt{g}\,  \hat \nabla_\mu J^\mu = \int  d^dx\,
  \partial_\mu (\sqrt{g} J^\mu) + (d+w_J) \int d^dx \sqrt{g}
  \, \w_\mu J^\mu \, ,
\end{equation}
and, in 4 dimensions, for the terms above, the coefficient of the last term vanishes and therefore they are all total derivatives.

Notice that in $d\neq4$ one needs to add a
dilaton $\phi$ in order to make the terms invariant and in this case
they are no longer total derivatives. Indeed, in this case one has an
integration by parts formula of the type
\begin{equation}
  \int d^d x \sqrt{g}\, \phi  \hat \nabla_\mu J^\mu = \int d^dx \,
  \partial_\mu (\sqrt{g} \phi J^\mu) - \int d^dx \sqrt{g}\,
  J^\mu \hat \nabla_\mu \phi \, ,
\end{equation}
where $w_\phi = - (d-4)$.

\subsection{Pure gauge limit (integrable Weyl geometry)}

There exists a special limit of action (\ref{action}) when
the dilatation gauge field is ``pure gauge'' or an exact one-form
which implies $F_{\mu\nu}=0$. Conversely, one may impose $F_{\mu\nu}=0$
which means that {\it locally} $\w_\mu$ is an exact one-form.
In both cases this means that $\w_\mu$ is not a dynamical gauge field
and can be integrated out from (\ref{action}). In $d=4$ dimensions
the last term in (\ref{action}) is also a total derivative so it can be
ignored. In this case the action simplifies into
\begin{equation}\label{fg}
  S_g=\int d^4x \sqrt{g} \,\Big[ a_0 R^2 +
  c_0 \mathring C_{\rho\sigma\mu\nu} \mathring C^{\rho\sigma\mu\nu}\Big]\, ,
\end{equation}
where the Weyl term is independent of $\w_\mu$, while $R$ depends on
it as in (\ref{riem-3}). The action above is equivalent to the following one
\begin{equation}
S_g = \int d^4 x \sqrt{g} \left[-a_0 \left( 2 \phi^2 R +\phi^4 \right)+ c_0 \mathring C_{\rho\sigma\mu\nu} \mathring C^{\rho\sigma\mu\nu} \right] \, ,
\end{equation} 
with a scalar field $\phi$ satisfying the equation of motion $\phi^2 = -R$ and the condition $\phi\neq 0$ (and thus $R\neq 0$). The equation of motion for $\w_\mu$ is algebraic and can be solved to give
\begin{equation}
\w_\mu = \frac2{d-2} \partial_\mu \ln \phi\, ,
\end{equation}
consistent with the fact that $\w_\mu$ is pure gauge. Putting the solution above back into the action one finds in Riemannian notation
\begin{equation}
\label{pure-gauge}
S_g = \int d^4x \sqrt{g} \left\{- a'_0 \left[\frac{d-2}{8(d-1)} \phi^2 \mathring R + \frac12 \partial_\mu \phi \partial^\mu \phi \right] + c_0 \mathring C_{\rho \sigma \mu \nu} \mathring C^{\rho \sigma \mu \nu}\right\} \, ,
\end{equation}
where $a_0' = \tfrac{16(d-1)}{d-2} a_0$.
For the sake of generality we left the dimension arbitrary inside the Lagrangian. However, one should take $d=4$ to be consistent with the fact that we neglected the Euler term. Note that in this case the symmetry that is actually left is not Weyl
gauge symmetry but its (local) restriction without a Weyl
gauge field (local Weyl symmetry). For an extensive discussion about
this case and relation to the general case, see \cite{Ghilencea:2022lcl,SMW2}. We show later on that one can obtain the same action when considering the full conformal group as a gauge/local symmetry. In view of this we conclude that from Weyl gravity one can obtain conformal gravity in the pure gauge limit (also called integrable Weyl geometry).

\subsection{Minimal couplings to matter fields}

In this section we review the classical couplings that the
dilatation gauge field $\w_\mu$ can have to matter fields and also
present some less known couplings.

Consider first a scalar field $\varphi$ with Weyl weight $q_\varphi$; its
covariant derivative is then
$\hat D_\mu \varphi = (\partial_\mu + q_\varphi \w_\mu)\varphi$ and the
corresponding action is Weyl gauge invariant
\begin{equation}
  S_{\varphi} = \int d^d x \sqrt{g}\,  g^{\mu \nu} \hat D_\mu \varphi \hat D_\nu \varphi \, ,
\end{equation}
if the charge is  $q_\varphi = - (d-2)/2$. Hence,
the dilatation gauge field $\w_\mu$ can couple to scalars, in particular
to the Higgs field of the SM, see Section 2.5 in \cite{SMW} for details.

Regarding the couplings of $\w_\mu$ to a Dirac field $\psi$ it is
well-known that this coupling vanishes \cite{Kugo,Hayashi:1976uz}, for a recent
review see \cite{SMW} (Section 2.3).  Hence the Dirac action with
(Weyl) gauged dilatation symmetry is identical to the Dirac action in
the Riemannian space-time.

A gauge boson $A_\mu^\alpha$ with an internal gauge symmetry algebra
does not couple to gauged dilatations field $\w_\mu$ simply because
the corresponding generators commute. Indeed, the contribution to the
gauge curvatures is of the form
\begin{equation}
  R_{\mu \nu}^A \supset A_\mu^\alpha \w_\nu f_{\alpha \#}{}^A = 0 \, ,
\end{equation}
and it automatically vanishes.\footnote{A mixing at the level of the kinetic terms of the Weyl gauge boson and the hipercharge boson is allowed by Weyl gauge symmetry. This was analysed in \cite{SMW}. However, this coupling is not influenced by the torsion or non-metricity picture and therefore has no relevance in the torsion/non-metricity duality discussed in this paper.} Therefore, the SM with
a vanishing Higgs mass parameter and kinetic term as above (adjusted
to account for SM gauge group) and with an action ``upgraded'' to the
curved-space time, is gauged dilatations invariant and thus can
naturally be embedded in Weyl geometry without any new degrees of
freedom beyond SM and Weyl geometry \cite{SMW}.

Finally, we consider the minimal coupling of a Rarita-Schwinger (or
gravitino) field $\Psi_\mu$ to the dilatation field. The action that
we consider has the form
\begin{equation}
  S_{\Psi} = \int d^dx \sqrt{g} \left(\bar \Psi_\mu \gamma^{\mu \nu \rho}
    \hat D_{\nu} \Psi_\rho
    - \hat D_\nu \bar \Psi _\mu \gamma^{\mu \nu \rho} \Psi_\rho\right) \, ,
\end{equation}
where $\gamma^{\mu \nu \rho}$ denotes the antisymmetrised product of three $\gamma$-matrices, $\gamma^{\mu \nu \rho} = \gamma^{[\mu} \gamma^\nu \gamma^{\rho ]} = \tfrac1{3!} (\gamma^\mu \gamma^\nu \gamma^\rho \pm \mathrm{permutations})$, with $\gamma^\mu$ defined from the tangent space $\gamma-$matrices, $\gamma^a$, with the help of the vielbein $\gamma^\mu = e_a^\mu \gamma^a$. 
In the absence of a dilaton/compensator\footnote{In supergravity the gravitino field is the gauge field associated to local supersymmetry transformations and its transformation under dilatations is fixed by the superconformal algebra and is different from what we consider here, requiring a dilaton in order to write a kinetic term for it.}, invariance of
the kinetic term fixes the Weyl weight to be $q_{\Psi} = -
(d-3)/2$. The Weyl gauge covariant derivative in the action above acts
as
\begin{equation}
  \hat D_\mu \Psi_\nu = \left( \partial_\mu  + q_{\Psi} \w_\mu
  + \frac14 \Om_\mu{}^{ab} \gamma_{ab} \right) \Psi_\nu \, .
\end{equation}
Calculations similar to the case of a Dirac field
show that the coupling of fermions to the dilatation gauge field
$\w_\mu$ (via the charge $q_\Psi$) cancels out independently of the
value of this charge.
One is then left with the following Weyl gauge invariant action:
\begin{equation}
  \label{RS}
  \mathcal{L}_\Psi = \sqrt{g} \left(\bar \Psi_\mu \gamma^{\mu \nu \rho}
    \partial_\nu \Psi_\rho - \partial_\nu \bar \Psi_\mu \gamma^{\mu \nu \rho}
    \Psi_\rho + \frac14 \bar \Psi_\mu \Om_\nu{}^{ab}
    \{\gamma^{\mu \nu \rho}, \gamma_{ab} \} \Psi_\rho\right) \, .
\end{equation}
After some algebra, in the case of Weyl gravity, one finds that the
last term above reduces to
\begin{equation}
  \mathcal{L}_\Psi^\w = 4 \sqrt{g}\, \bar\Psi_\mu \w^{[\mu} \gamma^{\nu]}
  \Psi_\nu \, .
\end{equation}
Hence, we find in this case that there is a non-trivial coupling of
the spin 3/2 fermion to the dilatation gauge field $\w_\mu$ via the
torsion/non-metricity part of the spin connection.  The projective
transformation
$\Om_\mu{}^{ab} \mapsto \Om_\mu{}^{ab} + \w_\mu\eta^{ab}$ that maps
vectorial torsion into vectorial non-metricity leaves invariant the
fermion coupling $(1/4) \Om_\mu{}^{ab} \gamma_{ab}$.

\section{Comparison with the gauging of the full conformal group}
\label{sec3}

Let us finally consider the case of gauging the full conformal group \cite{KTN1}
and compare to the results so far of gauging only the Weyl
group. The conformal group includes
the Weyl group as a subgroup, but in addition also contains special
conformal transformations. The gauge algebra in this case
  consists of the following commutators 
\begin{equation}
\begin{split}
 &  [P_{a}, M_{bc}] =  \eta_{a b} P_{c} - \eta_{ac}P_b\, , \qquad   [D,P_a] = P_a\, ,
  \qquad [P_a, P_b] = 0 \, , \qquad    [D, M_{ab}] = 0 \, , \\
& [M_{ab}, M_{cd}] = \eta_{a c}M_{d b} - \eta_{bc}M_{da} - \eta_{ad} M_{cb} + \eta_{bd} M_{ca}\, ,\\
 & [K_{a}, M_{bc}] = \eta_{ab} K_{c} - \eta_{ac} K_b\, ,  \qquad  [P_a,K_b] = 2 (\eta_{ab} D + M_{ab}) \, ,  \qquad [D,K_a] = - K_a \, ,
  \end{split}
\end{equation} 
involving the extra generator $K_a$. A first observation is that dilatations are no longer factorised from the rest of the algebra by a semi-direct product structure (like in the Weyl case). The reason is that the vector space generated by $\{P_a, M_{ab}, K_a \}$, though invariant under the adjoint action of $D$, is not a Lie subalgebra (because of the commutator $[P_a, K_b]$). This will have important consequences as we shall see shortly. One can extract the structure constants from the algebra above to obtain
\begin{align}
& f_{[ab][cd]}{}^{[ef]} =2\left(\eta_{cb} \delta^{ef}_{ad} - \eta_{db} \delta^{ef}_{ac}
  - \eta_{ca} \delta^{ef}_{bd} + \eta_{da} \delta^{ef}_{bc} \right) \, ,\\
   &f_{a[bc]}{}^d =  \eta_{ab} \delta_{c}^d - \eta_{ac} \delta_b^d  \, , \qquad f_{\hat a[bc]}{}^{\hat d} =  \eta_{ab} \delta_{c}^d - \eta_{ac} \delta_b^d \, , \qquad  f_{a \hat b}{}^\#=2\eta_{ab}\, , \nonumber\\
& f_{a \hat b}{}^{[cd]} = 2 (\delta_a^{c}\, \delta_b^{d} -\delta_a^{d}\, \delta_b^{c} ) \, , \qquad   f_{\#a}{}^b = \delta_a^b \, , \qquad f_{\# \hat a}{}^{\hat b} = - \delta_a^b \, ,
\end{align}
where we have introduced the notation $\hat a, \ldots$ for vectorial indices related to the generator of special conformal transformations $K_a$. Aside from the fields already introduced in the case of the Weyl
group, we now have $f_\mu{}^a$ as the gauge field of special conformal
transformations.  By the methods that we have described in detail in our paper, we can write the gauge variations of the fields
\begin{align}
  \delta_\epsilon e_\mu^a
   &=- \partial_\mu \xi^a - \xi_b \Om_\mu{}^{ab}-\xi^a \w_\mu
     + \lambda^a{}_b e_\mu^b + \lambda_D e_\mu^a \, , \label{conf-1}\\
  \delta_\epsilon \Om_\mu{}^{ab}
   & = -\partial_\mu \lambda^{ab}- 4 \xi^{[a} f_\mu{}^{b]}
     - 2 \lambda^{[a}{}_c \Om_\mu{}^{b]c}
     + 4 \lambda_K^{[a} e_\mu{}^{b]} \, ,\label{conf-2}\\
  \delta_\epsilon \w_\mu
   & = -\partial_\mu \lambda_D +2 \xi_a f_\mu{}^a- 2 \lambda_K^a e_{\mu a}\, ,
     \label{conf-3}\\
  \delta_\epsilon f_\mu{}^a
   & = - \partial_\mu \lambda_K^a+ \lambda^a{}_b f_\mu{}^b
     + \w_\mu \lambda_K^a- \Om_\mu{}^{ab} \lambda_{K b} - \lambda_D f_\mu{}^a \, ,
     \label{conf-4}
\end{align}
where the variation $\delta_\epsilon$ with respect to $\epsilon = \epsilon^A T_A $ now has an additional term
\begin{equation}
\label{parameter-conformal}
  \epsilon = \xi^a P_a + \frac12 \lambda^{ab} M_{ab} + \lambda_D D
  + \lambda_K^a K_a \, .
\end{equation}
From eq.\eqref{conf-3} we see that the gauge field of dilatations can
be set zero (or any other value) by a special conformal transformation. The corresponding term $-2 \lambda_K^a e_{\mu a}$ arises precisely because of the commutator $[P_a, K_b] \supset 2 \eta_{ab}D$ which spoils the semi-direct product structure. Hence, in the case
of conformal gravity $\w_\mu$ cannot be a dynamical field. This is in stark contrast with respect to Weyl quadratic gravity where it is dynamical and plays an essential role in the whole structure of the theory.

The field strengths of Weyl gravity, \eqref{rotations} and \eqref{dilatations}, are not covariant with respect to special conformal transformations
\begin{align}
\delta_K F_{\mu \nu} &= -4 \left( \partial_{[\mu} \lambda_K^a - \w_{[\mu} \lambda_K^a + \lambda_K^b \Om_{[\mu}{}^a{}_b \right) e_{\nu] a} = -4 \partial_{[\mu} \lambda_{K \nu]} \, ,\\
\delta_K R^{ab}{}_{\mu \nu} & =  8 \left(\partial_{[\mu} \lambda_K^{[a}-\w_{[\mu} \lambda_K^{[a} + \lambda_K^{b} \Om_{[\mu}{}^{[a}{}_b  \right) e_{\nu]}{}^{b]} + 4 \lambda_K^{[a} R_{\mu \nu}(P^{b]})\, .
\end{align}
Hence, new covariant field strengths need to be introduced for the conformal case
\begin{align}
R_{\mu \nu}(P^a) &= 2(\partial_{[\mu} + \w_{[\mu}) e_{\nu]}{}^a + 2 \Om_{[\mu}{}^{ab}\, e_{\nu] b} =2 \hat  D_{[\mu} e_{\nu]}{}^a \, , \label{fs-conf-1}\\
R_{\mu \nu} (M^{ab}) & = 2 \partial_{[\mu} \Om_{\nu]}{}^{ab} + 2 \Om_{[\mu}{}^a{}_c\, \Om_{\nu]}{}^{cb} + 8 f_{[\mu}{}^{[a} e_{\nu ]}{}^{b]}  \equiv \mathcal{R}^{ab}{}_{\mu\nu}\, ,\label{fs-conf-2}\\
R_{\mu \nu}(D) & = 2 \partial_{[\mu} \w_{\nu]} - 4 f_{[\mu}{}^a e_{\nu] a} \equiv \mathcal F_{\mu \nu}\, , \label{fs-conf-3}\\
R_{\mu \nu} (K^a) & = 2(\partial_{[\mu} - \w_{[\mu}) f_{\nu]}{}^a + 2 \Om_{[\mu}{}^{ab} f_{\nu] b} \equiv \mathcal{K}_{\mu \nu}{}^a \, , \label{fs-conf-4}
\end{align}
where we introduced the conformal ``Riemann'' tensor ${\mathcal R}^{ab}{}_{\mu \nu}$, the notation $\mathcal F_{\mu\nu}$, $\mathcal K_{\mu \nu}{}^a$ for the dilatation and the special conformal field strength respectively. We also defined $f_{\mu \nu}:=f_\mu{}^a e_{\nu a} $. Above, one can see that additional terms depending on $f_\mu{}^a$ arise. In particular, the field strength of dilatations is no longer Abelian since the semi-direct product structure is lost. Notice that $R_{\mu \nu}(P^a)$ is the same as in the case of the Weyl group and hence the solution $\Om_\mu{}^a{}_b$ of the corresponding constraint is the one given in eq.\,\eqref{spin-invariant} which, in particular, does not depend on the gauge field of special conformal transformations. The variations of the field strengths under $\epsilon$ in eq.\,\eqref{parameter-conformal} are found to be   
\begin{align}
\delta_\epsilon R_{\mu \nu}(P^a) &= -\xi^b R_{\mu \nu}(M^{a}{}_b)  - \xi^a  R_{\mu \nu}(D) + \lambda^a{}_b R_{\mu \nu}(P^b) + \lambda_D R_{\mu \nu} (P^a)\, ,\\
\delta_\epsilon R_{\mu \nu} (M^{ab}) & = - 4\xi^{[a} R_{\mu\nu}(K^{b]}) -2 \lambda^{[a}{}_c R_{\mu \nu}(M^{b]c}) + 4 \lambda_K^{[a} R_{\mu \nu}(P^{b]})\, ,\\
\delta_\epsilon R_{\mu \nu}(D) & = 2\xi_aR_{\mu \nu}(K^a)\, , \\
\delta_\epsilon R_{\mu \nu} (K^a) &  = \lambda^a{}_b R_{\mu \nu}(K^b)  - \lambda_D R_{\mu \nu}(K^a)+ \lambda_K^a R_{\mu \nu} (D) + \lambda_{K}{}_b R_{\mu \nu}(M^{ba}) \, . 
\end{align}
The relations above determine the action of the gauge covariant derivative $\mathcal D_\mu$ (involving also translations) on the various field strengths. It is easy to see that we have
\begin{align}
\mathcal{D}_\mu R_{\nu \rho} (P^a) &= \partial_\mu R_{\nu \rho} (P^a) - e_\mu^b R_{\nu \rho}(M^a{}_b) - e_\mu^a R_{\nu \rho}(D)+  \omega_{\mu}{}^a{}_b R_{\nu \rho} (P^b) + b_\mu R_{\nu \rho} (P^a)\, ,\\
\mathcal{D}_\mu R_{\nu \rho} (M^{ab}) &= \partial_\mu R_{\nu \rho} (M^{ab}) + 2 \omega_{\mu c}{}^{[a} R_{\nu \rho}(M^{b]c}) -4 e_\mu^{[a} R_{\nu \rho}(K^{b]})+ 4 f_\mu{}^{[a} R_{\nu \rho} (P^{b]})\, ,\\
\mathcal{D}_\mu R_{\nu \rho} (D) & = \partial_\mu R_{\nu \rho} (D)+2 e_\mu^a R_{\nu \rho}(K_a)\, , \\
\mathcal{D}_\mu R_{\nu \rho} (K^a) & = \partial_\mu R_{\nu \rho}(K^a) + \omega_{\mu}{}^a{}_b R_{\nu \rho}(K^b)- \w_\mu R_{\nu \rho}(K^a)+ f_{\mu b} R_{\nu \rho}(M^{ba}) + f_\mu{}^a R_{\nu \rho}(D)\, .
\end{align}
Assuming the constraint on the curvature of translations and using the notation introduced in eqs.\,\eqref{fs-conf-2}-\eqref{fs-conf-4} one arrives at the following Bianchi identities
\begin{align}
& \mathcal R^a{}_{[\mu \nu \rho]} = -e_{[\mu}^a \mathcal F_{\nu \rho]} \, , \qquad D_{[\mu} \mathcal R^{ab}{}_{\nu \rho]}  = 4 e_{[\mu}^a \mathcal K_{\nu \rho]}{}^a\, , \qquad \partial_{[\mu} \mathcal F_{\nu \rho]}  = -2 e_{[\mu}^a \mathcal K_{\nu \rho] a}\, , \label{cf-bianchi1}\\
 & D_{[\mu} \mathcal K_{\nu \rho]}{}^a  =\w_{[\mu}\mathcal K_{\nu \rho]}{}^a - f_{[\mu}{}^a \mathcal F_{\nu \rho]} - f_{[\mu}{}^b \mathcal R_{|b|}{}^a{}_{\nu \rho]} \, . \label{cf-bianchi2}
\end{align}
\subsection{Quadratic action}
In order to correctly identify the independent terms in the gauged conformal action we need to understand the symmetry properties of the tensors that we introduced. The four tensor $\mathcal R_{\rho \sigma \mu \nu} : = e_{a\rho} e_{b \sigma} \mathcal R^{ab}{}_{\mu \nu}$ is antisymmetric in both the first pair and the last pair of indices. When exchanging the first pair with the last pair of indices one has the identity
\begin{equation}
\begin{split}
{\mathcal R}_{\rho \sigma \mu \nu } - {\mathcal R}_{ \mu \nu \rho \sigma} &= -\mathcal F_{\mu \rho} g_{\nu \sigma} + \mathcal F_{\mu \sigma} g_{\nu \rho} + \mathcal F_{\nu \rho}g_{\mu \sigma} - \mathcal F_{\nu \sigma} g_{\mu \rho} \, ,
\end{split}
\end{equation}
which follows from the first Bianchi identity \eqref{cf-bianchi1}. There is only one independent trace, the conformal ``Ricci" $\mathcal R_{\mu \nu}:= g^{\rho \sigma} \mathcal R_{\rho \mu \sigma \nu}$ and its contraction $\mathcal R : = g^{\mu \nu} \mathcal R_{\mu \nu}$. The antisymmetric part of $\mathcal R_{\mu \nu}$ is related to $\mathcal F_{\mu \nu}$ via the identity
\begin{equation}
\mathcal R_{\mu \nu} - \mathcal R_{\nu \mu} = (d-2) \mathcal F_{\mu \nu} \, ,
\end{equation} 
which one can obtain again from the first Bianchi identity in \eqref{cf-bianchi1} or from the relation above by contraction with the metric. 
The most general action quadratic in the field strengths and invariant under parity can be inferred like in the Weyl group case by noting that we have the similar identities
\begin{align}
  \mathcal R_{\rho \sigma \mu \nu } \mathcal R^{ \rho \sigma \mu \nu} &= \mathcal R_{\rho \sigma \mu \nu }
     \mathcal R^{  \mu \nu \rho \sigma}
    + 2(d-2) \mathcal F_{\mu \nu} \mathcal F^{\mu \nu} \, , \label{idc-1}\\
  \mathcal R_{\rho \sigma \mu \nu } \mathcal R^{\mu \sigma \nu \rho } & =
     - \frac12 \mathcal R_{\rho \sigma \mu \nu } \mathcal R^{\rho \sigma \mu \nu }
     +\frac{d-2}{2} \mathcal F_{\mu \nu} \mathcal F^{\mu \nu} \, ,\label{idc-2}
 \end{align}
 and
\begin{align} 
  \label{idc-3}
\mathcal  R_{\mu \nu} \mathcal R^{\mu \nu} &= \mathcal R_{\mu \nu} \mathcal R^{\nu \mu} + \frac{(d-2)^2}{2} \mathcal F_{\mu \nu}\mathcal F^{\mu \nu} \, ,
    \\
  \mathcal  R_{\mu \nu} \mathcal F^{\mu \nu} &= \frac12 (d-2) \mathcal F_{\mu \nu} \mathcal F^{\mu \nu} \label{idc-4} \, ,
\end{align}
matching the ones in Weyl gravity \eqref{id-1}-\eqref{id-4} with the substitutions $R \to \mathcal R$ and $F \to \mathcal F$. What happens with the terms involving the field strength of special conformal transformations $R_{\mu \nu}(K^a)$? It turns out that the only term invariant under dilatations that can be written is
\begin{equation}
\sqrt{g}\, R_{\mu \nu}(P^a) R^{\mu \nu}(K_a) \, ,
\end{equation}
but {\it it is not} invariant under special conformal transformations. Moreover,
the standard kinetic term for the gauge field of special conformal
transformations 
\begin{equation}
\sqrt{g}\, R_{\mu \nu}(K^a) R^{\mu \nu}(K_a)\, ,
\end{equation}
is not invariant under dilatations because the Weyl charge of the field
strength $R_{\mu \nu}(K^a) $ is equal to $-1$ (and neither under special conformal transformations). Therefore, $f_\mu{}^a$ is
an auxiliary field in conformal gravity \cite{Wheeler2}. In order to find a convenient basis of quadratic terms for the action it is useful to define a Weyl conformal tensor $\mathcal C_{\rho \sigma \mu \nu}$ by extracting the traces of $\mathcal R_{\rho \sigma \mu \nu}$. We have the analogous formula
\begin{equation}
  \mathcal C_{ \rho \sigma \mu \nu}= \mathcal R_{ \rho \sigma \mu \nu}
  + \frac1{d-2} \left(\mathcal R_{ \sigma \mu} g_{\nu \rho} - \mathcal R_{ \rho \mu} g_{\nu \sigma}
    + \mathcal R_{ \rho \nu} g_{\mu \sigma} - \mathcal R_{ \sigma \nu} g_{\mu \rho} \right) 
  +  \mathcal R \frac{(g_{\mu \rho} g_{\nu \sigma} - g_{\mu \sigma} g_{\nu \rho})}{(d-1) (d-2)} \, ,
\end{equation}
and it turns out that the tensor above does not depend on $f_\mu{}^a$ or $\w_\mu$ at all. Indeed, one can check that we have
\begin{equation}
\mathcal C_{\rho \sigma \mu \nu} = C_{\rho \sigma \mu \nu} = \mathring C_{\rho \sigma \mu \nu}\, ,
\end{equation}
with $C_{\rho \sigma \mu \nu}$ defined in the Weyl case \eqref{weyl-tensor}. Furthermore, we also decompose $\mathcal R_{\mu \nu}$ and $f_{\mu \nu}$ into algebraically irreducible components (antisymmetric, symmetric traceless and trace) according to
\begin{align}
f_{\mu \nu} &= f_{[\mu \nu]} + f^0_{(\mu \nu)} + \frac1d g_{\mu \nu} f\, , \\
\mathcal R_{\mu \nu} &= \mathcal R_{[\mu \nu]} + \mathcal R^0_{(\mu \nu)} + \frac1d g_{\mu \nu} \mathcal R \, ,
\end{align}
with $g^{\mu \nu} f^0_{(\mu \nu)} = g^{\mu \nu} \mathcal R^0_{(\mu \nu)}= 0 $ and the trace $f : = f_\mu{}^a e_a^\mu = g^{\mu \nu} f_{\mu \nu}$. With these considerations in mind we find that the most general action for the gauge theory of the conformal group can be written as
\begin{equation}
 \label{action-conformal}
S = \int d^4 x \sqrt{g} \left[a_1 \mathcal R^2 + a_2 \mathcal R^0_{(\mu \nu)} \mathcal R^{0(\mu \nu)} + a_3 \mathcal C_{\mu \nu \rho \sigma} \mathcal C^{ \mu \nu \rho \sigma} + a_4 \mathcal R_{[\mu \nu]} \mathcal R^{[\mu \nu]} \right] \, ,
\end{equation}
which depends at this point on $e_\mu^a$, $b_\mu$ and $f_\mu{}^a$ (with $\Om_\mu{}^a{}_b$ fixed by the constraint to depend only on the vielbein and the dilatation gauge field). Recall also the identity $\mathcal R_{[\mu \nu]} = \tfrac{d-2}{2} \mathcal F_{\mu \nu}$ and thus the action above depends on the field strength of dilatations. The equation of motion for $f_\mu{}^a$ is purely algebraic and can be solved. In order to do this, it is useful to extract the dependence on $f_\mu{}^a$ of the various terms in the action. The connection between the field strengths of conformal gravity and those of Weyl gravity are useful
\begin{align}
\mathcal R_{\rho \sigma \mu \nu} & = R_{\rho \sigma \mu \nu} + 2(f_{\mu \rho} g_{\nu \sigma} -f_{\mu \sigma} g_{\nu \rho} - f_{\nu \rho} g_{\mu \sigma} +f_{\nu \sigma} g_{\mu \rho})\, ,\\
\mathcal R_{\mu \nu} & = R_{\mu \nu} + 2(d-2) f_{\nu \mu} + 2 g_{\mu \nu} f\, ,\\
\mathcal R & = R + 4(d-1) f \, ,
\end{align}
and further for the irreducible components of $\mathcal R_{\mu \nu}$ in terms of the ones of $f_{\mu \nu}$ we have
\begin{align}
\mathcal R_{[\mu \nu]} & = R_{[\mu \nu]} - 2(d-2)f_{[\mu \nu]}\, , & \mathcal R^0_{(\mu \nu)} & = R_{(\mu \nu)}^0 + 2(d-2) f^0_{(\mu\nu)}\, .
\end{align}
Varying the action \eqref{action-conformal} with respect to the independent components $f$, $f_{(\mu \nu)}^0$ and $f_{[\mu \nu]}$  we obtain the corresponding equations of motion
\begin{align}
a_1 \mathcal R & = 0\, , & a_2 \mathcal R^0_{(\mu \nu)} & = 0\, , & a_4 \mathcal R_{[\mu \nu]} & = 0\, .
\end{align}
For non-zero coefficients $a_1 \neq 0 $, $a_2 \neq 0$ and $a_4 \neq 0$ the equations above can be written compactly as
\begin{equation}
\mathcal R_{\mu \nu} = e_a^\rho e_{b \nu} R_{\rho \mu}(M^{ab}) = 0 \, .
\end{equation}
In practice, the relation above is usually imposed by hand as a constraint similar to the one on $R_{\mu \nu}(P^a)$ \footnote{There are cases when this relation is actually different from the equation of motion of $f_\mu{}^a$, but as we shall explicitly show in the next section the resulting theory is the same either solving the constraint or the equations of motion.}. Either way, the solution can be found to be
\begin{equation}
f_{\mu \nu} = - \frac1{2(d-2)} R_{\nu \mu} + \frac{1}{4(d-1)(d-2)} g_{\mu \nu} R\, ,
\label{solfd}
\end{equation}
which on the components reads
\begin{align}
f_{[\mu \nu]} & =  \frac1{2(d-2)} R_{[\mu \nu]}\, , & f_{(\mu \nu)}^0 & = - \frac1{2(d-2)} R_{(\mu \nu)} + \frac1{2d(d-2)} g_{\mu \nu} R\, , & f & = - \frac1{4(d-1)} R \, .
\end{align}
If some of the coefficients $a_1$, $a_2$ or $a_4$  vanish  the corresponding (irreducible) component of $f_{\mu \nu}$ remains undetermined, but also drops out of the action. Therefore, the action is completely determined by the symmetry and, for arbitrary coefficients, after integrating $f_\mu{}^a$ out, one is left with the well-known result of \footnote{For actions with conformal symmetry in the context of MAG see \cite{Paci:2023twc}.}
\begin{equation}
S = \int d^4 x \sqrt{g} \, a_3 \mathring C_{\rho \sigma \mu \nu} \mathring C^{\rho \sigma \mu \nu}\, , \label{ungauged}
\end{equation}
which only depends on the vielbein $e_\mu^a$. It is important to stress that after eliminating $f_\mu{}^a$ from its equation of motion, the dilatation gauge field also drops out of the action. This could have been argued {\it apriori} since we are writing an action invariant under special conformal transformations and after eliminating $f_\mu{}^a$ the only field transforming non-trivially is $\w_\mu$. Therefore, it has to drop from the action \cite{VanNieuwenhuizen:1981ae}. One does not need to set it to zero by gauge fixing special conformal transformations.

The above action has an (ungauged) local conformal symmetry. Therefore, the standard gauging of the conformal group yields a
somewhat curious theory where the extra gauge fields beyond the
Poincar\'e subgroup are non-dynamical\footnote{Note that it is for
  this reason that one could construct Poincar\'e gravity and
  supergravity as gauge fixed theories with conformal/superconformal
  symmetry \cite{Kaku:1978ea, KTN2, Freedman}.}. Comparing with the results in the previous section we find that Weyl quadratic gravity is more general  as it contains conformal gravity as its pure gauge limit.   
  
In conclusion, the main difference between Weyl and conformal
(quadratic) gravity is that the dilatation gauge field is dynamical in
the former and is ``pure gauge'' (or outright absent from the action)
in the latter. In other words, conformal gravity is not a true gauge
theory since the corresponding gauge bosons $f_\mu{}{}^a$ and $\w_\mu$ are non-dynamical. Therefore, only gauged dilatations and the
associated Weyl geometry can give a true (and anomaly-free) gauge
theory of a four-dimensional space-time symmetry of the action beyond Poincar\'e. Weyl
geometry is the unique underlying geometry that realises a gauge
theory of dilatations, even in the absence of matter.

\subsection{Conformal scalar field}

Adding matter fields to the setup above does not change the main features already outlined: the gauge field of special conformal transformations $f_\mu{}^a$ is still an auxiliary field, the gauge field of dilatations drops out of the action and the final action is completely fixed by the symmetry. We exemplify this by adding a scalar field in the theory.
 An invariant for the kinetic term can be constructed by making use of the gauge covariant derivative to build the conformal d'Alembertian. Let us assume that the scalar field $\phi$ transforms covariantly under dilatations and is invariant under special conformal variations
\begin{align}
\delta_D \phi &=  w \lambda_D \phi\, , & \delta_K \phi &= 0 \, ,
\end{align}
then its covariant derivative is given by
\begin{equation}
 \hat D_a \phi := e_a{}^\mu(\partial_\mu + w b_\mu)\phi \, .
\end{equation}
This term transforms non-trivially under special conformal transformations due to the presence of the dilatation gauge field. In fact the total variation $\delta_\epsilon$ (excluding translations, as we are interested in $\hat D_a$) is given by
\begin{equation}
\delta_\epsilon \hat D_a \phi = (w-1)\lambda_D \hat D_a \phi + \lambda_a{}^b \hat D_b \phi - 2 w \lambda_{Ka} \phi \, ,
\end{equation}
and hence taking the second gauge covariant derivative acts as
\begin{equation}
\hat D_b \hat D_a \phi= e_b^\mu \left[\partial_\mu \hat D_a \phi + (w-1) b_\mu \hat D_a \phi + \omega_{\mu ac} \hat D^c \phi - 2 w f_{\mu a} \phi \right] \, .
\end{equation}
Thus, from above it follows that the conformal d'Alembertian $\hat \Box$ acting on a scalar field has the expression \cite{Freedman}
\begin{equation}
\hat \Box \phi = g^{\mu \nu} \hat \nabla_\mu \hat \nabla_\nu \phi = \eta^{ab} \hat D_a \hat D_b \phi = e^{a \mu} \left[\partial_\mu \hat D_a \phi + (w-1) b_\mu \hat D_a \phi + \omega_{\mu ac} \hat D^c \phi - 2 w f_{\mu a} \phi \right]\, ,
\end{equation}
where one uses the definition $\hat \nabla_\mu V^\nu = e_a^\nu \hat D_\mu V^a $.
For generic Weyl weight $w$ the operator $\hat \Box \phi$ is not good for constructing an invariant action. This is due to the fact that it transforms as
\begin{equation}
\delta_\epsilon \hat \Box \phi= (w-2) \lambda_D \hat \Box \phi - 2(2w +d-2) \lambda_K^a \hat D_a \phi \, ,
\end{equation}
under the same variation excluding translations. Hence for the invariance of the kinetic term (both under dilatations and special conformal transformations) the Weyl weight has to be given by
\begin{equation}
w = -\frac{d-2}{2} \, .
\end{equation}
The most general (two-derivative) action that one can write for the scalar field $\phi$ coupled to conformal gravity, encoded in the tensor $\mathcal R_{\mu \nu \rho \sigma}$ and its contractions, is
\begin{equation}
S_\phi = -\int d^4 x \sqrt{g} \left[\frac{\epsilon}{2} \phi \hat \Box \phi + b_1 \phi^2 \mathcal R + b_2 \phi^4 \right]\, ,
\end{equation} 
where $\epsilon = \pm 1$ is related to the fact whether $\phi$ is a ghost field or not. Let us analyze what happens with the above action if one imposes the standard constraint
\begin{equation}
\mathcal{R}_{\mu \nu} = 0 \, ,
\end{equation}
with the solution given in eq.\,\eqref{solfd}. In this case one can show that the conformal d'Alembertian reduces to
\begin{equation}
\hat \Box \phi = e^{a\mu} \mathcal{\mathring D}_\mu \partial_a \phi + \left(w+ \frac{d-2}{2} 			\right) b^a (2 \partial_a + w b_a) \phi + \frac{w}{2(d-1)} \mathring R \phi \, ,
\end{equation}
which is valid for any Weyl weight $w$. 
If one further imposes $w= -\frac{d-2}{2}$ the result simplifies to
\begin{equation}
\left.\begin{split}
&\mathcal R_{\mu \nu} = 0\\
& w = -\frac{d-2}{2}
\end{split} \right\}
\qquad \implies \qquad 
 \hat \Box \phi = \mathring \Box \phi - \frac{d-2}{4(d-1)} \mathring R \phi \, ,
\end{equation}
where we used the identity 
\begin{equation}
e^{a \mu} \mathcal{\mathring D}_\mu \partial_a \phi = \mathring \Box \phi \, .
\end{equation}
Hence the Lagrangian of (quadratic) conformal gravity coupled to a scalar field (after an integration by parts) reads\footnote{Notice that higher derivative terms in the scalar field are also compatible with local conformal symmetry. For instance the following term is invariant: 
$\mathcal L_{h.d.} = c\left[\mathring R - 2(d-1) \mathring \Box \phi - (d-1)(d-2) (\mathring\nabla \phi)^2 \right]^2\, .$}
\begin{equation}
\mathcal L = a \mathring C_{\mu \nu \rho \sigma} \mathring C^{\mu \nu \rho \sigma} + \frac{\epsilon}{2} \left[ (\mathring \nabla \phi)^2 + \frac{d-2}{4(d-1)} \mathring R \phi \right]- b \phi^4 \, ,
\label{conformal-scalar}
\end{equation} 
where we have normalized canonically the kinetic term of the scalar field. Notice that the dilatation gauge field drops out of the action automatically and is not necessary to set it to zero. Giving a vev to $\phi$ will spontaneously break the (now ungauged) local scale invariance of the action above generating the Einstein term (with the proper sign if the scalar field is a ghost) and a cosmological constant. Unlike the Weyl gravity case where the dilatation gauge field becomes massive by eating the scalar field, in this case one is left with a Goldstone ghost boson.

Let us now do the exercise of solving the equation of motion of $f_\mu{}^a$ for the total Lagrangian density
\begin{equation}
\mathcal L = a_1 \mathcal R^2 + a_2 \mathcal R^0_{(\mu \nu)} \mathcal R^{0( \mu \nu)} + a_3 \mathcal C_{\mu \nu \rho \sigma} \mathcal C^{\mu \nu \rho \sigma } + a_4 \mathcal R_{[\mu \nu]} \mathcal R^{[\mu \nu]} -\frac{\epsilon}{2} \phi \hat \Box \phi - b_1 \phi^2 \mathcal R - b_2 \phi^4 \, .
\end{equation}
Compared to the no-matter case, only the trace part $f$ gets modified. The equations of motion read 
\begin{align}
8a_1 (d-1) \mathcal R - [\epsilon w + 4 (d-1) b_1] \phi^2 & = 0\, , & \mathcal R_{(\mu \nu)}^0 & = 0\, , & \mathcal R_{[\mu \nu]} & = 0\, ,
\end{align}
and can be written compactly as
\begin{equation}
\mathcal R_{\mu \nu} =  \frac{\alpha}{d} g_{\mu \nu}\phi^2 \, ,
\end{equation}
with $\alpha$ depending on the couplings $a_1$, $b_1$ of the original action (and also on $w, \epsilon$ and the dimension $d$). The solution is found to be
\begin{equation}
f_{\mu \nu} = - \frac{1}{2(d-2)} R_{\mu \nu} + \frac{1}{4(d-1)(d-2)}g_{\mu \nu} R + \frac{\alpha}{4d(d-1)} g_{\mu \nu} \phi^2 \, .
\end{equation}
It is easy to see by plugging in the above into the original Lagrangian that one is again lead to the same result as in eq.\,\eqref{conformal-scalar} with a redefinition of the couplings. Since the geometric Lagrangian and the matter field Lagrangian are independently invariant under conformal transformations one can safely work with the standard constraint and obtain the same action. When the space-time gauge group is larger, like in the superconformal case, it can be convenient to modify the constraint in such a way as to remain invariant under as many gauge symmetries of the theory as possible. Finally, the action that we obtained in eq.\,\eqref{conformal-scalar} is the same as the pure gauge limit of Weyl gravity \eqref{pure-gauge}, thus showing that conformal gravity corresponds to integrable Weyl geometry.

\section{Generalised Weyl gravity}
\label{sec4}

We consider in this section Weyl gravity with a more general connection containing, aside from the dilatation gauge field, a general (dilatation covariant) torsion. This can be introduced into the theory by modifying the constraint on the curvature of translations as has been alluded before.

\subsection{Dilatation covariant torsion and action}
In order to have a more general connection, we consider the following constraint on the curvature of translations
\begin{equation}
\label{gen-constraint}
R_{\mu \nu}(P^a) = \mathcal T_{\mu \nu}{}^a\, ,
\end{equation}
where the torsion $\mathcal T_{\mu \nu}{}^a$ has to transform covariantly under dilatations \eqref{field-1}, with corresponding charge equal to $+1$, such that, after converting to space-time indices $\mathcal T_{\mu \nu}{}^\rho$ is invariant. The constraint above can be solved in the usual way to give
\begin{equation}
\Gamma_{\mu \nu}^\rho = \mathring \Gamma_{\mu \nu}^\rho + \delta_\mu^\rho \w_
\nu - g_{\mu \nu}\w^\rho + \frac12 (\mathcal T_{\mu \nu}{}^\rho + \mathcal T^\rho{}_{ \mu \nu} - \mathcal T_\nu{}^\rho{}_\mu) \, ,
\end{equation}
such that the total torsion $T_{\mu \nu}{}^a$ has two components, one that transforms like a gauge field and the other that transforms covariantly under dilatations
\begin{equation}
  T_{\mu \nu}{}^a = -2 \w_{[\mu} e_{\nu]}^a + \mathcal T_{\mu \nu}{}^a \, .
\end{equation}

Before we continue we should specify what $\mathcal T$ is. There are basically two distinct ways to think of it. First one can consider the MAG approach where the torsion is an independent field. In this case \eqref{gen-constraint} is not really a constraint, but just a notation, and the connection itself is dynamical \cite{Hehl:1994ue, Lasenby:2015dba}. The other way is to have $\mathcal T$ specified by the theory. In the spirit of this paper, given a larger symmetry, by imposing the constraint of vanishing fully covariant translation curvature would yield a relation like \eqref{gen-constraint}. A notable example in this sense is supersymmetry where $\mathcal T$ is given as a gravitino bilinear $\mathcal T_{\mu \nu}{}^a \sim \bar \psi_\mu \gamma^a \psi_\nu$.

In the first case, the action quadratic in curvatures one can write is just \eqref{actionP} where the quadratic terms in $\mathcal T$ drop out as they are not Weyl invariant. Compared to the standard action for Weyl quadratic gravity \eqref{action}, the presence of the torsion term in \eqref{gen-constraint} enters the action indirectly from the fact that now the Bianchi identity for the Lorentz curvature can no longer be used to discard certain combinations appearing in the action. Specifically, using the action of the covariant derivative $\mathcal D_\mu$ on $R_{\nu \rho}(P^a)$  in eq.\,\eqref{der-1}, yields the Bianchi identity
\begin{equation}
R^a{}_{[\mu \nu \rho]} = - e_{[\mu}^a F_{\nu \rho]} + \mathcal H_{\mu \nu \rho}{}^a \, ,
\end{equation} 
where $F_{\mu \nu} = \partial_\mu \w_\nu - \partial _\nu \w_
\mu$ is the usual dilatation field strength in Weyl gravity and we defined
\begin{equation}
\label{H-1}
\mathcal H_{\mu \nu \rho}{}^a = \hat D_{[\mu} \mathcal T_{\nu \rho]}{}^a = \partial_{[\mu} \mathcal T_{\nu \rho]}{}^a + \Om_{[\mu}{}^{ab} \mathcal T_{\nu \rho] b} + \w_{[\mu} \mathcal T_{\nu \rho]}{}^a  \, .
\end{equation} 
Inserting in \eqref{Rcomb}, we find the analog of the equations \eqref{symm-1}-\eqref{symm-2} to be
\begin{equation}
\begin{split}
  R_{ \rho \sigma \mu \nu} - R_{ \mu \nu \rho \sigma} &= -F_{\mu \rho} g_{\sigma \nu}
  + F_{\mu \sigma} g_{\rho \nu} + F_{\nu \rho} g_{\sigma \mu}
  - F_{\nu \sigma} g_{\rho \mu} \\
  & - \frac32 \left(\mathcal H_{\mu \nu \rho \sigma} - \mathcal H_{\sigma \mu \nu \rho} + \mathcal H_{\nu \rho \sigma \mu} - \mathcal H_{\rho \sigma \mu \nu} \right)\, ,\\
  R_{\mu \nu} - R_{\nu \mu} & = (d-2) F_{\mu \nu} - 3 \mathcal H_{\mu \nu} \, ,
\end{split}
\end{equation}
where we defined the trace $\mathcal H_{\mu \nu} := \mathcal H_{\mu \nu \rho}{}^a e_a^\rho$.

In a similar way, instead of \eqref{id-1}-\eqref{id-4}, for the quadratic terms in the curvature one now has the identities
\begin{equation}
\begin{split}
R_{\rho \sigma \mu \nu} R^{\rho \sigma \mu \nu}  & = R_{\rho \sigma \mu \nu} R^{\mu \nu \rho \sigma} +2(d-2) F_{\mu \nu} F^{\mu \nu} -12 \mathcal H_{\mu \nu} F^{\mu \nu}\\
& + 6 \mathcal H_{\mu \nu \rho \sigma} \mathcal H^{\mu \nu \rho \sigma} - 6\mathcal H_{[\mu \nu \rho \sigma]} \mathcal H^{[\mu \nu \rho \sigma]}\, , \\
R_{\rho \sigma \mu \nu} R^{\mu \sigma \nu \rho} & = - \frac12 R_{\mu \nu \rho \sigma} R^{\mu \nu \rho \sigma} + \frac{d-2}{2} F_{\mu \nu} F^{\mu \nu} - 3 \mathcal H_{\mu \nu} F^{\mu \nu} + \frac32 \mathcal H_{\mu \nu \rho \sigma} \mathcal H^{\mu \nu \rho \sigma}\, , \\
R_{\mu \nu \rho \sigma} \mathcal H^{\mu \nu \rho \sigma} & = \mathcal H_{\mu \nu \rho \sigma} \mathcal H^{\mu \nu \rho \sigma}  + 2 \mathcal H_{[\mu \nu \rho \sigma]} \mathcal H^{[\mu \nu \rho \sigma]} -\mathcal H_{\mu \nu} F^{\mu \nu} \, , \\
R_{\mu \nu \rho \sigma} \mathcal H^{\nu \rho \sigma \mu} & =\mathcal H_{\mu \nu \rho \sigma} \mathcal H^{\mu \nu \rho \sigma}  -\mathcal H_{\mu \nu} F^{\mu \nu} \, ,
\end{split}
\end{equation}
and for the traces
\begin{equation}
\begin{split}
R_{\mu \nu} R^{\mu \nu} & = R_{\mu \nu} R^{\nu \mu} + \frac{(d-2)^2}{2} F_{\mu \nu} F^{\mu \nu} - 3(d-2) \mathcal H_{\mu \nu} F^{\mu \nu} + \frac92 \mathcal H_{\mu \nu} \mathcal H^{\mu \nu} \, ,\\
R_{\mu \nu} F^{\mu \nu} &= \frac{d-2}{2} F_{\mu \nu}F^{\mu \nu} - \frac32 \mathcal H_{\mu \nu} F^{\mu \nu} \, , \\
R_{\mu \nu} H^{\mu \nu} &= \frac{d-2}{2} \mathcal H_{\mu \nu}F^{\mu \nu} - \frac32 \mathcal H_{\mu \nu} \mathcal H^{\mu \nu} \, .
\end{split}
\end{equation}
Before writing the resulting action let us decompose $\mathcal H_{\mu \nu \rho \sigma }$ into irreducible pieces 
\begin{equation}
\mathcal H_{\mu \nu \rho \sigma} = \mathcal H^0_{\mu \nu \rho \sigma} + \mathcal H_{[\mu \nu \rho \sigma]} + \frac{3}{d-2} \mathcal H_{[\mu \nu} g_{\rho] \sigma} \, ,
\end{equation} 
where $\mathcal H^0_{\mu \nu \rho \sigma}$ is traceless $\mathcal H^0_{\mu \nu \rho \sigma} g^{\rho \sigma} = 0$ and has zero completely antisymmetric part $\mathcal H^0_{[\mu \nu \rho \sigma]} = 0$. The three components above are related to the field strengths of the irreducible components of the covariant torsion $\mathcal T_{\mu \nu}{}^\rho$. Applying the general decomposition in eq.\,\eqref{dec2} yields
\begin{equation}
\mathcal T_{\mu \nu \rho} = -2 \tau_{[\mu}g_{\nu]\rho} + \chi_{\mu \nu \rho} + \mathcal T_{[\mu \nu \rho]} \, ,
\end{equation}
with $\chi_{\mu \nu \rho} = \chi_{[\mu \nu] \rho}$ a traceless two-form satisfying $\chi_{[\mu \nu \rho]} = 0$. As we did in the standard Weyl case, we can choose the square of the Weyl tensor $C_{\mu \nu \rho \sigma}$ defined in \eqref{weyl-tensor} and the Euler term $G$ defined in \eqref{Euler} as part of the basis. Therefore, the quadratic action of generalised Weyl gravity can be written as
\begin{equation}
\label{generalised-action}
\begin{split}
S &= \int d^4 x \sqrt{g} \Big[  a_0 R^2 + b_0 F_{\mu \nu} F^{\mu \nu} + c_0 C_{\mu \nu \rho \sigma} C^{\mu \nu \rho \sigma} + d_0 G \\
& + a_1 \mathcal H_{\mu \nu} F^{\mu \nu} + b_1 \mathcal H^0_{\mu \nu\rho \sigma} \mathcal H^{0 \mu \nu \rho \sigma} + c_1 \mathcal H_{[\mu \nu \rho \sigma]} \mathcal H^{[\mu \nu \rho \sigma]} + d_1 \mathcal H_{\mu \nu} \mathcal H^{\mu \nu}  \Big] \, .
\end{split}
\end{equation}

In the case where the Weyl covariant torsion in \eqref{gen-constraint} is given by the theory (as is the case in supersymmetry \cite{KTN1,KTN2}) there is no reason to forbid higher order terms in the torsion if they are compatible with the symmetry. Indeed, even if not explicitly written, such terms were already present in superconformal constructions \cite{KTN1,KTN2,VanNieuwenhuizen:1981ae}. To have a systematic approach, note that tensors constructed like
\begin{align}
 \mathcal T^{(1)}_{\mu \nu \rho \sigma} &= \mathcal T_{\mu \nu}{}^\tau \mathcal T_{\rho \sigma \tau}\, ,  & \mathcal T^{(2)}_{\mu \nu \rho \sigma} &= \mathcal T_{\mu \tau \nu }\mathcal T_{\rho \sigma}{}^\tau\, ,  & H_{\sigma \mu \nu \rho }& =e_{a\sigma} \hat D_\mu \mathcal T_{\nu \rho}{}^a \, ,
\end{align}
have the same behavior under dilatations as the curvature tensor $R_{\mu \nu \rho \sigma}$ and therefore their squares (and mixtures thereof) can be written in the action.

Finally, let us also observe that the Weyl tensor $C_{\mu \nu \rho \sigma}$ of generalised Weyl gravity is no longer equal to its Riemannian counterpart $C_{\mu \nu \rho \sigma} \neq \mathring C_{ \mu \nu \rho \sigma}$ but depends additionally on the dilatation covariant torsion. It also depends on the Weyl gauge field but only algebraically (actually linearly). A projective transformation can transform the action \eqref{generalised-action} in a non-metric frame. The corresponding connection would have vectorial non-metricity and dilatation covariant torsion. This connection is given for instance in \cite{Charap}. In our notation it can be written as
\begin{equation}
\tilde \Gamma_{\mu \nu}^\rho = \mathring \Gamma_{\mu \nu}^\rho + \delta_\mu^\rho \w_\nu + \delta_\nu^\rho \w_\mu - g_{\mu \nu} \w^\rho  + \frac12 (\mathcal T_{\mu \nu}{}^\rho + \mathcal T^\rho{}_{ \mu \nu} - \mathcal T_\nu{}^\rho{}_\mu)\, ,
\end{equation}
and is still invariant under dilatations. One can build the same theory with the connection above. As we have seen explicitly, working with a non-metric connection is cumbersome and in the case of Weyl gravity not necessary. Using either the metric formulation with torsion or, as is natural for a gauge theory, the manifestly covariant picture, is much more efficient. In view of the additional components of torsion that we introduced in Weyl gravity, a natural question arises whether one can define a more general duality between torsion and non-metricity beyond the vectorial case implemented by the projective transformation \eqref{projective}. This is considered in the next subsection.

\subsection{Torsion vs non-metricity: general case}

We shall consider here the possibility of a more general map between torsion and non-metricity in the context of generalised Weyl gravity.\footnote{One could consider the same question in the context of Metric Affine Gravity (MAG). However, we always work with a connection fixed by the constraint on the curvature of translations.}
Recall that in the vectorial case discussed in detail before, the existence of a torsion/non-metricity map was just a result of the Weyl gauge symmetry. Even though, based on the tangent space formalism, we managed to find a generalisation to a connection with an arbitrary torsion, a similar one is not known for the non-metricity side. Therefore, we have to find a different route to obtain a generalisation of the torsion/non-metricity map.

Before that, let us recall the decomposition of a general affine connection in terms of the Levi-Civita $\mathring \Gamma$ and the non-metricity and torsion tensors. More details can be found in the Appendix.
A general connection is fully specified once a distorsion tensor, $N_{\mu \nu \rho}$, is fixed, 
  \begin{equation}
    \label{genGamma}
  \mathsf \Gamma_{\mu \nu}^\rho = \mathring \Gamma_{\mu \nu}^\rho + N_{\mu \nu}{}^\rho \, .
\end{equation}
The definition above is natural in view of the fact that the difference of two connections is always a tensor and thus distorsion measures the departure of a general affine connection from the Levi-Civita connection.

The distorsion tensor can be uniquely decomposed into disformation $S_{\mu \nu}{}^\rho$ and contorsion $K_{\mu \nu}{}^\rho$ 
\begin{equation}
N_{\mu \nu}{}^\rho = S_{\mu \nu}{}^\rho - K_{\mu \nu}{}^\rho\, ,
\end{equation}
with $S_{\mu \nu}{}^\rho=S_{\nu \mu}{}^\rho$ symmetric in the first two indices and $K_{\mu \nu}{}^\rho = -K_\mu{}^\rho{}_\nu$ antisymmetric in the last two indices.
The relation above can be inverted (thus showing its uniqueness) to give
\begin{align}
  S_{\mu \nu \rho}  &= N_{\mu (\nu \rho)} + N_{\nu (\rho \mu)} - N_{\rho (\mu \nu)} \, , \label{disformation-1}\\
   K_{\mu \nu \rho}  &= -N_{[\mu \nu] \rho} - N_{[\rho \mu] \nu} + N_{[\nu \rho] \mu} \, . \label{contorsion-1}
\end{align}
The non-metricity and torsion tensors are related to the disformation, contorsion and distortion via 
\begin{align}
  \label{QSN}
  Q_{\mu \nu \rho} = -2 S_{\mu (\nu \rho)} = -2 N_{\mu(\nu \rho)} \, , \\
  \label{TKN}
  T_{\mu \nu}{}^\rho  = -2K_{[\mu \nu]}{}^\rho = 2N_{[\mu \nu]}{}^\rho \, . 
\end{align}
Finally, $Q_{\mu \nu \rho}$ measures departures from metricity as it satifies $\tilde \nabla_\mu g_{\nu \rho} = Q_{\mu \nu \rho}$ and $T_{\mu \nu}{}^\rho$ measures departures from symmetry of the connection $T_{\mu \nu}{}^\rho = 2\Gamma_{[\mu \nu]}{}^\rho$.

With these definitions in mind, we want to specify a map between distorsion tensors and based on the experience gained in the vectorial case we ask the following for the generalised map:
\begin{itemize}
\item it reduces to the projective transformation presented before in the case of vectorial torsion/non-metricity;
\item it maps pure non-metricity into pure torsion; in other words we start from a torsion-free connection which is non-metric and end up with a metric connection with torsion;
\item it leaves invariant the minimal coupling of fermions to the associated spin connection;
\item it preserves the number of degrees of freedom described by the torsion and non-metricity tensor respectively.

\end{itemize}

In general, torsion and non-metricity have different number of degrees of freedom\footnote{Torsion has $\frac{d^2(d-1)}{2}$ independent components and non-metricity has $\frac{d^2(d+1)}{2}$ components (see Table 1 and 2 in the Appendix).} and therefore we can neither expect an equivalence (bijective) map from an arbitrary torsion to an arbitrary non-metricity nor the other way around. It is therefore useful to have a decomposition of torsion and non-metricity into irreducible parts under $so(1,d-1)$ in order to see more clearly which components may be mapped into one another. The following result, which is explained in the Appendix is  equivalent to the one found in the literature \cite{McCrea, Hehl:1994ue}
\begin{align}
  \label{Qdec}
   Q_{\mu \nu \rho} & = q_\mu g_{\nu \rho} + \frac{2d}{(d-1)(d+2)}
  \left( g_{\mu(\nu} q'_{\rho)} - \frac1d q'_\mu g_{\nu \rho} \right)
   + \tilde Q_{\mu \nu \rho} + \mathcal{Q}_{\mu \nu \rho} \, , \\
  \label{Tdec}
  T_{\mu \nu \rho} & = -2 t_{[\mu} g_{\nu] \rho} + \tilde T_{\mu \nu \rho}
  + \Theta_{\mu \nu \rho}  \, ,
\end{align}
where $q_\mu$, $q'_\mu$ and $t_\mu$ are the vectorial components which can be defined by contraction with the metric \eqref{nm-trace}, \eqref{torsion-trace}, $\mathcal Q_{\mu \nu \rho}$ and $\mathcal T_{\mu \nu \rho}$ are the completely symmetric and antisymmetric pieces of $Q_{\mu \nu \rho}$ and $T_{\mu \nu \rho}$ respectively \eqref{symmQ}, \eqref{antiT}, while the remainders, $\tilde Q_{\mu \nu \rho}$ and $\tilde T_{\mu \nu \rho}$ can be written in terms of traceless rank 3 tensors $\chi_{\mu \nu \rho}^{Q}$ and $\chi_{\mu \nu \rho}^T$ with two antisymmetric indices \eqref{tildeQ},\eqref{tildeT}.  Notice that $\tilde Q_{\mu \nu \rho}$ and $\tilde T_{\mu \nu \rho}$ correspond to the same irreducible representation of dimension $\frac{d(d-2)(d+2)}{3}$ (see Table 1 and 2 in the appendix). Therefore they constitute natural candidates for a generalisation of the vectorial torsion/non-metricity map. 

Recall the projective transformation we had in the vectorial case, from eq.~\eqref{tildeom}.
This obviously satisfies
\begin{equation}
  \Om_{\mu}{}^{ab} = \tilde \Om_\mu{}^{[ab]} \, .
\end{equation} 
Therefore, this may be a good starting point for finding a generalised map between torsion and non-metricity. Hence we define our generalised torsion/non-metricity map as
\begin{equation}
\label{gen-map1}
\tilde \Om_\mu{}^{ab} \mapsto \Om_\mu{}^{ab} = \tilde \Om_\mu{}^{[ab]}\, .
\end{equation}
This map explicitly fulfills all requirements listed above including the last one when restricted to the common subspace in the decompositions of the non-metricity and torsion tensors (see \eqref{Qdec} and \eqref{Tdec}). 
Notice that the spin connection $\tilde \omega_\mu{}^a{}_b$ corresponding to a torsionless non-metric connection admits the decomposition (a particular case of \eqref{general-spin} )
\begin{equation}
  \tilde \Om_{\mu}{}^a{}_b = \mathring{\Om}_\mu{}^a{}_b + S_{\mu \nu}{}^\rho  e^\nu_b e_\rho^a
  \, , 
\end{equation}
in terms of $\mathring \Om_\mu{}^a{}_b$ (given in \eqref{spinLC}) corresponding to the Levi-Civita connection (via the vielbein postulate). Let us see the effect of the map \eqref{gen-map1}. From above one has
\begin{equation}
\tilde \Om_\mu{}^{[ab]} = \mathring \Om_\mu{}^{ab}+ S_{\mu \nu}{}^\rho e^{\nu [b} e_\rho^{a]} \equiv \Om_\mu{}^{ab}\, . \label{map-tilde}
\end{equation}
The new spin connection $\Om_\mu{}^{ab}$ is metric (as it is antisymmetric in the indices $a,b$) and therefore can be decomposed as (see \eqref{general-spin})
\begin{equation}
\Om_\mu{}^a{}_b = \mathring \Om_\mu{}^a{}_b - K_{\mu \nu}{}^\rho e_b^\nu e_\rho^a \, .
\end{equation}
Comparing the equation above with \eqref{map-tilde} allows one to express the contorsion of the new connection in terms of the disformation
\begin{equation}
K_{\mu \nu}{}^\rho e_\nu^b e_\rho^a = -S_{\mu \nu}{}^\rho e^{\nu [b} e_\rho^{a]} \, .
\end{equation}
This describes the action of the map from disformation to contorsion. It can also be written as 
\begin{align}
 K_{\mu \nu}{}^\rho & = - \frac12 \left(S_{\mu \nu}{}^\rho - S_\mu{}^\rho{}_\nu \right)\, , & T_{\mu \nu}{}^\rho &= \frac12 \left(Q_{\mu \nu}{}^\rho - Q_{\nu \mu}{}^\rho \right) \, .
\end{align}
In order to obtain the second equation above one uses \eqref{TKN} and \eqref{QSN}. With the help of the vielbein postulate \eqref{postulate}, or with the equation above, one can write the induced action on the space-time connections as
\begin{equation}
\label{gen-map2}
\tilde \Gamma_{\mu \nu}^\rho \mapsto \Gamma_{\mu \nu}^\rho = \tilde \Gamma_{\mu \nu}^\rho - \frac12\left(S_{\mu \nu}{}^\rho + S_\mu{}^\rho{}_\nu \right) = \mathring \Gamma_{\mu \nu}^\rho + \frac12 \left(S_{\mu \nu}{}^\rho - S_\mu{}^\rho{}_\nu \right) \, ,
\end{equation} 
where for the second equality sign we used the expansion of a torsionless non-metric connection $\tilde \Gamma_{\mu\nu}^\rho = \mathring \Gamma_{\mu \nu}^\rho + S_{\mu \nu}{}^\rho $.

In order to check the requirement of preserving the number of degrees of freedom we have to find explicitly the action of this map on the components of torsion/non-metricity as defined in \eqref{Qdec} and \eqref{Tdec}.
From the general decomposition \eqref{TKN} it is easy to see that the resulting torsion has the expression
\begin{equation}
T_{\mu \nu}{}^\rho = Q_{[\mu \nu]}{}^\rho = \left(q_{[\mu} - \frac1{d-1}q'_{[\mu} \right)g_{\nu] \rho} + \frac34 \chi_{\mu \nu \rho}^Q \, .
\end{equation}
Comparing now with the general decomposition of torsion \eqref{Tdec} yields the following action on components
\begin{equation}
\label{map-components}
  -\frac12\left(q^{\mu} - \frac{1}{d-1} q'^{\mu} \right) \mapsto t_\mu \, , \qquad \tilde q_\mu \mapsto 0\, , \qquad \frac34 \chi^Q_{\mu \nu \rho} \mapsto \chi^T_{\mu\nu \rho}  \, , \qquad \mathcal Q_{\mu \nu \rho} \mapsto 0 \, ,
\end{equation}
where $\tilde q_\mu$ is an orthogonal combination to the vector in the first term. One could further refine this map such that it acts as $-\frac12 q^\mu \mapsto t_\mu$ as was the case with the projective transformation \eqref{projective} or just assume that $q'_\mu = 0$. Notice also that in the case of Weyl non-metricity one has $\w_\mu = - \frac12 q_\mu$ and $q'_\mu = 0$.
Therefore, in order to ensure the same number of degrees of freedom on both sides of the map, we shall assume the equivalence relation to be of the form
\begin{equation}
\label{gen-equivalene}
Q_{\mu \nu \rho} = q_\mu g_{\nu \rho} + \chi^Q_{\mu (\nu \rho)} \qquad \longleftrightarrow \qquad T_{\mu \nu \rho} = q_{[\mu}g_{\nu] \rho} + \frac34 \chi^Q_{\mu \nu \rho} \, ,
\end{equation} 
involving the vector $q_\mu$ and the traceless two-form $\chi^{T}_{\mu \nu \rho} = \frac34 \chi^{Q}_{\mu \nu \rho}$.

For completeness, we give here the effect of the map \eqref{gen-map1} on the curvature tensors 
\begin{equation}
\tilde R^{ab}{}_{\mu \nu} \mapsto R^{ab}{}_{\mu \nu} = \tilde R^{ab}{}_{\mu \nu} - \tilde R^{(ab)}{}_{\mu \nu} - \left[S_\mu{}^{(ac)} S_{\nu}{}^{(db)}- S_\nu{}^{(ac)} S_{\mu}{}^{(db)} \right]\eta_{cd}\, ,
\end{equation}
where we used the identity $\tilde \Om_\mu{}^{(ab)} = S_\mu{}^{(ab)}$. 

We now want to consider the above map in the context of generalised Weyl gravity that we introduced earlier in this section. On the torsion side we have the identifications
\begin{align}
q_\mu &= - 2 \w_\mu \, , & \chi^T_{\mu \nu}{}^a = \mathcal T_{\mu \nu}{}^a \, .
\end{align}
The extra torsion component has been introduced into the theory via the constraint on the gauge curvature of translations \eqref{gen-constraint}. In our case, the resulting first Cartan structure equation can be written as
\begin{equation}
\label{car-1}
\mathring D_\mu e_\nu^a - \mathring D_\nu e_\mu^a + \w_\mu e_\nu^a - \w_\nu e_\mu^a = \chi^T_{\mu \nu}{}^a \, .
\end{equation}
In order for the equivalence to hold we need that the above equation be written equivalently as
\begin{equation}
\label{car-2}
\tilde D_\mu e_\nu^a - \tilde D_\nu e_\mu^a = 0 \, ,
\end{equation}
for a non-metric spin connection defined by the non-metricity \eqref{gen-equivalene}. Indeed, it is easy to see that the corresponding spin connection which allows the equivalent rewriting of the Cartan structure equation is  
\begin{equation}
\tilde \Om_\mu{}^{ab} = \mathring \Om_\mu{}^{ab} + 2 e_\mu{}^{[a} e^{b]\nu} \w_\nu + \w_\mu \eta^{ab} + \chi^Q_{\nu (\mu \rho)} \, ,
\end{equation}
where the identification $\chi^{T}_{\mu \nu \rho} = \frac34 \chi^{Q}_{\mu \nu \rho}$ is used when comparing eqs.\,\eqref{car-1} and \eqref{car-2}. Furthermore, since we are dealing with the same degrees of freedom on both sides and the same symmetries the resulting actions should also be equivalent. We sketched the action for general dilatation covariant torsion in eq.\,\eqref{generalised-action}. One just needs to set $\mathcal T_{\mu \nu}{}^a = \chi^T_{\mu \nu}{}^a$. A similar discussion in the non-metric frame can be done.  In the next section we shall show that the matter couplings (and in
particular the gravitino coupling) to torsion and non-metricity are left
invariant by this map and therefore it is compatible with generalised Weyl gravity. However it should be stressed that if other symmetries are added to the theory then the equivalence might not hold anymore.

\subsection{Minimal coupling to matter fields}

In view of the more general
torsion/non-metricity map that we introduced, it is worth
looking at the coupling to the distortion tensor, \eqref{genGamma}, via the (spin) connection.  The standard kinetic term of
a scalar field or of a gauge field does not yield a coupling to
distortion simply because the affine connection is
absent. Furthermore, a Dirac spinor would
couple only to the completely antisymmetric part of torsion (see for instance \cite{Shapiro,Hehl:1976kj}), and therefore, does not couple to the distorsion tensors which are relevant for the generalised torsion/non-metricity equivalence introduced in the previous section.
The
more interesting case is that of the Rarita-Schwinger field which
can be analyzed starting from eq.\,\eqref{RS}.
By standard Clifford algebra calculations, the last term in this relation can be written as
\begin{equation}
  \mathcal{L}_\Psi^N = 3 \sqrt{g} \, \bar \Psi_\mu N_\nu{}^{[\mu \nu}
  \gamma^{\rho]} \Psi_\rho \, ,
\end{equation}
where we ignored a term of the form
$\tfrac12 \sqrt{g} \, \bar \Psi_\mu N_{[\nu \alpha \beta]} \gamma^{\mu
  \nu \rho \alpha \beta} \Psi_\rho$ which contributes only in
dimensions $d\geq 5$. Above we have worked with a general spin connection \eqref{general-spin} and expressed the result in terms of the distortion tensor. Then, particular cases of torsion and non-metricity are found from above by setting the distortion to be equal to (minus) the contorsion, that is $N_{\mu \nu}{}^\rho = -K_{\mu \nu}{}^\rho$ and by setting the distorsion equal to the disformation, {\it i.e. } $N_{\mu \nu}{}^{\rho} = S_{\mu \nu}{}^\rho$ respectively. Furthermore, by making use of the decomposition given in
\eqref{Tdec}, one can show that the coupling of fermions to
torsion has the form
\begin{equation}
  \mathcal{L}_\Psi^T = \sqrt{g}\, \bar \Psi^\mu \left[2(d-2)
    t_{[\mu}\gamma_{\rho]} + \chi^T_{\mu \rho \nu} \gamma^\nu
    - \frac12 \Theta_{\mu \rho \nu} \gamma^\nu \right] \Psi^\rho \, .
\label{coupling-T}
\end{equation}
Hence the Rarita-Schwinger field can probe all the components of
torsion.  This is important, as it allows going beyond the vectorial
non-metricity-torsion duality/equivalent
descriptions and to essentially probe couplings of generalised Weyl gravity. \\
\noindent
Similarly, for the non-metricity tensor one obtains the following  coupling
\begin{equation}
  \mathcal{L}^Q_{\Psi} = \sqrt{g} \, \bar \Psi^\mu \Big[-(d-2)\Big(q^{[\mu}
      - \frac{1}{d-1} q'^{[\mu} \Big) \gamma^{\rho]}
    + \frac34 \chi^Q_{\mu \rho \nu} \gamma^\nu \Big] \Psi^\rho \, .
\label{coupling-Q}
\end{equation}
\noindent
Hence, in this case one can only couple the spin 3/2 fermion to a
linear combination of the vectors $q_\mu$ and $q'_\mu$ and also to the
two-form $\chi^Q_{\mu \nu \rho}$. Notice that these are precisely the
components of non-metricity that are mapped into torsion by the map
defined in the previous section. Indeed, if one sets
$\Theta_{\mu \nu \rho} = 0$, the map acting on components as \eqref{map-components}
yields
\begin{equation}
\mathcal{L}^Q_\Psi
 \mapsto \mathcal{L}_\Psi^T \, .
\end{equation} The fact
that the fermion coupling plays well with our generalised
torsion-non-metricity duality was already anticipated by eq.\,\eqref{gen-map1}. Finally, the Lagrangians in eqs.\,\eqref{coupling-T}-\eqref{coupling-Q} are invariant under dilatations when the Weyl charge of $\Psi_\mu$ is equal to $-(d-3)/2$.

\section{Conclusions}

We considered gravity theories from a gauge theoretic point of view. The Coleman-Mandula theorem restricts the list of possible space-time gauge symmetries to three possibilities: the Poincar\'e group, the Weyl group and the full conformal group. Supersymmetric extensions are also allowed but we do not deal with them in this work. We analysed in detail all three possibilities, starting from the corresponding gauge algebras on the tangent space, with special emphasis on the Weyl case. 
We have provided, a complete construction of Weyl gravity starting from the gauge algebra of the Weyl group and the corresponding structure constants, including both tangent space and space-time formalism in the non-metric, torsion and manifest gauge covariant formulations. Parts of the space-time formalism (the non-metric one) already exists in the literature (for instance in \cite{Tann,Jia,DG1}) including  the gauge covariant derivative $\hat \nabla$ (also introduced in \cite{Dirac}), however the fact that is not implemented by an affine connection and its relation to torsion \eqref{general-formula} are not found elsewhere.
In our previous work \cite{Condeescu}, we have shown that Weyl quadratic gravity admits two equivalent formulations, one in terms of vectorial non-metricity and the other in terms of vectorial torsion. The two are bridged by the manifestly gauge covariant formulation. In the current work we have further shown that the torsion/non-metricity equivalence is built-in Weyl gravity and corresponds to a redefinition of the generators of the gauge algebra \eqref{tildeM} which in turn translates to the projective transformation \eqref{projective} relating the connections. Hence the equivalence/duality is independent of a particular choice of Lagrangian. 

Our paper contains all the technical details and derivations necessary to understand \cite{Condeescu}. For instance, we explain various subtleties related to the implementation of general coordinate invariance in the gauge approach \eqref{gcte}, \eqref{cgcte}, the correct definition of the Euler term \eqref{GB-nonmetricity} and the Weyl tensor for a connection with (Weyl) vectorial non-metricity \eqref{real-weyl}. In addition, we have also considered the minimal couplings to matter fields. It has been known that spin $1/2$ fermions only couple to the completely antisymmetric part of torsion and do not couple at all to non-metricity components. This is in agreement with the fact that spin $1/2$ fields do not couple to the Weyl gauge field in either of the torsion or non-metricity pictures. Our most interesting finding here is  that spin $3/2$ fermions couple to \emph{all} components of torsion \eqref{coupling-T} and also to certain  components of non-metricity \eqref{coupling-Q}. Moreover, we also find that matter couplings, including spin $3/2$, are compatible with the torsion/non-metricity duality, hence invariant under the projective transformation (Section 2.7.). 

We considered also the gauging of the conformal group. One finds that in this case both the gauge field of dilatations as well as the gauge field of special conformal transformations are not dynamical. Hence the gauged version of the action \eqref{action-conformal} is equivalent to the ungauged one \eqref{ungauged} where one merely imposes a local symmetry without associated gauge bosons (the well known action of quadratic conformal gravity given by the Weyl tensor squared). We have further shown that conformal gravity, minimally coupled to a scalar field \eqref{conformal-scalar} (which can break it spontaneously) is obtained as a pure gauge limit (for dilatations) of Weyl gravity \eqref{pure-gauge}. We made a clear distinction between theories with  a local symmetry and theories with propagating gauge bosons implementing the symmetries. We call the latter true gauge theories and expect that in quantum gravity all symmetries to be true gauge symmetries. From this point of view Weyl quadratic gravity is the only true gauge theory implementing scale invariance while conformal gravity is a pure gauge limit of it.    

Finally, additional (covariant) torsion degrees of freedom can be introduced into Weyl theory via modifying the constraint on translations \eqref{gen-constraint}. We call this theory generalised Weyl gravity. An example of this would be a supersymmetric extension where the additional torsion arises from the fact that the space-time gauge group is enlarged. Without assuming a particular origin of the aforementioned torsion we  generalise the vectorial torsion/non-metricity equivalence to include an additional common component arising in the decompositions of torsion \eqref{Tdec} and non-metricity \eqref{Qdec}. This component is parametrised as a traceless three-tensor with two antisymmetric indices and zero completely antisymmetric part \eqref{tildeQ}, \eqref{tildeT}. The new map generalising the projective transformation has a simple description on the tangent space \eqref{gen-map1}. Starting from a non-metric spin connection (with only the two components non-zero) one simply defines the torsion spin connection by taking the antisymmetric part in the tangent space indices. This automatically preserves the fermionic couplings. As with the vectorial case, the new torsion/non-metricity map \eqref{gen-map2} is consistent with the first Cartan structure equation being equivalently written in the two frames \eqref{car-1},\eqref{car-2}. Since the components that can be mapped from non-metricity to torsion correspond to the same irreps of the Poincar\'e, we expect the duality to hold at the level of the action as well.     \\



\section*{Appendix}

\setcounter{section}{0}
\def\thesection{\Alph{section}}

\def\theequation{A-\arabic{equation}}
 \setcounter{equation}{0}
\def\thefigure{A-\arabic{figure}}

\section{Conventions}\label{appendix}
\label{appendix}

In this appendix we summarise our conventions. The calculations are done in arbitrary number of dimensions which we generically denote by $d$. Greek indices $\mu, \nu, \ldots = 0, 1, \ldots , d-1$ denote space-time indices, while latin ones $a, b, \ldots$ are tangent space indices. We work in the mostly minus signature convention {\it i.e.} the metric on the tangent space is $\eta_{ab} = \mathrm{diag} \ (1, -1, -1, \ldots, -1)$.

We work interchangeably on space-time and tangent space. The connection on the tangent space (aka spin connection) is denoted by $\genO_\mu{}^{ab}$ and is a space-time one-form with values in the Lie algebra of the structure group of the frame bundle. The affine connection is denoted by $\genG_{\mu \nu}^\rho$. Throughout the paper, the symbols above denote general quantities with arbitrary torsion and non-metricity. Bare symbols $\Gamma$ and $\Om$
denote metric connections with torsion, while $\tilde \Om_\mu{}^{ab}$ and $\tilde \Gamma_{\mu \nu}^\rho$ denote torsionless non-metric connections. Riemannian quantities (metric without torsion) are denoted by $\mathring \Om_\mu{}^{ab}$ and $\mathring \Gamma_{\mu \nu}^\rho$. The same symbols are used to denote the corresponding curvature tensors $\mathsf R$, $R$, $ \tilde R$, $\mathring R$ and tangent space covariant derivatives $\mathsf D$, $D$, $\tilde D$ and $\mathring D$.

Since in the whole paper we use many type of connections it is useful to make clear our conventions by a more detailed presentation.

\subsection{Summary of connections and derivatives}

The unique metric and torsion-free connection is given in terms of the Christoffel symbols
\begin{equation}
  \mathring \Gamma_{\mu \nu}^\rho = \frac12 g^{\rho \sigma}\left( \partial_\mu g_{\nu \sigma} + \partial_\nu g_{\mu \sigma} - \partial_\sigma g_{\mu \nu} \right)\, .
\end{equation}
This has the undesirable property that under a Weyl gauge transformation does not transform covariantly.
The Weyl connection, which is torsion-free, but non-metric, is given by
\begin{equation}
  \tilde \Gamma_{\mu \nu}^\rho = \mathring \Gamma_{\mu \nu}^\rho + (\delta_\mu^\rho \w_\nu + \delta_\nu^\rho \w_\mu - g_{\mu \nu} \w^\rho)\, ,
\end{equation}
and has the property that is invariant under Weyl gauge transformations \eqref{Weyl1} and \eqref{Weyl2}. From this, a metric connection with torsion can be defined
\begin{equation}
  \Gamma_{\mu \nu}^\rho = \tilde \Gamma_{\mu \nu}^\rho - \w_\mu \delta_\nu^\rho = \mathring \Gamma_{\mu \nu}^\rho + \delta_\mu^\rho \w_\nu - g_{\mu \nu} \w^\rho \, ,
\end{equation}
which transforms like a gauge field.

These connections define space-time covariant derivatives, $\mathring \nabla$, $\tilde \nabla$, $\nabla$, in the usual way. However, none of these derivatives is covariant with respect to Weyl gauge transformations. The Weyl gauge covariant derivative is defined using the Weyl charge of the object it acts on
\begin{equation}
  \hat \nabla_\mu X = \tilde \nabla_\mu X + \tilde q_X \w_\mu X = \nabla_\mu X + q_X \w_\mu X \, ,
\end{equation}
where $\tilde q_X$ denotes the Weyl charge in space-time, while $q_X$ denotes the Weyl charge on the tangent space. It is important to note that no affine connection can be associated to this covariant derivative.

On the tangent space, one usually associates to $\mathring \Gamma$ the spin connection $\mathring \Om$
\begin{equation}
  \mathring \Om_\mu{}^{ab} = 2e^{\nu[a} \partial_{[\mu} e_{\nu]}^{b]}
  - e^{\nu[a}e^{b]\sigma}e_{\mu c} \partial_\nu e_\sigma^c \, ,
\end{equation}
which does not transform covariantly under Weyl gauge transformations as well.
The non-metric spin connection, which, now transforms like a gauge field, is given by
\begin{equation}
  \tilde \Om_\mu{}^{ab} = \Om_\mu{}^{ab} + \w_\mu \eta^{ab} = \mathring{\Om}_\mu{}^{ab} +2 e_\mu^{[a} e^{b]\nu} \w_\nu + \w_\mu \eta^{ab} \, ,
\end{equation}
while the metric and Weyl-invariant spin connection is
\begin{equation}
  \Om_\mu{}^{ab} = \mathring{\Om}_\mu{}^{ab} +2 e_\mu^{[a} e^{b]\nu} \w_\nu\, .
\end{equation}
Tangent space derivative operators, $\mathring D$, $\tilde D$ and $D$, can be defined from each of these connections and again none of it is going to be Weyl covariant. The Weyl covariant derivative is defined as
\begin{equation}
  \hat D_\mu X = D_\mu  X + q_X \w_\mu X = \tilde D_\mu X + \tilde q_X \w_\mu X\, ,
\end{equation}
with $q_X$ and $\tilde q_X$ as explained above.

\subsection{General affine connections}
A general affine connection $\genG_{\mu \nu}^\rho$, corresponding to a $GL(d)$ frame bundle, can be decomposed
in terms of the Levi-Civita connection, $\mathring{\Gamma}_{\mu \nu}^\rho$, given explicitly in eq.\,\eqref{Levi-Civita}, and the
distortion tensor $N_{\mu \nu}{}^\rho$
\begin{equation}
\label{affine}
  \genG_{\mu \nu}^\rho = \mathring{\Gamma}_{\mu \nu}^\rho + N_{\mu \nu}{}^\rho
  = \mathring{\Gamma}_{\mu \nu}^\rho + S_{\mu \nu}{}^\rho - K_{\mu \nu}{}^\rho \, .
\end{equation}
The equation above can be seen as the definition of the distortion since the difference of two connections is always a tensor. After the second equality above, we have further decomposed the distortion tensor into
disformation $S_{\mu \nu \rho} = S_{(\mu \nu) \rho}$ symmetric in the first indices and contorsion
$K_{\mu \nu \rho} = K_{\mu [\nu \rho]} $ antisymmetric in the last two indices. As we shall shortly see, this decomposition is done in order to single out the contributions to non-metricity ($S$) and torsion ($K$).
For any distortion tensor
this decomposition is unique. Indeed, $K_{\mu \nu \rho}$ and $S_{\mu \nu \rho}$ can be expressed in terms of distortion by the following
\begin{align}
  K_{\mu \nu \rho} & = -N_{[\mu \nu] \rho} - N_{[\rho \mu] \nu} + N_{[\nu \rho] \mu} \, ,\\
  \label{disformation}
  S_{\mu \nu \rho} & = N_{\mu (\nu \rho)} + N_{\nu (\rho \mu)} - N_{\rho (\mu \nu)} \, .
\end{align}
An affine connection is symmetric if and only
if the distortion tensor is symmetric in its first two indices
$N_{\mu \nu \rho} = N_{\nu \mu \rho}$. This is equivalent to have the
contorsion (and therefore the torsion as well) to be zero and hence
corresponds to the case of non-metricity (if $N_{\mu \nu \rho} \ne 0$). Furthermore, an affine
connection is metric if and only if the distortion tensor is
antisymmetric in its last two indices
$N_{\mu \nu \rho} = -N_{\mu \rho \nu}$ (equivalent to setting the
disformation $S_{\mu \nu}{}^\rho$ to zero). 

The covariant (space-time) derivative corresponding to the connection $\genG$ is defined as
\begin{align}
  \label{covd}
 \nabla_\mu V^\nu &= \partial_\mu V^\nu + \genG_{\mu \rho}^ \nu V^\rho \, , & \nabla_\mu V_\nu &= \partial_\mu V_\nu - \genG_{\mu \nu}^ \rho V_\rho \, ,   
\end{align}
while the ``Riemann'' curvature tensor is given in the usual way by
\begin{equation}
  \mathsf R^\rho{}_\sigma{}_{\mu \nu} = \partial_\mu \genG^\rho_{\nu \sigma}
  - \partial_\nu \genG^\rho_{\mu \sigma} + \genG^\rho_{\mu \tau}
  \genG^\tau_{\nu \sigma} - \genG^\rho_{\nu \tau} \genG^\tau_{\mu \sigma} \, , \label{riemann-affine}
\end{equation}
and it corresponds to the commutator of covariant derivatives, thus satisfying \eqref{commutator-torsion}. 
The curvature tensor can also be decomposed into Levi-Civita part
$\mathring{R}^\rho{}_\sigma{}_{\mu \nu}$ and the rest depending on the
distortion. It is easy to see that we have
\begin{equation}
  \mathsf R^\rho{}_\sigma{}_{\mu \nu}  = \mathring{R}^\rho{}_\sigma{}_{\mu \nu}
  + 2 \mathring{\nabla}_{[\mu} N_{\nu]\sigma}{}^\rho
  +N_{\mu \tau}{}^\rho N_{\nu\sigma}{}^\tau
  - N_{\nu \tau}{}^\rho N_{\mu \sigma}{}^\tau \, .
\end{equation}
In general, the Riemann tensor defined above is antisymmetric only in
the last two indices. Therefore, we can define three different ``Ricci'' traces
\begin{align}
  \label{nm-traces}
  \mathsf R_{\mu \nu} &\equiv \mathsf R^{(1)}_{\mu \nu} = \mathsf R^\rho{}_{\mu \rho \nu} \, , &
  \mathsf R^{(2)}_{\mu \nu} &= \mathsf R_\mu{}^\rho{}_{\nu \rho} \, , &
  \mathsf R^{(3)}_{\mu \nu} &= \mathsf R^\rho{}_{\rho \mu \nu} \, ,
\end{align}
where $\mathsf R_{\mu \nu} =\mathsf R^{(1)}_{\mu \nu}$ corresponds to the usual Ricci tensor defined in Riemannian geometry. The tensors above are not all independent as they are related by the Bianchi identity \eqref{bianchi-1}
\begin{equation}
   \mathsf R^{(1)}_{\mu \nu} - \mathsf R^{(1)}_{\nu \mu} - \mathsf
  R^{(3)}_{\mu \nu} = 3 e_a^\nu \mathsf D_{[\mu} T_{\nu \rho]}{}^a \, ,
\end{equation}
where $\mathsf D_\mu$ denotes the tangent space covariant derivative for an arbitrary spin connection \eqref{gen-cd}. In particular, if the torsion is zero one has $\mathsf R^{(3)}_{\mu \nu} = 2 \mathsf R^{(1)}_{[\mu \nu]}$.

For a metric connection the curvature tensor is also antisymmetric in the first two indices and from the traces above only one is independent and this coincides with the Ricci tensor defined in Riemannian geometry $R_{\mu \nu} = \mathsf R^{(1)}_{\mu \nu} = \mathsf R^{(2)}_{\mu \nu}$ with $\mathsf R^{(3)}_{\mu \nu}=0$.
For this case one can prove the following combinatorial relations\footnote{For the general case one can always split the Riemann tensor in its symmetric and antisymmetric piece in the first indices and find similar relations.}
\begin{equation}
  \label{Rcomb}
  \begin{aligned}
    & R_{[\mu \nu \rho]\sigma} - R_{\sigma [\rho \mu \nu]} =
    2 R_{[\mu \nu \rho \sigma]} \, ,\\
    & R_{\mu \nu \rho \sigma} - R_{\rho \sigma \mu \nu} =
    3R_{[\mu \nu \rho] \sigma} - 3 R_{\rho [\sigma \mu \nu]} = \frac32 \left(
      R_{\mu [\nu \rho \sigma]} - R_{\nu [\mu \rho \sigma]} + R_{\sigma [\rho \mu \nu]} -  R_{\rho [\sigma \mu \nu]} \right) \, .
  \end{aligned}
\end{equation}
Contracting with the metric the last relation we find for the Ricci scalar
\begin{equation}
  R_{\nu \rho} - R_{\rho \nu} = 3 R^\mu{}_{[\rho \mu \nu]} \, .
\end{equation}
These relations are particularly useful in relation to the Bianchi identity \eqref{bianchi-1} in order to find a basis of independent terms when we deal with a connection with torsion. For example, in this case (excluding self-contractions) there are only three terms quadratic in the Riemann tensor which are independent. They can be chosen to be
\begin{equation}
  \label{Riemann2}
  R_{\rho \sigma \mu \nu} R^{\rho \sigma \mu \nu} \, , ~~ R_{\rho \sigma \mu \nu} R^{\rho \nu \sigma \mu} \, , ~~ R_{\rho \sigma \mu \nu} R^{\mu \nu \rho \sigma } \, .
\end{equation}
Considering also traces of the Riemann tensor we find three more independent terms
\begin{equation}
  \label{Ricci2}
  R^2 \, , ~~ R_{\mu \nu} R^{\mu \nu} \, , ~~ R_{\mu \nu} R^{\nu \mu} \, .
\end{equation}

Rather than working with the full Riemann tensor, sometimes it is convenient to work with the Weyl tensor which, for metric connections was already defined in \eqref{weyl-tensor}. For a general affine connection, a completely traceless tensor can be defined as
\begin{align}
  &\mathsf W_{\rho  \sigma \mu \nu }= \mathsf R_{\rho \sigma \mu \nu} - \mathsf R_{(\rho \sigma) \mu \nu}+\frac{1}{(d-1)(d-2)} \mathsf R (g_{\mu \rho} g_{\nu \sigma} - g_{\mu \sigma} g_{\nu \rho}) \label{real-weyl}\\
  &+\frac1{2(d-2)} \left[\left(\mathsf R^{(1)}_{\sigma \mu} +
    \mathsf R^{(2)}_{\sigma \mu}\right)g_{\nu \rho} -
    \left(\mathsf R^{(1)}_{\rho \mu}
    + \mathsf R^{(2)}_{\rho \mu}\right)g_{\nu \sigma}
    + \left(\mathsf R^{(1)}_{\rho \nu}
    + \mathsf R^{(2)}_{\rho \nu}\right)g_{\mu \sigma}
    - \left(\mathsf R^{(1)}_{\sigma \nu}
    + \mathsf R^{(2)}_{\sigma \nu}\right)g_{\mu \rho} \right] \, . \nonumber 
\end{align}
Indeed, it can be easily checked that all the traces of the above tensor vanish
\begin{align}
\mathsf W^\rho{}_{\mu \rho \nu} & = 0\, , & \mathsf W_\mu{}^\rho{}_{\nu \rho} & = 0\, , & \mathsf W^\rho{}_{\rho \mu \nu} & = 0\, , & \mathsf W_{\mu \nu}{}^\rho{}_\rho & = 0 \, .
\end{align} 
For general non-metricity (and torsion) the tensor above can be further decomposed in such a way as to be symmetric under the exchange of the first pair of indices with the last pair. One can define the new symmetric tensor as the mean $\tfrac12 \left( \mathsf W_{\rho \sigma \mu \nu }  + \mathsf W_{ \mu \nu \rho \sigma }\right) $.
Moreover, in the case of a metric connection the definition \eqref{real-weyl} reduces to the usual one \eqref{weyl-tensor}.
Finally, in the case of Weyl non-metricity it is easy to show that one has
\begin{equation}
\tilde W_{\rho \sigma \mu \nu } = C_{\rho \sigma \mu \nu} = \mathring C_{\rho \sigma \mu \nu} \, .
\end{equation}
All the above formalism can be translated to the tangent space using the vielbein or its inverse. One can
define a ``spin'' connection $\genO_\mu{}^a{}_b$ associated to
$\genG_{\mu \nu}^\rho$ via the vielbein postulate
\begin{equation}
  \partial_\mu e_\nu^a + \genO_\mu{}^a{}_b\, e_\nu^b - \genG^\rho_{\mu \nu}\,
  e_\rho^a = 0 \, .
  \label{postulate}
\end{equation}
Notice that $\genO_\mu{}^{a}{}_b$ defined above is not, in general,
antisymmetric in the tangent space indices
$\genO_\mu{}^{ab} \neq \genO_\mu{}^{ba}$ signaling that fact that it is valued in $gl(d)$ and not in $so(1,d-1)$. As a consequence, some care is
required when manipulating tangent space derivatives since
\begin{equation}
\label{gen-cd}
  \begin{split}
    \mathsf D_\mu V^a &=\partial_\mu V^a + \genO_\mu{}^a{}_b\, V^b \, ,\\
    \mathsf D_\mu V_a &= \partial_\mu V_a - V_b\, \genO_\mu{}^b{}_a  \, .
  \end{split}
\end{equation}
The spin connection can also be decomposed in terms of a Levi-Civita
part $\mathring{\Om}_\mu{}^a{}_b$ and a part related to the distortion
via the vielbein
\begin{equation}
  \genO_{\mu}{}^a{}_b = \mathring{\Om}_\mu{}^a{}_b + N_{\mu \nu}{}^\rho e^\nu_b e_\rho^a =
  \mathring{\Om}_\mu{}^a{}_b + S_{\mu \nu}{}^\rho  e^\nu_b e_\rho^a
  - K_\mu{}_\nu{}^\rho  e^\nu_b e_\rho^a\, . \label{general-spin}
\end{equation}
Metricity on the tangent space ({\it i.e.} $D_\mu \eta_{ab} = 0$) is
equivalent to antisymmetry of the spin connection
$\genO_\mu{}^{ab} = - \genO_{\mu}{}^{ba}$ (hence it is valued in the
Lorentz algebra).

\subsection{Non-metricity}
\label{A2}
Non metricity is the failure of the connection to preserve the
metric. It is measured by the non-metricity tensor $Q_{\mu \nu \rho}$
\begin{equation}
  Q_{\mu \nu \rho} = \tilde \nabla_\mu g_{\nu \rho} \equiv Q_{\mu \rho \nu} \, . \label{non-metricity}
\end{equation}
Note that our definition for the non-metricity tensor is with all indices down. When raising indices care must be taken as $Q^{\mu \nu \rho} \neq \tilde \nabla^\mu g^{\nu \rho}$.

There is a unique (torsion-free) connection with non-metricity $Q_{\mu \nu \rho}$. It is given by
\begin{equation}
\label{nm-connection}
  \tilde \Gamma_{\mu \nu}^\rho = \mathring{\Gamma}_{\mu \nu}^\rho
  -\frac12 g^{\rho \sigma} \big( Q_{\mu \nu \sigma} + Q_{\nu \sigma \mu}
  - Q_{\sigma \mu \nu} \big) \, ,
\end{equation}
where $\mathring{\Gamma}_{\mu \nu}^\rho$ is the Levi-Civita connection.
Comparing to \eqref{affine} and \eqref{disformation} we find that 
the disformation tensor can be expressed in terms of the non-metricity tensor as
\begin{align}
  \label{SQ}
  S_{\mu \nu}{}^\rho &= -\frac12 g^{\rho \sigma} \big( Q_{\mu \nu \sigma}
  + Q_{\nu \sigma \mu} - Q_{\sigma \mu \nu} \big) \; ,  
\end{align}
and, as noted before, is explicitly symmetric in the pair $(\mu , \nu)$. Conversely, one can express  the non-metricity in terms of disformation or distortion
\begin{equation}
  \label{QS}
  Q_{\mu \nu \rho} = -2 S_{\mu (\nu \rho)} = -2 N_{\mu(\nu \rho)} \, .
\end{equation}

The non-metricity tensor is symmetric in its last indices while the first index
is unconstrained and therefore the number of independent degrees of freedom encoded
in arbitrary dimensions is
\begin{equation}
  N_Q = d \times \frac{d(d+1)}2 = \frac{d^2(d+1)}2 \, .
\end{equation}
It is useful to understand how this tensor decomposes under the $so(1,d-1)$ algebra.
Clearly we can define two independent traces which are going to transform as vectors under $so(1,d-1)$. We choose them as
\begin{equation}
\label{nm-trace}
  \begin{aligned}
    q_\mu & = \frac1d  Q_{\mu \nu \rho} g^{\nu \rho} \, ,\\
 q'_\nu & = g^{\mu \rho} \left(Q_{\mu \nu \rho} - q_\mu g_{\nu \rho} \right) \, .
  \end{aligned}
\end{equation}
Another irreducible component (with respect to the Lorentz group) is the traceless totally symmetric piece defined by
\begin{equation}
  \mathcal{Q}_{\mu \nu \rho} = Q_{(\mu \nu \rho)} -q_{(\mu}g_{\nu \rho)}
  -\frac{2}{d+2} q'_{(\mu} g_{\nu \rho)} \, . \label{symmQ} 
\end{equation}
The remainder defines the last irreducible component, which can be written in terms of a traceless rank 3 tensor
with two antisymmetric indices
\begin{equation}
  \label{tildeQ}
  \tilde Q_{\mu \nu \rho} = \chi^Q_{\mu (\nu \rho)}  \, ,
\end{equation}
where $\chi^Q_{\mu \nu \rho} =- \chi^Q_{\nu \mu \rho}$,
$\ g^{\mu \rho} \chi^Q_{\mu \nu \rho}=0$ and
$\chi^Q_{[\mu \nu \rho]} = 0$. Explicitly, the three tensor
$\chi^Q_{\mu \nu \rho}$ can be also written as
\begin{equation}
  \chi_{\mu \nu \rho}^Q = \frac43 \left(Q_{[\mu \nu]\rho} - q_{[\mu} g_{\nu] \rho}
    +\frac{1}{d-1} q'_{[\mu}g_{\nu]\rho} \right) \, .
\end{equation}
After putting everything together, one arrives at the following decomposition (valid in any number of
dimensions)
\begin{equation}\label{dec1}
  Q_{\mu \nu \rho} = q_\mu g_{\nu \rho} + \frac{2d}{(d-1)(d+2)}
  \left( g_{\mu(\nu} q'_{\rho)} - \frac1d q'_\mu g_{\nu \rho} \right)
  + \tilde Q_{\mu \nu \rho} + \mathcal{Q}_{\mu \nu \rho} \, ,
\end{equation}
which match precisely the irreducible decompositions in
\cite{McCrea, Hehl:1994ue} written in the language of forms. It is not
difficult to compute the number of degrees of freedom for each of
these components and the result is summarised in Table \ref{nonmdof}.
\begin{table}[h]
  \centering
  \begin{tabular}{cccccc}
    \# of dimensions & $q_\mu$ & $q'_\mu$ & $\tilde Q_{\mu \nu \rho}$ &
$\mathcal{Q}_{\mu \nu \rho}$ & Total  \\[2mm]
    \hline
    & & & & & \\
    d & d & d & $\frac{d(d-2)(d+2)}3 $ &  $\frac{d(d-1)(d+4)}6 $
                                 & $\frac{d^2(d+1)}2$ \\[2mm]
    \hline
    & & & & & \\
    4 & 4 & 4 & 16 & 16 & 40 \\
  \end{tabular}
  \caption{Non-metricity degrees of freedom.}   \label{nonmdof}
\end{table}

\noindent
Finally, we also display the decomposition of the disformation
  tensor in terms of the components of non-metricity
\begin{equation}
  S_{\mu \nu \rho} = \frac12 g_{\mu \nu} q_\rho- q_{(\mu} g_{\nu) \rho}
  + \frac1{(d-1)(d+2)} \left[ 2q'_{(\mu} g_{\nu) \rho}
    - (d+1) g_{\mu \nu} q'_\rho\right]+ \tilde Q_{\rho \mu \nu}
  - \frac12 \mathcal{Q}_{\mu \nu \rho} \, .
\end{equation}

Due to the presence of two traces in $Q_{\mu \nu \rho}$, the
decomposition of the non-metricity tensor is not unique as one may
choose any independent linear combinations of the two traces as the
vector components of the non-metricity. However, $\tilde Q_{\mu \nu \rho}$
and $\mathcal{Q}_{\mu \nu \rho}$ remain unchanged.

\subsection{Torsion}

\label{A3}

The torsion tensor is defined as the antisymmetric part of the affine connection and can be expressed in terms of distortion or contorsion via
\begin{equation}
  T_{\mu \nu}{}^\rho = 2 \Gamma_{[\mu \nu]}^\rho = 2N_{[\mu \nu]}{}^\rho = -2K_{[\mu \nu]}{}^\rho \, . \label{torsion}
\end{equation}
It is easy to see that the identity above can be inverted for the contorsion tensor in order to obtain
\begin{equation}
  \label{ctr}
  K_{\mu \nu \rho} = -\frac12 \left(T_{\mu \nu \rho} + T_{\rho \mu \nu}
   - T_{\nu \rho \mu} \right) \, ,
\end{equation}
which is antisymmetric in its last two indices translating
  into the fact that a connection with torsion preserves the metric.
Therefore, a general connection with torsion $T$ (equivalently a metric connection) can be written as
\begin{equation}
\label{torsion-connection}
  \Gamma_{\mu \nu}^\rho = \mathring{\Gamma}_{\mu \nu}^\rho
  + \frac12 \left( T_{\mu \nu}{}^\rho  + T^\rho{}_{\mu \nu}
    - T_\nu{}^\rho{}_\mu \right) \, .
\end{equation}

Taking into account the antisymmetry in its first two indices, it follows that the number of
independent degrees of freedom for the torsion tensor is
\begin{equation}
  N_T = d \times \frac{d(d-1)}2 = \frac{d^2(d-1)}2 \, .
\end{equation}

As for the non-metricity, we can decompose the torsion in $so(1,d-1)$
components, with the difference that now the decomposition is unique.
We can first single out the (only) trace
\begin{equation}
\label{torsion-trace}
  t_\mu = -\frac1{d-1} T_{\mu \nu \rho} g^{\nu \rho} \, .
\end{equation}
Then, another obvious component is the totally antisymmetric piece of torsion which we denote by
\begin{equation}
  \Theta_{\mu \nu \rho} = T_{[\mu \nu \rho]} \, ,   \label{antiT}
\end{equation}
and, finally, after subtracting the trace and the totally antisymmetric part one is left with a traceless two-form (with zero completely antisymmetric part)
\begin{align}
  \tilde T_{\mu \nu \rho} &= \chi^T_{\mu \nu \rho} \, , \label{tildeT}
\end{align}
thus satisfying $\chi^T_{\mu \nu \rho} = -\chi^T_{\nu \mu \rho}$, $\chi^T_{\mu \nu \rho} g^{\nu \rho} = 0$ and $\chi^T_{[\mu \nu \rho]}=0$.
By putting together the above, we can write the decomposition of a general torsion as
\begin{equation}\label{dec2}
  T_{\mu \nu \rho} = -2 t_{[\mu} g_{\nu] \rho} + \tilde T_{\mu \nu \rho}
  + \Theta_{\mu \nu \rho}  \, .
\end{equation}
The number of degrees of freedom for torsion in arbitrary dimension and the particular case of four dimensions are summarised in the Table \ref{torsiondof}.
\begin{table}[h]
  \centering
  \begin{tabular}{ccccc}
    \# of dimensions & $t_\mu$ & $\tilde T_{\mu \nu \rho}$ & $\Theta_{\mu \nu \rho}$ & Total\\[2mm]
    \hline
    && & &\\
    d & d & $\frac{d(d-2)(d+2)}3$ & $\frac{d(d-1)(d-2)}6$ & $\frac{d^2(d-1)}2$\\[2mm]
    \hline
    &&&& \\
    4 & 4 & 16 & 4 &24 
  \end{tabular}
  \caption{Torsion degrees of freedom.} \label{torsiondof}
\end{table}

\noindent
Finally, we also display the decomposition of the contorsion
  tensor in terms of the components of torsion
\begin{equation}
  K_{\mu \nu \rho}= 2 g_{\mu [\nu} t_{\rho]} + \tilde T_{\nu \rho \mu}
  - \frac12 \Theta_{\mu \nu \rho} \, ,
\end{equation} 
which is useful in the main part of the paper.
\

\

\subsection{Glossary of symbols and notations}
\label{glossary}
{\bf Riemannian objects}\\[5pt]
 $\mathring \Gamma_{\mu \nu}^\rho$: the Levi-Civita connection defined in \eqref{Levi-Civita}.\\
 $\mathring \nabla_\mu$: covariant derivative associated to the Levi-Civita connection.\\
 $\mathring R^\rho{}_\sigma{}_{\mu \nu}$: Riemannian curvature tensor defined by the general formula \eqref{riemann-affine} with $\genG = \mathring \Gamma$.\\
 $\mathring R_{\mu \nu}$: Ricci tensor of Riemannian geometry $\mathring R_{\mu \nu}:= \mathring R^\rho{}_{\mu \rho \nu}$.\\
$\mathring R$: Ricci scalar of Riemannian geometry $\mathring R : = g^{\mu \nu} R_{\mu \nu}$.\\
$\mathring C_{\mu \nu \rho \sigma}$: Weyl tensor of Riemannian geometry.\\
$\mathring \Om_\mu{}^{ab}$: spin connection associated to the Levi-Civita connection (see \eqref{spinLC}).\\
$\mathring D_\mu$: tangent space covariant derivative defined by  $\mathring \Om_\mu{}^{ab}$. 
\flushleft{\bf (Metric) Torsion objects}\\[5pt]
$ \Gamma_{\mu \nu}^\rho$: metric connection with torsion (see \eqref{affine-1} for Weyl gravity).\\
 $ \nabla_\mu$: covariant derivative associated to the metric torsion connection $\Gamma$.\\
 $ R^\rho{}_\sigma{}_{\mu \nu}$: curvature tensor (with torsion) defined by the general formula \eqref{riemann-affine} with $\genG =\Gamma$.\\
 $ R_{\mu \nu}$: Ricci tensor associated to a curvature tensor with torsion $ R_{\mu \nu}:= R^\rho{}_{\mu \rho \nu}$.\\
$R$: Ricci scalar associated to a curvature with torsion $ R : = g^{\mu \nu} R_{\mu \nu}$.\\
$C_{\mu \nu \rho \sigma}$: Weyl tensor associated to a curvature tensor with torsion defined in \eqref{weyl-tensor}.\\
$\Om_\mu{}^{ab}$: spin connection associated to the torsion connection (see \eqref{spin-invariant}).\\
$ D_\mu$: tangent space covariant derivative defined by  $ \Om_\mu{}^{ab}$. 
\flushleft{\bf (Torsionless) Non-metric objects}\\[5pt]
$ \tilde \Gamma_{\mu \nu}^\rho$: torsionless connection with non-metricity (see \eqref{affine-1} for Weyl gravity).\\
 $ \tilde \nabla_\mu$: covariant derivative associated to the torsionless non-metric connection $\tilde \Gamma$.\\
 $ \tilde R^\rho{}_\sigma{}_{\mu \nu}$: curvature tensor (with non-metricity) defined by the general formula \eqref{riemann-affine} with $\genG =\tilde \Gamma$.\\
 $ \tilde R_{\mu \nu}$: Ricci tensor associated to a curvature tensor with non-metricity $ \tilde R_{\mu \nu}:= \tilde R^\rho{}_{\mu \rho \nu}$.\\
$\tilde R$: Ricci scalar associated to a curvature with non-metricity $ \tilde R : = g^{\mu \nu} \tilde R_{\mu \nu}$.\\
$\tilde C_{\mu \nu \rho \sigma}$: Weyl tensor associated to a curvature tensor with non-metricity (it is defined by a similar formula as \eqref{weyl-tensor} but with the non-metric curvatures $\tilde R, etc.$).\\
$\tilde \Om_\mu{}^{ab}$: spin connection associated to the non-metric connection (see \eqref{tildeom}).\\
$ \tilde D_\mu$: tangent space covariant derivative defined by  $ \tilde\Om_\mu{}^{ab}$. 
\flushleft{\bf Gauge covariant derivatives}\\[5pt]
$\hat \nabla_\mu$: space-time gauge covariant derivative (see \eqref{general-formula} for Weyl gravity). {\it N.B.} it is not implemented by an affine connection! Its explicit expression depends, in general, on the gauge symmetry considered (Poincar\'e, Weyl, conformal, etc.).\\
$\hat D_\mu$: tangent space gauge covariant derivative (see \eqref{Dhat} for Weyl gravity and \eqref{hat-nabla} for the relation to the space-time counterpart). \\
$\mathcal D_\mu$: tangent space gauge covariant derivative including gauged translations used only to obtain the (gravity) Bianchi identities (of Poincar\'e, Weyl and conformal) from the standard gauge theory formula \eqref{gauge-bianchi}.
\flushleft{\bf General affine objects}\\[5pt]
$ \genG_{\mu \nu}^\rho$: general affine connection with both torsion and non-metricity.\\
$ \nabla_\mu$: covariant derivative associated to the general affine connection $\genG$ used only in the Appendix \eqref{covd}. Throughout the text the same symbol is used for a torsion covariant derivative.  \\
 $ \mathsf R^\rho{}_\sigma{}_{\mu \nu}$: curvature tensor with torsion and non-metricity defined by the general formula \eqref{riemann-affine}.\\
 $ \mathsf R_{\mu \nu}$: Ricci tensor associated to a curvature tensor $ \mathsf R_{\mu \nu}:= \mathsf R^\rho{}_{\mu \rho \nu}$.\\
$\mathsf R$: Ricci scalar associated to a curvature with torsion $ \mathsf R : = g^{\mu \nu} \mathsf R_{\mu \nu}$.\\
$\genO_\mu{}^{ab}$: spin connection associated to a general affine connection (see \eqref{general-spin}).\\
$ \mathsf D_\mu$: tangent space covariant derivative defined by  $ \genO_\mu{}^{ab}$. 

\def\theequation{B-\arabic{equation}}
 \setcounter{equation}{0}
\def\thefigure{B-\arabic{figure}}

\begin{center}
  --------------------------------
\end{center}

\noindent
{\bf Acknowledgments: } We are very grateful to D. Ghilencea for a highly enjoyable collaboration and for many useful discussions and suggestions which had a significant impact on the final form of our paper. This work was supported by Project Nucleu PN 23210101/2023.

\end{document}